\documentclass[fleqn,10pt]{wlscirep}
\usepackage[utf8]{inputenc}
\usepackage[T1]{fontenc}
\usepackage{amsmath}
\usepackage{url}
\usepackage[dvipsnames]{xcolor}
\usepackage{graphicx}
\usepackage{subcaption}
\newcommand{\rev}[1]{#1} %
\newcommand{\revaug}[1]{#1}
\newcommand{\revsep}[1]{#1} %
\newcommand{\revproof}[1]{#1} %

\newcommand {\slr}[1]{\ensuremath{{#1}_{\odot}}}
\newcommand {\dif}[3][]{\frac{d^{#1}#2}{d#3^{#1}}}
\newcommand {\pdif}[3][]{\frac{\partial^{#1}#2}{\partial#3^{#1}}}
\makeatletter
\def\mart{\@ifnextchar[{\mart@@}{\mart@}}
\def\mart@@[#1]#2{\sqrt[#1]{\mathstrut{#2}}}
\def\mart@#1{\sqrt{\mathstrut{#1}}}
\makeatother

\title{\revaug{Automated Optimization of Solar Energetic Particle Acceleration and Transport Parameters Against Five-Spacecraft Observations of the 9 October 2021 Event}}

\author[1,*]{Hiroyuki Tako Ishikawa}
\author[1]{Naho Fujita}
\author[1,2]{Yuta Kato}
\author[1]{Miki Kurihara}
\author[3]{Takashi Minoshima}
\author[2,4]{Daikou Shiota}
\author[2]{Kazumasa Iwai}
\author[5]{Marco Pinto}
\author[5,6]{Ant\'{o}nio Pessanha Gomes}
\author[7]{Go Murakami}
\author[2]{Kanya Kusano}
\author[1,2]{Chihiro Mitsuda}

\affil[1]{Space Data Frontiers Research Center, Fujitsu Research, Fujitsu Limited, Kanagawa, Japan}
\affil[2]{Institute for Space--Earth Environmental Research, Nagoya University, Aichi, Japan}
\affil[3]{Center for Mathematical Science and Advanced Technology, Japan Agency for Marine-Earth Science and Technology, Kanagawa, Japan}
\affil[4]{Radio Research Institute, National Institute of Information and Communications Technology, Tokyo, Japan}
\affil[5]{Laboratory of Instrumentation and Experimental Particle Physics, Lisboa, Portugal}
\affil[6]{Instituto Superior T\'{e}cnico, Universidade de Lisboa, Lisboa, Portugal}
\affil[7]{Institute of Space and Astronautical Science, Japan Aerospace Exploration Agency, Kanagawa, Japan}

\affil[*]{ishikawa.h.tako@fujitsu.com}


\begin{abstract}
Solar energetic particle (SEP) events threaten space infrastructure, yet physics-based prediction models depend on numerous poorly constrained free parameters.
We introduce the Fujitsu Solar Energetic particle Intelligent Tuning and Evaluation Network (SEITEN), which couples the \texttt{SUSANOO} MHD model of CME propagation to \revaug{\texttt{SEPIF}, a} diffusive shock acceleration and focused transport model, \revaug{through the shock conditions \texttt{SUSANOO} supplies}, and automatically optimizes the free parameters of the particle model against the SEP event of 9 October 2021.
The mean free path and the injection efficiency vary independently along each magnetic field line and heliolongitudinal bin, giving a 19-dimensional problem solved by quasi-Monte Carlo sampling followed by a Tree-structured Parzen Estimator.
The optimized model reproduces proton flux profiles at five spacecraft with a \revaug{mean absolute error} of \revsep{$\approx 0.45$} \revproof{in $\log_{10}$ flux}, and the inferred mean free paths show a clear east--west structure\revaug{; these are effective parameters conditional on the assumed transport geometry}.
Six independent runs show that whether mean free paths or injection efficiencies dominate depends on the MHD shock strength, and that injection efficiencies differ by orders of magnitude between MHD configurations despite nearly identical fit quality.
\revaug{Joint MHD--SEP parameter estimation is therefore a necessary next step} towards generalized, operational SEP forecasting.
\end{abstract}

\begin{document}

\flushbottom
\maketitle
\thispagestyle{empty}

\section*{Introduction} \label{sec:intro}

Solar energetic particle (SEP) events present a significant radiation hazard to aviation, spaceborne electronics, and human space exploration, particularly for missions venturing beyond Earth's magnetosphere such as the near-future expeditions to the Moon and Mars\rev{~\cite{2016LRSP...13....3D}}. 
The more human society relies on space infrastructure, the more essential space weather forecasting becomes, \revaug{particularly the accurate prediction of} the onset, peak intensity, and duration of SEP events. 
While empirical models and machine learning approaches offer rapid forecasts\rev{ \cite{2023AdSpR..72.5161W, 2025SSRv..221...82P}}, physics-based understanding is still essential for the complex chain of mechanisms, from particle acceleration at solar flares and coronal mass ejection (CME)-driven shocks to transport through the heliosphere, and to predict the full spatial and temporal distribution of particle fluxes \revaug{in an event of a severity that modern technological society has not yet faced, but eventually will}.

In recent years, sophisticated physics-based modelling chains that couple magnetohydrodynamic (MHD) simulations of the solar corona and solar wind with particle transport codes, such as PARADISE\rev{~\cite{2020PhDT.........8W, 2022A&A...659A.187W}} coupled with EUHFORIA \revaug{\cite{2018JSWSC...8A..35P, 2019A&A...624A..47W}}, iPATH~\revproof{\cite{2017JGRA..12210938H, 2018ApJ...854L..19H, 2020RAA....20..145D}} coupled with AWSoM \revaug{\cite{2021ApJ...919..146L}}, M-FLAMPA \cite{2004ApJ...616L.171S}\rev{ operating within the SOFIE model coupled with AWSoM-R~\cite{Zhao2024sofie}, and EPREM~\cite{Schwadron2010eprem} within the EMMREM framework}, have been developed.
These models solve the focused transport equation to simulate particle transport processes including pitch-angle diffusion, cross-field diffusion, and adiabatic cooling.
\revaug{The same transport model has since been coupled to a different MHD background: the iPATH acceleration and transport modules have been driven by EUHFORIA to model individual multi-spacecraft events, energetic storm particle events observed by Solar Orbiter, and, with its successor model, inverse velocity dispersion signatures~\cite{2022A&A...668A..71D, 2024A&A...681A..92D, 2025A&A...696A.199D}. Such pairings of one particle transport model with alternative MHD backgrounds bear directly on the present study, in which we quantify how much the inferred SEP parameters depend on the MHD background rather than on the transport physics.}
\rev{More recently, forecasting-oriented variants have also emerged, such as PARASOL~\cite{2025JSWSC..15....3A}, which augments the PARADISE transport model with a semi-analytical foreshock description derived from self-consistent acceleration simulations.}
\rev{Complementary observation-driven frameworks, such as SPREAdFAST~\cite{2017JSWSC...7A..32K, Kozarev2022spreadfast}, characterize CME-driven coronal shocks directly from extreme-ultraviolet imaging and couple them to particle acceleration and transport models to forecast SEP fluxes in the inner heliosphere.}
However, a fundamental challenge remains in the operational application of these models. 
They rely on numerous free parameters that are often hard to constrain by observations. 
In particular, the properties of the seed particle population and the transport coefficients significantly influence the resulting SEP time profiles\rev{~\cite{2016LRSP...13....3D}}. For instance, slight variations in the assumed injection energy or scattering mean free path can lead to orders-of-magnitude differences in predicted fluxes. 
Consequently, current physics-based simulations often require ad-hoc manual tuning of these parameters to reproduce past events observed near Earth, hindering their generalization for predictive purposes.

The SEP event on 9 October 2021 serves as an ideal testbed for addressing these challenges, as it was detected by spacecraft at five distinct locations in the inner heliosphere: BepiColombo\rev{~\cite{Benkhoff2021}}, Solar Orbiter\rev{~\cite{Muller2020}}, Parker Solar Probe (PSP)\rev{~\cite{Fox2016}}, STEREO-A\rev{~\cite{Kaiser2008}}, and SOHO (near-Earth)\rev{~\cite{Domingo1995}}.
These spacecraft span heliocentric distances from $\sim$0.3 to 1~au and cover approximately $50^{\circ}$ in heliolongitude (Fig.~\ref{fig:spacecraft_config}(a)), providing a uniquely broad multi-point view of the same event.

Palmerio et al.\cite{2024JSWSC..14....3P} successfully modelled the onset of this event by coupling the WSA-Enlil MHD model with the SEPMOD test-particle code.
They addressed the limitation of missing early-stage acceleration in MHD models by introducing a ``fixed source'' injection at the inner boundary, scaled empirically by flare and CME parameters. 
However, \revaug{the spatial form of these fixed sources is prescribed rather than inferred: the injection is applied over a circular patch centred on the mapped flare site or on the CME nose, with an axisymmetric profile whose width is set by the patch radius. Such a prescription cannot represent the non-uniformity of the actual injection region along the shock front.} 
Furthermore, the determination of injection parameters in such models generally relies on empirical scalings rather than optimization based on the physical transport processes themselves.

In this study, we introduce Fujitsu Solar Energetic particle Intelligent Tuning and Evaluation Network (SEITEN), a new framework that overcomes the ambiguity of free parameters in physics-based SEP modelling by applying automated parameter optimization.
The acronym is also a transliteration of the Japanese word \textit{seiten} \rev{(pronounced ``say-ten'')}, meaning ``clear sky''\rev{, reflecting the dual aspiration of this framework: to clear up the behaviour of SEPs, and to help secure the all-clear, the state in which they pose no threat.}
SEITEN couples the global MHD solar wind and CME propagation model, \texttt{SUSANOO} \cite{2014SpWea..12..187S, 2016SpWea..14...56S}, with \rev{the Solar Energetic Particle Injection and Focused transport model (\texttt{SEPIF}),} based on diffusive shock acceleration (DSA) and the focused transport equation \cite{2026EP&S...78...64M}. \revaug{The coupling is one-way and acts through the shock source conditions: \texttt{SUSANOO} supplies the time-dependent properties of the CME-driven shock, which define where and how strongly particles are injected, while the interplanetary transport itself is solved on analytic Parker-spiral field lines rather than on the MHD solution (see the Model Coupling subsection of the Methods for the scope of the coupling and its consequences).}
Unlike scatter-free approaches, our model explicitly accounts for pitch-angle scattering and momentum-dependent transport along the magnetic field lines.
We apply SEITEN to the 9 October 2021 event, utilizing multi-spacecraft observations to constrain the model.
\rev{Throughout this study, we focus on the proton component of the SEP event.}
A key innovation of our approach is the implementation of a two-stage optimization algorithm, combining quasi-Monte Carlo (QMC) and the Tree-structured Parzen Estimator (TPE), a class of Bayesian optimization, to search the high-dimensional parameter space efficiently and robustly.
This allows us to introduce spatial refinement into key parameters rather than assuming uniform values. 
The parallel mean free path coefficient ($\xi$) is determined independently for each magnetic field line, while the injection efficiency ($\epsilon$) is treated as a variable for each heliolongitudinal bin. \revaug{The optimization acts on these SEP-model parameters only; the CME and solar wind inputs to the MHD model are held fixed, and the two MHD configurations examined here are compared rather than optimized.} %
By minimizing the error between simulated and observed proton fluxes across five widely separated spacecraft, we aim both to reproduce the event accurately and to constrain the injection and transport parameters as effective quantities conditional on the assumed transport geometry, providing a pathway towards more generalized initial condition settings for future operational forecasting.

\rev{The remainder of this paper is organized as follows. We first describe the 9 October 2021 event and the multi-spacecraft observations used to constrain the model. We then present the optimization \revproof{results}, covering the multi-point flux reproduction, the convergence behaviour, the best-fit parameter values, and a multi-run reproducibility analysis. In the Discussion, we interpret the optimized mean free path and injection efficiency parameters physically, compare our approach with previous multi-spacecraft modelling of this event, and outline prospects for operational forecasting. The main conclusions are summarized thereafter. To keep the narrative focused on the results, the full description of the models (\texttt{SUSANOO}, \texttt{SEPIF}, and their coupling) and of the parameter optimization procedure is provided in the Methods section at the end of the paper, and implementation details are collected in the Supplementary Information.}

\section*{Event Overview and Observations} \label{sec:event}

\subsection*{Target SEP Event: 9 October 2021} \label{subsec:event}
The SEP event analysed in this work was triggered by an eruption that commenced on 9 October 2021 at approximately 06:30~UT.
The eruption originated from NOAA active region (AR) 12882,
located at N18E08 in Stonyhurst coordinates at the time of the event.
It was associated with an M1.6-class flare observed by the
Geostationary Operational Environmental Satellite (GOES),
with a start time of 06:19~UT, peak time of 06:38~UT, and end time of 06:53~UT,
as well as a CME and an extreme ultraviolet (EUV) wave.
\revaug{The CME formed a full halo in SOHO/LASCO despite its near-central-meridian
source, yet subtended only $\sim130^\circ$ from STEREO-A ($\sim39^\circ$ east
of Earth), a combination that constrains its propagation to lie close to the
Sun--Earth line and hence identifies the eruption as
Earth-directed\cite{2024JSWSC..14....3P}.}
\revaug{A Graduated Cylindrical Shell (GCS)\cite{Thernisien2011} reconstruction of this CME
gives a nose direction of $\theta = 9^\circ$ in latitude and $\phi = 5^\circ$ in Stonyhurst
longitude, a speed of 770~km~s$^{-1}$, half-angular widths
$(R_\mathrm{max}, R_\mathrm{min}) = (49^\circ, 28^\circ)$, and an axis tilt of
$\gamma = 40^\circ$ taken positive counterclockwise from solar west\cite{2024JSWSC..14....3P}.}

The SEP event was detected at five well-separated locations in the inner heliosphere:
STEREO-A, the Solar and Heliospheric Observatory (SOHO) near Earth,
BepiColombo, PSP, and Solar Orbiter (SolO).
At the time of the event, these spacecraft were located at heliocentric distances of
approximately 0.33~au (BepiColombo), 0.68~au (SolO), 0.77~au (PSP), 0.96~au (STEREO-A), and 1.0~au (SOHO), respectively.
For a detailed description of this event and the associated observations, the reader is referred to Palmerio et al.\cite{2024JSWSC..14....3P}.
\revaug{The large-scale interplanetary context of this event, including a stream interaction region that preceded the CME-driven shock at the observers, was analysed by Lario et al.\cite{2022ApJ...934...55L}, and Wijsen et al.\cite{2023ApJ...950..172W} modelled the SEP event with \texttt{EUHFORIA} and \texttt{PARADISE} including the effect of that structure on the shock.}

\begin{figure}[htbp]
\centering
\begin{subfigure}[b]{0.40\linewidth}
\includegraphics[width=\linewidth]{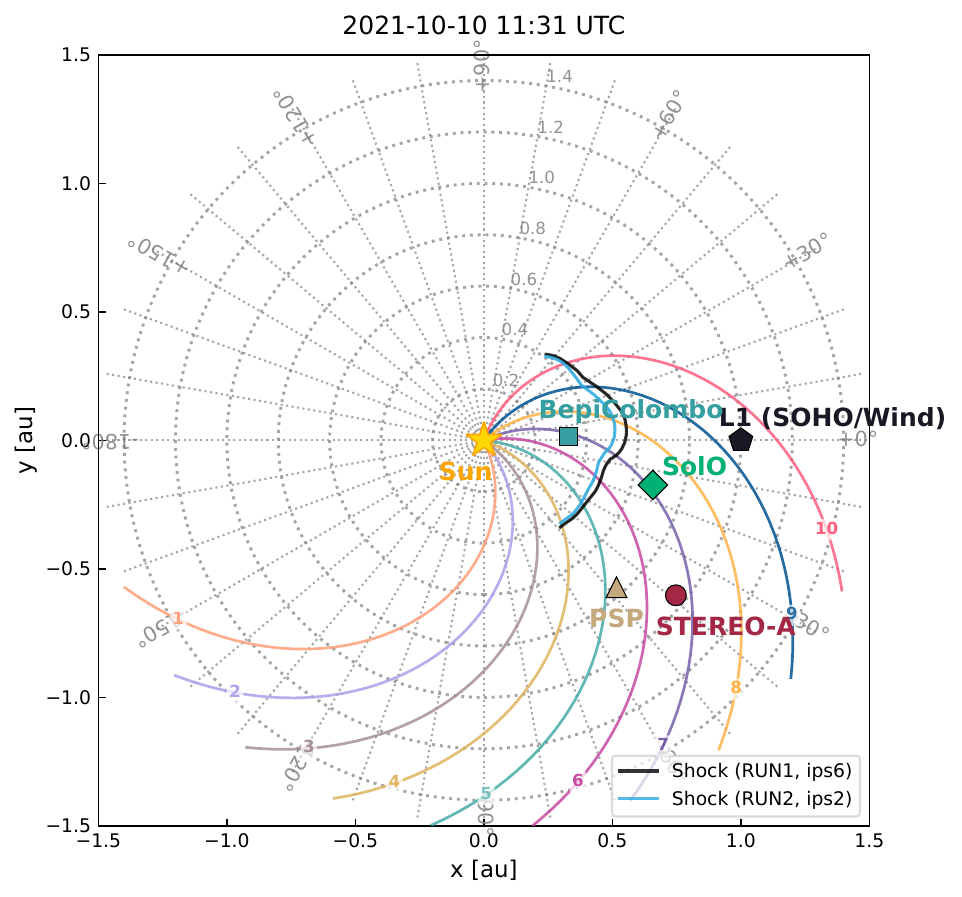}
\caption{}
\label{fig:spacecraft_config_a}
\end{subfigure}\hfill
\begin{subfigure}[b]{0.58\linewidth}
\includegraphics[width=\linewidth]{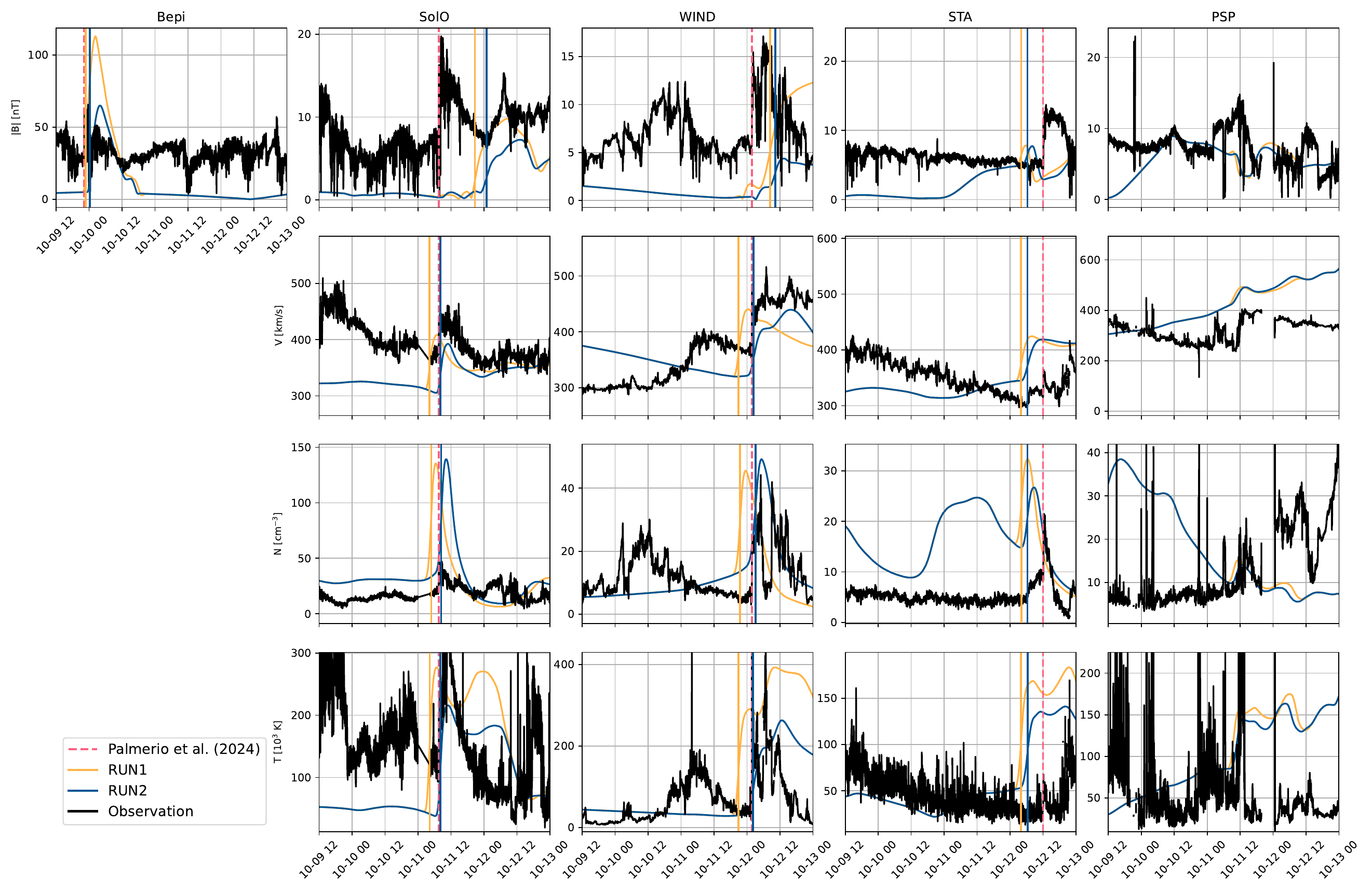}
\caption{}
\label{fig:spacecraft_config_b}
\end{subfigure}
\caption{%
\rev{(a) }Configuration of the five observing spacecraft and the CME-driven shock in the ecliptic plane at 10~October 2021 $\sim$11:30~UT, viewed from the solar north pole.
The Sun is at the origin; dotted concentric circles mark heliocentric distances at 0.2~au intervals, and dotted radial lines indicate heliolongitudes at $10^\circ$ intervals.
The 10 coloured curves are the Parker-spiral magnetic field lines used in the \texttt{SEPIF} simulation, with footpoint longitudes $\phi_j(0) \in [-82.5^\circ, +52.5^\circ]$ and solar wind speed $V_{sw} = 400$~km~s$^{-1}$; \rev{the labels $j = 1, \dots, 10$ index the mean free path coefficients $\xi_j$. The three easternmost lines ($\xi_1$--$\xi_3$; footpoints far from all spacecraft, hence negligible influence on the simulated fluxes) are held fixed at $10^6$, and the remaining seven ($\xi_4$--$\xi_{10}$) are free.
The CME-driven shock surfaces are overplotted for the two MHD configurations (black: RUN1; sky-blue: RUN2); they remain close along the Sun--Earth direction (RUN1 slightly ahead), and the resulting CME arrival times at Earth differ by several hours \rev{(see Methods)}.}
The five markers indicate the locations at which the SEP observations analysed here were made: STEREO-A, \revaug{L1 (SOHO/\revproof{WIND})}, BepiColombo, Parker Solar Probe (PSP), and Solar Orbiter (SolO).
\rev{(b) Simulated solar-wind plasma from the two \texttt{SUSANOO} MHD configurations, RUN1 (orange) and RUN2 (blue), compared with in-situ observations (black). Columns (left to right): BepiColombo (Bepi), Solar Orbiter (SolO), WIND, STEREO-A (STA), and PSP; rows (top to bottom): magnetic field magnitude $|\boldsymbol{B}|$ [nT], solar-wind speed $|v|$ [km\,s$^{-1}$], proton number density $N$ [cm$^{-3}$], and proton temperature $T$ [$10^3$\,K]. Pink dashed vertical lines mark the reference shock arrival times reported by Palmerio et al.\cite{2024JSWSC..14....3P}; orange and blue solid vertical lines mark the peak of each simulated speed profile (RUN1 and RUN2, respectively). Plasma data are unavailable at Bepi during the cruise phase, and no identifiable CME signal is present at PSP in either the observations or the simulations. We note that at the L1 point (near-Earth), the SEP data are obtained from SOHO COSTEP/EPHIN, while the magnetic field and plasma data are obtained from WIND (MFI and SWE), following Palmerio et al.\cite{2024JSWSC..14....3P}.}
}
\label{fig:spacecraft_config}
\end{figure}

\subsection*{Data from \revaug{Spacecraft}} \label{subsec:data}
\subsubsection*{\revproof{Solar Terrestrial Relations Observatory Ahead (STEREO-A)}} \label{subsubsec:stereo}
The Solar Terrestrial Relations Observatory Ahead\cite{Kaiser2008} is a spacecraft placed in a heliocentric orbit
slightly interior to that of Earth, located approximately \revaug{39}$^\circ$ east of Earth
and at about 1~au from the Sun at the time of the 9 October 2021 event.
STEREO-A was originally launched as part of a two-spacecraft mission designed
to provide stereoscopic imaging of CMEs and in-situ measurements of the
solar wind and energetic particles.

For the present analysis, we employ the following instruments onboard STEREO-A.
Energetic particle observations are obtained from two instruments within the \revaug{In situ Measurements of Particles And CME Transients (IMPACT)} \cite{Luhmann2008} suite:
the Low Energy Telescope \revproof{(LET)}\cite{Mewaldt2008},
which \revaug{provides proton measurements at a few MeV},
and the High Energy Telescope \revproof{(HET)}\cite{vonRosenvinge2008},
which extends the energy coverage to higher energies and provides proton
measurements at tens of MeV.

\revaug{Magnetic field data are obtained from the Magnetic Field Experiment
(MFE)~\cite{Acuna2008}, which is part of the IMPACT~\cite{Luhmann2008} suite.
Solar wind plasma data are provided by the Plasma and Suprathermal Ion Composition
investigation (PLASTIC)~\cite{Galvin2008}.}

\subsubsection*{\revproof{Solar and Heliospheric Observatory (SOHO)}}
The Solar and Heliospheric Observatory\cite{Domingo1995}
is a spacecraft jointly operated by ESA and NASA,
positioned at the Sun--Earth Lagrange L1 point approximately 1~au from the Sun.
In addition to its well-known role in solar imaging and helioseismology,
SOHO carries a suite of in-situ particle instruments that have provided
continuous monitoring of the near-Earth SEP environment since its launch in 1995.

For this study, energetic particle data are obtained from the
Comprehensive Suprathermal and Energetic Particle Analyser
COSTEP\cite{MullerMellin1995} instrument.
COSTEP measures the intensities of protons and helium ions
over a broad energy range (approximately 0.04--53~\revaug{MeV} for protons),
enabling the characterization of both the onset and the peak phase of SEP events.
In addition, coronagraph imagery from the C3 telescope of the Large Angle and
Spectrometric Coronagraph \revproof{(LASCO)}\cite{Brueckner1995} onboard SOHO
is used to observe the CME propagation through the solar corona
from the Sun--Earth line perspective.

\subsubsection*{BepiColombo}
BepiColombo \cite{Benkhoff2021} is a joint ESA/JAXA mission
to Mercury, launched in October 2018.
At the time of the 9 October 2021 event, the spacecraft was in its cruise phase
at a heliocentric distance of approximately 0.33~au,
making it the closest observer to the Sun among the five spacecraft
considered in this study.
Its longitudinal separation from Earth was approximately 2$^\circ$.

Energetic particle measurements are obtained from the
BepiColombo Environment Radiation Monitor
BERM\cite{Pinto2022}, which monitors the radiation environment
encountered by the spacecraft during its journey to Mercury.
BERM measures protons between $\sim$1.5 and $\sim$100~MeV. 
During this event, BERM observed protons in three energy bins. We use Proton Bin~1 ($\sim$1.5--5.9~MeV) and Bin~3 ($\sim$5.9--9.1~MeV), whose instrument response functions are relatively clean and well characterized.
Proton Bin~2, whose nominal energy range lies between those of Bin~1 and Bin~3, is excluded because its response function has a significant low-energy tail that complicates the assignment of an effective energy, and because it is associated with the largest calibration uncertainties among the three bins \cite{Pinto2022}.
\revproof{Gomes et al.\cite{Gomes2026} describe the calibration of the BERM bins used here, including the bow-tie determination of their effective energies and geometric factors as well as the cross-calibration against the Solar Intensity X-ray and particle Spectrometer (SIXS).}
During cruise, BERM is mounted on the MPO radiator panel, which by thermal design points away from the Sun. As a consequence, BERM samples a restricted range of pitch angles dominated by particles arriving from the anti-sunward hemisphere (i.e., backscattered or already isotropized populations), and is comparatively less sensitive to the field-aligned beam streaming outward from the Sun along the interplanetary magnetic field \cite{Pinto2022}. This look-direction effect should be borne in mind when comparing the measured intensities to model predictions, particularly during phases when the particle distribution is strongly anisotropic.

\revaug{Magnetic field data are provided by the Mercury Planetary Orbiter Magnetometer
(MPO-MAG)~\cite{Heyner2021}, a fluxgate magnetometer designed to measure
the magnetic field vector at the spacecraft location.
Plasma instruments are not operational during the cruise phase, so only the magnetic field magnitude $|B|$ is used in Fig.~\ref{fig:spacecraft_config}(b).}

\subsubsection*{Parker Solar Probe (PSP)}
Parker Solar Probe\cite{Fox2016} is a NASA mission designed to
study the solar corona and inner heliosphere at unprecedentedly close distances to the Sun.
At the time of the 9 October 2021 event, PSP was located at approximately 0.77~au
from the Sun, with a longitudinal separation of about 48$^\circ$ east of Earth,
making it the most longitudinally separated observer from the Earth--Sun line
among the five spacecraft considered here.

Energetic particle data are provided by the Integrated Science Investigation
of the Sun IS$\odot$IS\cite{McComas2016} suite, which measures energetic
electrons and ions over a wide energy range and is designed to study
particle acceleration and transport in the near-Sun environment.
\revaug{For the present analysis we use the Level-2 60-second proton rate product of the High Energy Telescope (HET) of the Energetic Particle Instrument--High Energy (EPI-Hi) within IS$\odot$IS. Only the A telescope is used; no averaging or combination over telescope look directions is applied. The three channels employed are listed in Supplementary Table~S3.}

\revaug{Magnetic field measurements are provided by the FIELDS instrument suite
\cite{Bale2016}, which characterizes the electric and magnetic fields,
plasma waves, and radio emissions in the inner heliosphere.
Solar wind plasma data are obtained from the Solar Probe Cup
(SPC)~\cite{Case2020}, which is part of the Solar Wind Electrons Alphas
and Protons (SWEAP)~\cite{Kasper2016} investigation.}

\subsubsection*{Solar Orbiter (SolO)}
Solar Orbiter\cite{Muller2020} is a joint ESA/NASA mission
designed to investigate the Sun and inner heliosphere from a combination
of in-situ and remote-sensing instruments.
At the time of the 9 October 2021 event, SolO was located at
approximately 0.68~au from the Sun,
with a longitudinal separation of about 15$^\circ$ east of Earth.

Energetic particle measurements are obtained from the
Energetic Particle Detector \revproof{(EPD)}\cite{RodriguezPacheco2020},
a suite of four sensors that together cover a broad energy range
from suprathermal to high-energy particles,
including electrons and ions such as protons, helium, and heavier species.
\revaug{Of these sensors, the Electron Proton Telescope (EPT) provides the lowest of the four proton channels used here and the High Energy Telescope (HET) provides the remaining three. In both cases the sunward-pointing telescope (viewing direction \texttt{sun}) is used.}

\revaug{Magnetic field data are obtained from the Magnetometer
(MAG)~\cite{Horbury2020} onboard SolO,
which provides high-resolution measurements of the interplanetary
magnetic field vector.
Solar wind plasma data are provided by the Proton and Alpha Sensor (PAS),
which is part of the Solar Wind Analyser (SWA)~\cite{Owen2020} suite.}

\subsubsection*{\revaug{WIND}}
\revaug{The WIND spacecraft~\cite{OgilvieDesch1997} is a NASA mission launched on
1 November 1994 as part of the Global Geospace Science (GGS) initiative
within the International Solar--Terrestrial Physics (ISTP) programme.
After an initial phase in a series of near-Earth and lunar-swingby orbits,
it was placed in a halo orbit about the Sun--Earth L1 Lagrange point, where
it has provided nearly continuous in-situ monitoring of the solar wind
upstream of the Earth's magnetosphere for almost three decades.
At the time of the 9 October 2021 event, WIND was located near L1, at a
longitudinal separation of essentially $0^\circ$ from Earth and about 1~au
from the Sun, making it the near-Earth reference point for the
time-of-arrival comparison in this study.}

\revaug{In Fig.~\ref{fig:spacecraft_config}(b) we use the following instruments onboard WIND.
The magnetic field magnitude is obtained from the Magnetic Field
Investigation (MFI)~\cite{Lepping1995}, a pair of fluxgate magnetometers
that measure the interplanetary magnetic field vector.
Solar wind plasma data (the solar-wind speed, proton number density, and
proton temperature) are provided by the Solar Wind Experiment
(SWE)~\cite{Ogilvie1995}, which measures the velocity distribution of
solar-wind thermal ions.}

\section*{Results} \label{sec:results}
Because the MHD background directly influences the shock geometry and, consequently, the SEP transport solution, we examine the sensitivity of the optimized results to the choice of \texttt{SUSANOO} initial conditions.
Two shock configurations are considered, each selected by a different criterion from a set of candidate runs (see the \texttt{SUSANOO} model description in Methods):
\begin{itemize}
    \item \textbf{RUN1}: the configuration that best reproduces the interplanetary scintillation (IPS) observations, and
    \item \textbf{RUN2}: the configuration that best reproduces the CME arrival time at Earth.
\end{itemize}
A detailed study of the MHD parameter space is beyond the scope of this work; rather, we report the optimization results for both configurations to quantify how much the residual MHD uncertainty propagates into the SEP parameter estimates.
For each configuration, an independent 1480-trial optimization (1000 QMC + 480 TPE) was carried out on the 19-parameter model described in the \rev{Methods (Parameter Optimization Method subsection)}, using all five spacecraft simultaneously.
\revaug{Three such sessions were run for each configuration, differing only in the random seed, and the resulting six-session ensemble is analysed in the Multi-Run Reproducibility subsection. The single session shown in Figure~\ref{fig:best_fit} and Table~\ref{tab:best_params} is the first session run for each configuration, reported as a representative rather than as the best-scoring member of the ensemble; \revsep{the lowest objective values attained across the ensemble are 0.431 for RUN1 and 0.436 for RUN2.} The parameter-level conclusions of this work are drawn from the ensemble statistics rather than from any individual trial.}

\subsection*{Multi-Point Flux Reproduction} \label{subsec:flux_reproduction}

\begin{figure}[ht]
\centering
\begin{subfigure}[t]{0.49\linewidth}
\includegraphics[width=\linewidth,height=0.82\textheight,keepaspectratio]{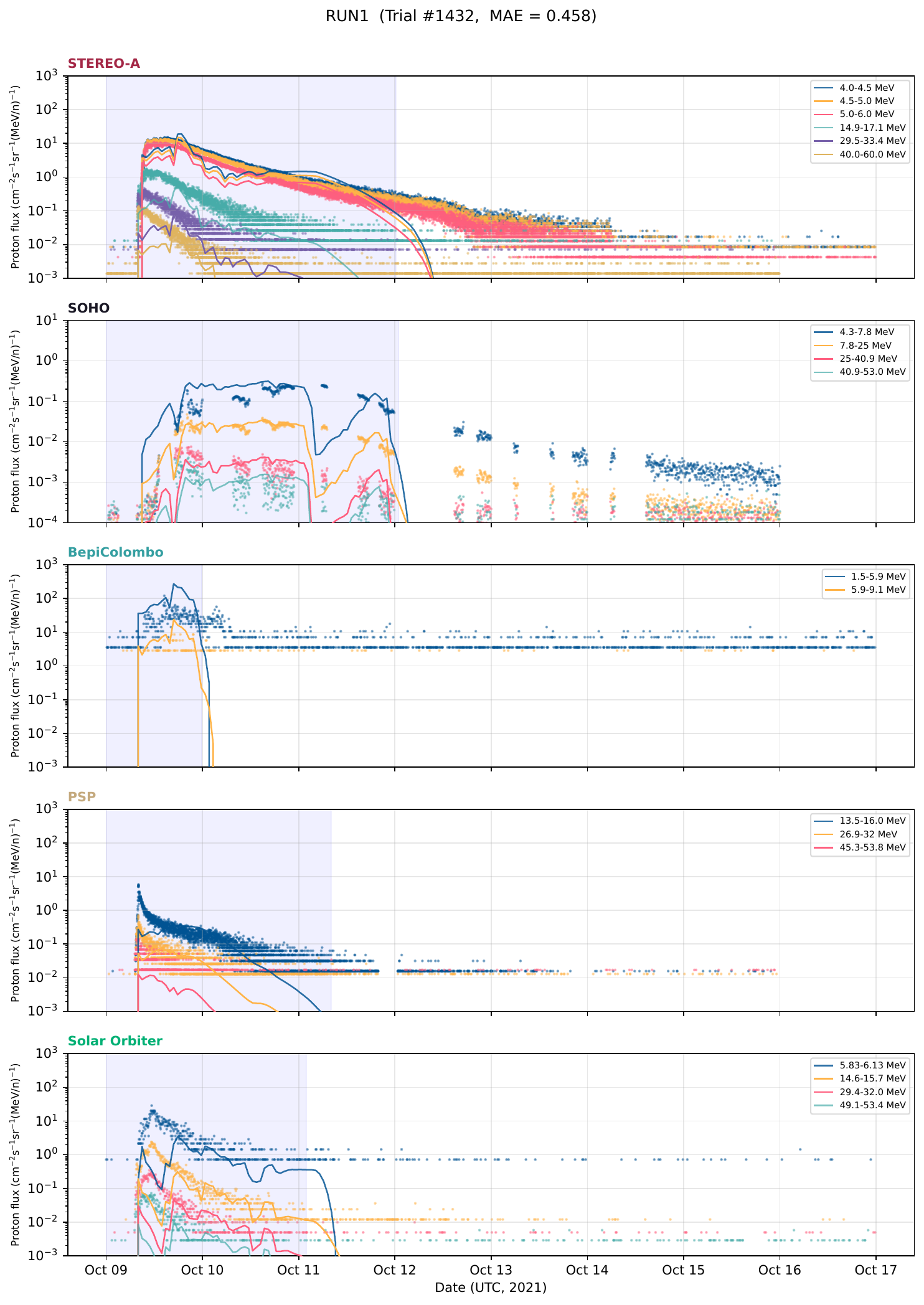}
\caption{RUN1}
\label{fig:best_ips6}
\end{subfigure}\hfill
\begin{subfigure}[t]{0.49\linewidth}
\includegraphics[width=\linewidth,height=0.82\textheight,keepaspectratio]{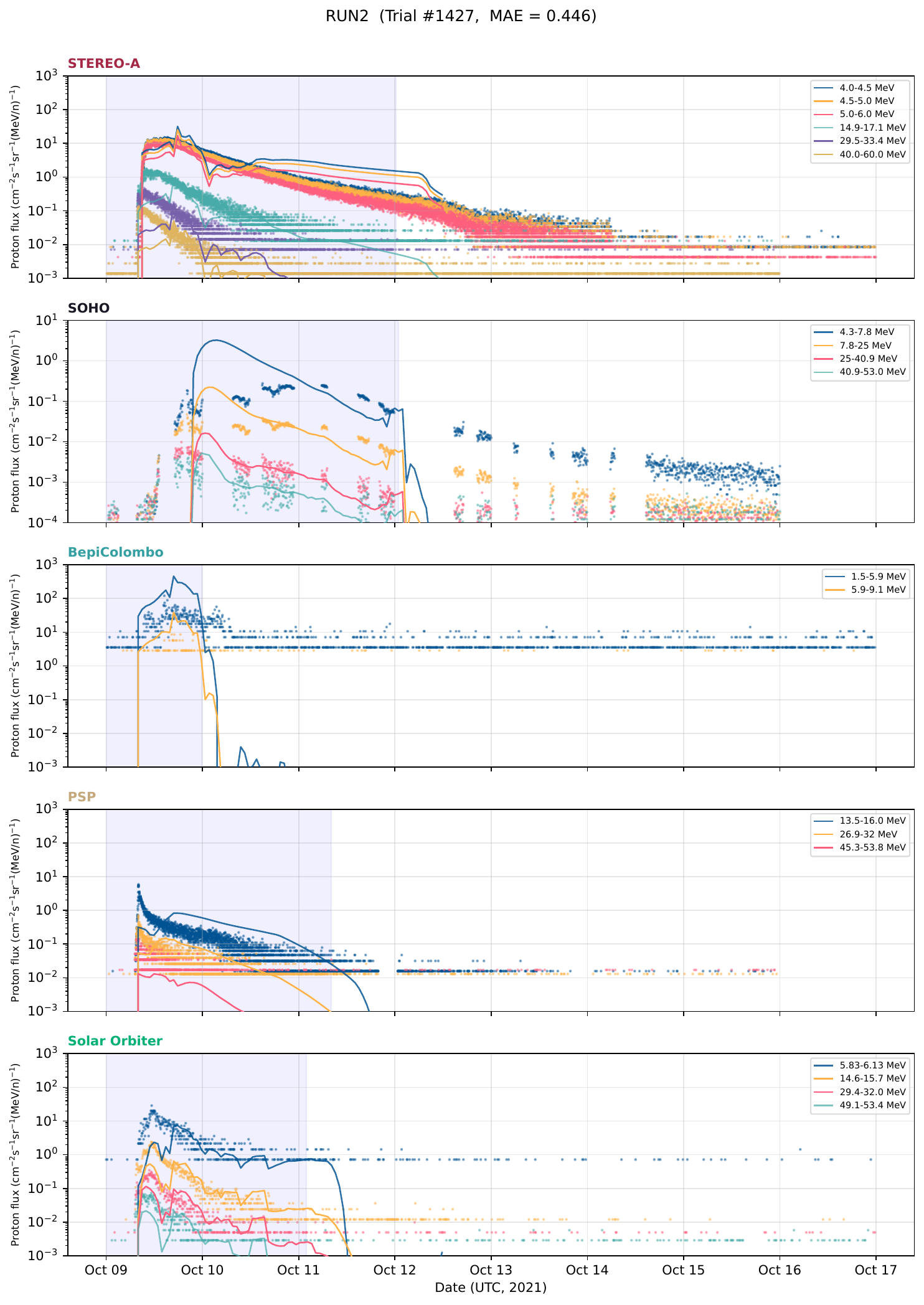}
\caption{RUN2}
\label{fig:best_ips2}
\end{subfigure}
\caption{\rev{Best-fit simulated (solid curves) and observed (data points) proton differential intensity time profiles for (a) the RUN1 MHD configuration (trial \#1432, \revaug{objective-function} MAE $= \revsep{0.458}$) and (b) the RUN2 MHD configuration (trial \#1427, \revaug{objective-function} MAE $= \revsep{0.446}$).} From top to bottom, the panels show STEREO-A, SOHO (near-Earth), BepiColombo, Parker Solar Probe, and Solar Orbiter. Different colours correspond to different energy channels as listed in \rev{Supplementary Table~S3}. The blue shaded regions indicate the \revaug{observation-selection windows} \rev{(Supplementary Table~S3)}. \revaug{Only data inside the shaded windows contribute to the objective function; the remainder of each panel, in particular the post-shock decay phase, is shown for diagnostic purposes only.}
\label{fig:best_fit}}
\end{figure}

Figure~\ref{fig:best_fit}\rev{(a) and (b)} show the best-fit simulated differential intensity time profiles overlaid on the observations for the RUN1 and RUN2 configurations, respectively.
Each panel displays the proton differential intensity as a function of time for a single spacecraft, with different colours representing different energy channels.
Solid curves denote the simulation and data points with error bars denote the observations.
The shaded blue bands mark the \revaug{observation-selection windows}, listed in \rev{Supplementary Table~S3}.

\revaug{The value of the objective function (equation~(\ref{eq:mae_final})) for the best trial}, hereafter simply the MAE, is \revsep{0.458} for RUN1 (trial \#1432 out of 1480 valid trials) and \revsep{0.446} for RUN2 (trial \#1427 out of \revaug{1480} valid trials), corresponding to a typical simulation-observation discrepancy of a factor of \revsep{$\sim$2.9} in differential intensity. \revproof{(Unless stated otherwise, objective values quoted in this paper incorporate a correction to one Solar Orbiter energy channel, described in the Objective Function subsection of the Methods.)}
The two configurations thus yield virtually identical overall quality of fit, despite employing different MHD backgrounds.
\revaug{The objective function is a weighted average of the five per-spacecraft MAE values, with unit weight for STEREO-A, SOHO, PSP and Solar Orbiter and a weight of 0.7 for BepiColombo; the rationale for this weighting, and the effect of removing it, are given in the Objective Function subsection of the Methods.}

\revsep{The simulated intensities are integrated over pitch angle, whereas most of the instruments used here resolve the direction of arrival. For STEREO-A the LET and HET products used are direction-integrated, and the COSTEP-EPHIN and BERM measurements are single-aperture, so no alternative direction is available. In the case of BERM the aperture views the anti-sunward hemisphere during cruise, so the simulated pitch-angle integrated intensity is not strictly the same physical quantity as the measurement, a caveat that applies particularly during the anisotropic rising phase. For Solar Orbiter the sunward EPT and HET telescopes were used, and for PSP the A telescope of EPI-Hi HET. The difference between viewing directions is modest over the intervals that enter the objective, the median absolute difference in $\log_{10}$ intensity between the telescope used and the alternatives being below 0.18 for Solar Orbiter and between 0.13 and 0.30 for PSP. It remains below 0.31 for Solar Orbiter during the rising phase, but reaches 0.75--1.67 at PSP over the few samples concerned there. Averaging the available telescopes and re-scoring the best-fit solutions changes the objective function by $-0.018$ (RUN1) and $-0.012$ (RUN2), that is, the agreement improves, as expected if the simulated pitch-angle integrated intensity is the appropriate comparison. We note that the mean over the available telescopes is not a true omnidirectional intensity, since the angular coverage is neither complete nor uniform between them.}
Both configurations reproduce the gross features of the event at all five spacecraft:
the rapid rise at STEREO-A and SOHO, the early onset at Parker Solar Probe, the moderate peak at Solar Orbiter, and the short-duration pulse at BepiColombo.
The simulated peak intensities generally agree with the observations to within a factor of $\sim$3--5 across the energy range 1.5--60~MeV.

Notable discrepancies are, however, apparent in both configurations.
First, the decay phase after the peak is reproduced too steeply in the simulations, particularly at STEREO-A and SOHO, where the observed flux decreases more slowly than the model prediction beyond the shock passage, resulting in the model underestimating the flux during this phase.
As discussed in the Temporal Evolution of the Mean Free Path subsection, this behaviour is likely related to the time-independent mean free path formulation and possible contributions from corotating interaction regions.
Second, the onset timing at PSP and SolO in the RUN2 configuration appears to be better aligned with observations than in the RUN1 case, consistent with the different shock arrival profiles of the two MHD backgrounds.
\rev{At \revproof{SOHO and STEREO-A}, the RUN1 configuration, which was selected for its agreement with the IPS observations rather than the Earth arrival time, drives a faster CME whose shock reaches Earth several hours \emph{earlier} than observed (the discrepancy noted in the \revproof{Event Overview and Observations}), producing a correspondingly early simulated SEP onset there. The RUN2 configuration instead matches the observed bulk shock arrival time at Earth; both configurations leave residual differences in the simulated SEP onset and peak times at these \revproof{two} locations, with RUN2 in particular showing a noticeable offset in both the onset timing and the peak intensity at SOHO.}

\revaug{The late simulated onset at SOHO originates in the parallel mean free path on the field line that connects to Earth, and not in the inward extrapolation of the shock or the shock source prescription. At the time of onset, Earth is connected to the outermost western field line $j = 10$ in every session examined, so the arrival of the first particles is governed by $\xi_{10}$. We verified this causally with a controlled experiment in which $\xi_{10}$ alone was changed and all other parameters, including the MHD configuration, were held at the RUN2 best-fit values. Raising $\xi_{10}$ from 54.9 to $2.86\times10^{3}$ advanced the simulated onset at SOHO, defined as the first time the 4.3--7.8~MeV intensity exceeds $10^{-4}$, from 21:44~UT to 09:56~UT on 9 October, an advance of 11.8~hours, and reproduced the onset of an independent session that had converged on the larger value. Across all six sessions the simulated SOHO onset varies monotonically with $\xi_{10}$ and shows no corresponding dependence on $\xi_8$ or $\xi_9$, and the relation holds in both MHD configurations, which share the same inward extrapolation procedure. Physically, a large $\xi_{10}$ makes the Earth-connected field line only weakly scattering over the outer part of the path, so the leading component streams to 1~au rapidly; a small $\xi_{10}$ makes the transport there diffusive and delays that leading component. Consistently, the time at which the intensity reaches $10^{-2}$, which reflects the bulk of the event rather than its leading edge, is unchanged at 21:44~UT in both runs of the controlled experiment: the leading edge is set by parallel transport while the main increase is set by the shock and the injection. This behaviour is the direct observable counterpart of the weak constraint on $\xi_{10}$ discussed in the Longitude Profile of the Mean Free Path and Shock Obliquity subsection, and it means that the simulated onset time at SOHO should be regarded as a fitted rather than a predicted quantity in this study. That weak constraint has a specific origin in the objective function. At SOHO the evaluation mask removes the rising-phase samples at which the simulated intensity is still below the threshold, and those are precisely the samples that would otherwise constrain $\xi_{10}$; the removal is more severe in RUN2, where the earliest sample that enters the objective at SOHO lies late on 9 October. The trial dependence of the mask, discussed in the Objective Function subsection of the Methods, and the weakly determined onset described here are therefore two aspects of the same limitation rather than independent ones.}

\subsection*{Optimization Convergence} \label{subsec:convergence}

Supplementary Fig.~S1 shows the optimization history for both MHD configurations, plotting the objective function value \revaug{(the weighted MAE of equation~(\ref{eq:mae_final}))} as a function of trial number.
Both runs exhibit a rapid improvement during the initial QMC stage (trials 1--1000), where the Sobol' sequence uniformly explores the 19-dimensional parameter space.
\revsep{The course of the optimization is described here in terms of the objective as minimized during the runs, since the per-trial output needed to recompute it is not available (see the Objective Function subsection of the Methods).}
\revaug{The objective value of the very first trial is 0.92 (RUN1) and 0.84 (RUN2), and within the first 100 trials the running best falls to 0.586 (RUN1) and 0.554 (RUN2).}
By the end of the QMC stage at trial 1000, the best MAE reaches 0.495 (RUN2) and 0.491 (RUN1).

The subsequent TPE stage (trials 1001--1480) yields further improvement by exploiting the landscape learned from the QMC warmup.
\revaug{In the two sessions reported here}, the most significant improvements occur in the TPE stage, \revsep{which lowers the best MAE to 0.427 (RUN2) and 0.429 (RUN1), corresponding to the corrected values of 0.446 and 0.458 quoted above}.
Notably, the overall best trials are found late in the optimization (trial \#1427 and \#1432 out of 1480), indicating that the TPE sampler continues to discover competitive solutions even after extensive exploration.

The top 5 trials in each configuration cluster within a narrow MAE range: \revsep{0.446--0.480} for RUN2 and \revsep{0.458--0.482} for RUN1.
This tight clustering indicates that the optimizer has identified a cluster of competitive low-error solutions.
Although the coefficient of variation of the objective over the final 10 trials \revaug{(0.14 for RUN1 and 0.11 for RUN2, defined as the sample standard deviation divided by the mean)} suggests that the optimization has not yet fully converged in the strict sense, the tight clustering of the top-5 solutions and the diminishing marginal improvement indicate that any residual gain would be small relative to the systematic uncertainties inherent in the MHD background (see Discussion).

With 19 free parameters and \revaug{1480} valid trials, the trial-to-parameter ratio exceeds 75. \rev{The adequacy of the search is judged empirically, however, from the tight clustering of the top-5 solutions and the across-session reproducibility reported in the Multi-Run Reproducibility subsection.}

\subsection*{Best-Fit Parameter Values} \label{subsec:best_params}

Table~\ref{tab:best_params} summarizes the best-fit parameter values for both MHD configurations, converted from the normalized space to physical units via equation~(\ref{eq:denormalize}).
For reference, the approximate footpoint longitude of each magnetic field line and the centre longitude of each injection bin are indicated.

\paragraph{Mean free path coefficients $\xi_j$.}
The optimized $\xi$ values reveal a clear contrast between the eastern field lines ($j = 4$--6, footpoints at \revaug{$-37.5^{\circ}$ to $-7.5^{\circ}$}) and the western field lines ($j = 7$--10, footpoints at \revaug{$+7.5^{\circ}$ to $+52.5^{\circ}$}), where positive longitudes correspond to the solar west in the simulation frame.
In the RUN1 configuration, the optimizer drives the eastern field lines, to which only PSP is magnetically connected via the Parker spiral, towards $\xi \sim 10^{2\text{--}3}$ (relatively short mean free paths, implying strong scattering), whereas the western field lines connecting towards SolO, STEREO-A, BepiColombo, and SOHO converge to $\xi \sim 10^{5\text{--}6}$ (weak scattering).
\revaug{The RUN2 configuration shows the same east--west contrast on the near-nose western lines ($\xi_7, \xi_8 \sim 10^6$), but its two outermost western coefficients take much smaller single-trial values ($\xi_9 \sim 1.5 \times 10^3$ and $\xi_{10} \sim 55$, against $\sim 4 \times 10^5$ for both in RUN1). As discussed in the Multi-Run Reproducibility subsection of the Results, the simulated fluxes become insensitive to $\xi$ above a ceiling at $\xi \approx 10^4$, so single-trial differences that straddle this ceiling should be interpreted with care.}
A more robust assessment of the multi-run statistics in the Longitude Profile of the Mean Free Path and Shock Obliquity subsection shows that, after applying the ceiling clamp, both configurations yield similar median $\xi$ profiles for the western field lines.

The eastern $\xi$ values ($\xi_4$--$\xi_6$) nonetheless carry weaker observational constraints than their western counterparts.
Among the five spacecraft analysed, only PSP is magnetically connected to the eastern field lines, so each of $\xi_4$, $\xi_5$, and $\xi_6$ is constrained at most by a single spacecraft trajectory across part of the event.
The \revaug{rank-correlation} parameter importance for these eastern coefficients is correspondingly lower than for $\xi_7$ and $\xi_8$ (Fig.~\ref{fig:importance_dotplot}). \revaug{A low rank-correlation score is not by itself evidence of weak dependence, as the case of $\xi_6$ below shows; the weaker constraint on the eastern coefficients follows from the observational coverage.}
The east--west pattern reported here should therefore be interpreted as a robust qualitative feature (short mean free path on the PSP-connected eastern flank, long mean free path on the multi-spacecraft-connected western flank) rather than a quantitatively constrained longitudinal profile of $\xi$.

Overall, the optimized $\xi$ values span $\sim$4--5 orders of magnitude across the 7 active field lines in the single best-fit trial, with the eastern field lines settling at small values and the western field lines pushed to or above the sensitivity ceiling. The east-versus-west contrast in the scattering regime is reproduced consistently across all six independent sessions (see the Longitude Profile of the Mean Free Path and Shock Obliquity subsection).
\revaug{Converting the coefficients into physical units through $\lambda_\parallel = \xi\, r_g$, where $r_g$ is the local proton gyroradius (see the Free Parameters subsection of the Methods for the derivation and for the numerical conversion), the eastern field lines, to which PSP is magnetically connected, correspond at 10~MeV and 1~au to $\lambda_\parallel = 0.13$--$0.45$~au for the RUN1 best-fit values and $0.03$--$2.0$~au for RUN2. The RUN1 range and part of the broader RUN2 range overlap the ``consensus'' range of 0.08--0.3~au reported for gradual SEP events \cite{1982RvGSP..20..335P, 1994ApJ...420..294B}. The western field lines, which connect to \revproof{SOHO and STEREO-A} as well as to Solar Orbiter and BepiColombo, lie at or above the sensitivity ceiling $\xi \approx 10^4$, corresponding to $\lambda_\parallel \gtrsim 6$~au when evaluated at the same energy and heliocentric distance; the RUN2 single-trial values of $\xi_9$ and $\xi_{10}$ are the exceptions, and are poorly constrained (see the Longitude Profile of the Mean Free Path and Shock Obliquity subsection). Above that ceiling the simulated profiles no longer respond to $\xi$, so the optimization constrains only a lower bound on $\lambda_\parallel$, and the entries of Table~\ref{tab:best_params} that lie above the ceiling should be read as such. The RUN2 entries for $\xi_9$ and $\xi_{10}$ lie below it and are not lower bounds; they should be read as poorly determined.}
\revaug{Because the observations do not distinguish values of $\xi$ above the ceiling on the western field lines, a scatter-free (infinite mean free path) treatment, of the kind adopted in some operational SEP models \cite{2024JSWSC..14....3P}, is not excluded by the fit there. On the eastern side it is excluded: the fit requires scattering strong enough to bring $\lambda_\parallel$ into the consensus range. The adequacy of a globally scatter-free description is therefore longitude dependent in this event.}

\paragraph{Injection efficiency $\epsilon_{0,k}$.}
The injection efficiencies show substantial variation across the 12 active longitude bins, spanning nearly the full search range of $10^{-4}$--$3 \times 10^{-2}$.
The spatial distribution of $\epsilon_{0,k}$ differs markedly between the two MHD configurations, reflecting the different shock geometries.
As discussed in the Free Parameters subsection, the longitude-dependent $\epsilon_{0,k}$ was introduced as a diagnostic degree of freedom: if the MHD-derived shock parameters were perfectly accurate, a single global $\epsilon_0$ would suffice.
\revaug{Because $\epsilon_{0,k}$ can absorb residual error in the MHD-derived shock parameters, in the injection prescription, and in the transport physics omitted by the assumed geometry, its spatial scatter (spanning nearly the full search range and differing by up to two orders of magnitude between RUN2 and RUN1 for the same bin index) localizes where the coupled model departs from the observations at each longitude.}
Several bins in the RUN2 configuration yield $\epsilon_{0,k}$ near or above $10^{-2}$, above the range typically adopted in DSA models; the implications are taken up in the Discussion.

\begin{table}[ht]
\centering
\caption{Best-fit parameter values for the RUN1 and RUN2 MHD configurations. Mean free path coefficients ($\xi_j$) are dimensionless ratios; injection efficiencies ($\epsilon_{0,k}$) are dimensionless fractions. The approximate longitude refers to the initial footpoint at $t = 0$ of each field line (for $\xi$) or the centre of each longitudinal bin (for $\epsilon_0$) in the simulation frame. Fixed parameters ($\xi_1$--$\xi_3 = 10^6$; $\epsilon_{0,1}$--$\epsilon_{0,3}$, $\epsilon_{0,16}$--$\epsilon_{0,18} = 1.505 \times 10^{-2}$) are omitted. \label{tab:best_params}}
\begin{tabular}{lrcc}
\hline
\textbf{Parameter} & \textbf{Longitude ($^{\circ}$)} & \textbf{RUN1} & \textbf{RUN2} \\
\hline
\multicolumn{4}{c}{\textit{Mean free path coefficient} $\xi_j$ (field-line footpoint at $t=0$)} \\
\hline
$\xi_4$  & $-37.5$ & $7.4 \times 10^{2}$ & $3.3 \times 10^{3}$ \\
$\xi_5$  & $-22.5$ & $3.9 \times 10^{2}$ & $4.7 \times 10^{1}$ \\
$\xi_6$  & $-7.5$  & $2.2 \times 10^{2}$ & $1.8 \times 10^{2}$ \\
$\xi_7$  & $+7.5$  & $3.0 \times 10^{5}$ & $9.3 \times 10^{5}$ \\
$\xi_8$  & $+22.5$ & $1.8 \times 10^{5}$ & $9.8 \times 10^{5}$ \\
$\xi_9$  & $+37.5$ & $3.9 \times 10^{5}$ & $1.5 \times 10^{3}$ \\
$\xi_{10}$ & $+52.5$ & $4.4 \times 10^{5}$ & $5.5 \times 10^{1}$ \\
\hline
\multicolumn{4}{c}{\textit{Injection efficiency} $\epsilon_{0,k}$} \\
\hline
$\epsilon_{0,4}$  & $-55$ & $2.9 \times 10^{-2}$ & $1.2 \times 10^{-4}$ \\
$\epsilon_{0,5}$  & $-45$ & $1.7 \times 10^{-2}$ & $3.5 \times 10^{-3}$ \\
$\epsilon_{0,6}$  & $-35$ & $2.0 \times 10^{-3}$ & $1.3 \times 10^{-2}$ \\
$\epsilon_{0,7}$  & $-25$ & $5.4 \times 10^{-3}$ & $2.5 \times 10^{-2}$ \\
$\epsilon_{0,8}$  & $-15$ & $1.9 \times 10^{-4}$ & $7.4 \times 10^{-4}$ \\
$\epsilon_{0,9}$  & $-5$  & $8.8 \times 10^{-3}$ & $2.8 \times 10^{-2}$ \\
$\epsilon_{0,10}$ & $+5$  & $5.4 \times 10^{-3}$ & $1.8 \times 10^{-2}$ \\
$\epsilon_{0,11}$ & $+15$ & $2.4 \times 10^{-4}$ & $9.5 \times 10^{-3}$ \\
$\epsilon_{0,12}$ & $+25$ & $1.5 \times 10^{-2}$ & $1.7 \times 10^{-2}$ \\
$\epsilon_{0,13}$ & $+35$ & $1.9 \times 10^{-3}$ & $2.2 \times 10^{-2}$ \\
$\epsilon_{0,14}$ & $+45$ & $4.0 \times 10^{-3}$ & $2.7 \times 10^{-4}$ \\
$\epsilon_{0,15}$ & $+55$ & $3.2 \times 10^{-3}$ & $3.3 \times 10^{-3}$ \\
\hline
\end{tabular}
\end{table}

\paragraph{Sensitivity to the MHD background.}
The comparison between the RUN1 and RUN2 best-fit parameters reveals that, while the overall MAE is nearly identical (\revsep{$\Delta\mathrm{MAE} \approx 0.012$}), the individual transport and injection parameters can differ substantially.
For the western mean free path coefficients ($\xi_7$--$\xi_{10}$), the best-fit values appear to differ by up to four orders of magnitude between configurations (e.g., $\xi_{10}$: $4.4 \times 10^5$ in RUN1 vs.\ $5.5 \times 10^1$ in RUN2).
\revaug{However, as detailed in the Multi-Run Reproducibility subsection of the Results, the objective function discriminates poorly among the western $\xi$ values: any value $\gtrsim 10^4$ is essentially equivalent in terms of fit quality, and $\xi_{10}$ is flat across the entire search range. The apparent factor-of-thousands single-trial differences for these field lines should therefore not be interpreted as physically meaningful.}

The RUN1--RUN2 differences in the injection efficiencies $\epsilon_{0,k}$, by contrast, are not subject to such a ceiling and are \revaug{reproducible across sessions and interpretable as configuration-dependent responses of the optimizer}.
For instance, $\epsilon_{0,4}$ (the easternmost active bin at $-55^{\circ}$) differs by more than two orders of magnitude between the two configurations.
The Sensitivity of Injection Efficiency to CME Magnetic Flux subsection shows, from the six-session ensemble, that the RUN2 (slower shock) configuration requires elevated $\epsilon_{0,k}$ in three near-nose bins to compensate for the reduced shock speed and altered geometry, a physically interpretable response of the optimizer to a single-variable change in the MHD input.

We regard the RUN1--RUN2 spread, in particular for the injection efficiencies, as a lower bound on the systematic uncertainty introduced by the MHD background assumption.
This is because RUN1 and RUN2 share exactly the same ambient solar wind; only the CME magnetic flux differs.
Since recent work has shown that the ambient solar wind can influence CME propagation and the resulting shock parameters at least as strongly as the CME initial conditions \cite{2021SpWea..1902775R, 2023ApJ...950..172W, 2026arXiv260202300S}, the true systematic uncertainty in the SEP transport parameters, which would include variations in the background wind, is likely larger than the RUN1--RUN2 spread reported here.
A systematic quantification of this uncertainty through multiple independent runs per configuration is presented in the following subsection.

\subsection*{Multi-Run Reproducibility} \label{subsec:multirun}

To assess the reliability of the optimized parameter estimates and to quantify how the optimization landscape varies with the MHD background, we conducted an ensemble of six independent optimization sessions: three runs per shock configuration (RUN1 and RUN2), each initialized from a different random seed and run to \revaug{1480} valid trials.
The six sessions together provide \revaug{8880} valid trials and enable a clear separation of intra-configuration variability (run-to-run scatter for the same MHD background) from inter-configuration variability (systematic differences between RUN1 and RUN2).

\begin{figure}[htbp]
\centering
\includegraphics[width=\linewidth,height=0.85\textheight,keepaspectratio]{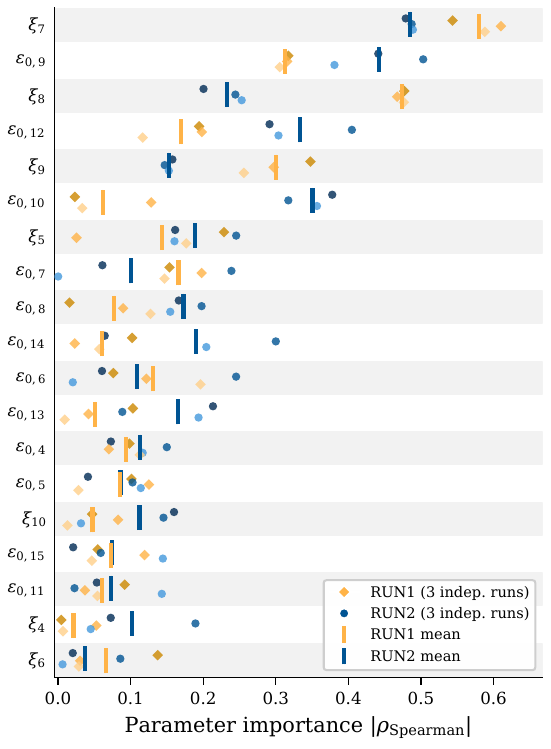}
\caption{\revaug{Rank-correlation parameter importance, $|\rho_{\mathrm{Spearman}}|$ between the parameter value and the objective function value,} for all 19 free parameters across six independent optimization sessions (three per shock configuration).
    Orange dots and bars correspond to RUN1 ($\mathit{MAGFLUX} = 1.2\times10^{21}$~Mx); blue dots and bars correspond to RUN2 ($\mathit{MAGFLUX} = 8.0\times10^{20}$~Mx).
    Bars show the \revaug{arithmetic} mean importance per configuration; dots show individual session scores.
    \revaug{The mean free path coefficients are clamped at $\xi = 10^4$ before the correlation is evaluated. The measure is not a variance decomposition and the 19 values do not sum to unity.}
    Parameters are ordered \revaug{from top to bottom} by decreasing average importance.}
\label{fig:importance_dotplot}
\end{figure}

\begin{figure}[htbp]
\centering
\begin{subfigure}[b]{\linewidth}
\centering
\includegraphics[width=\linewidth]{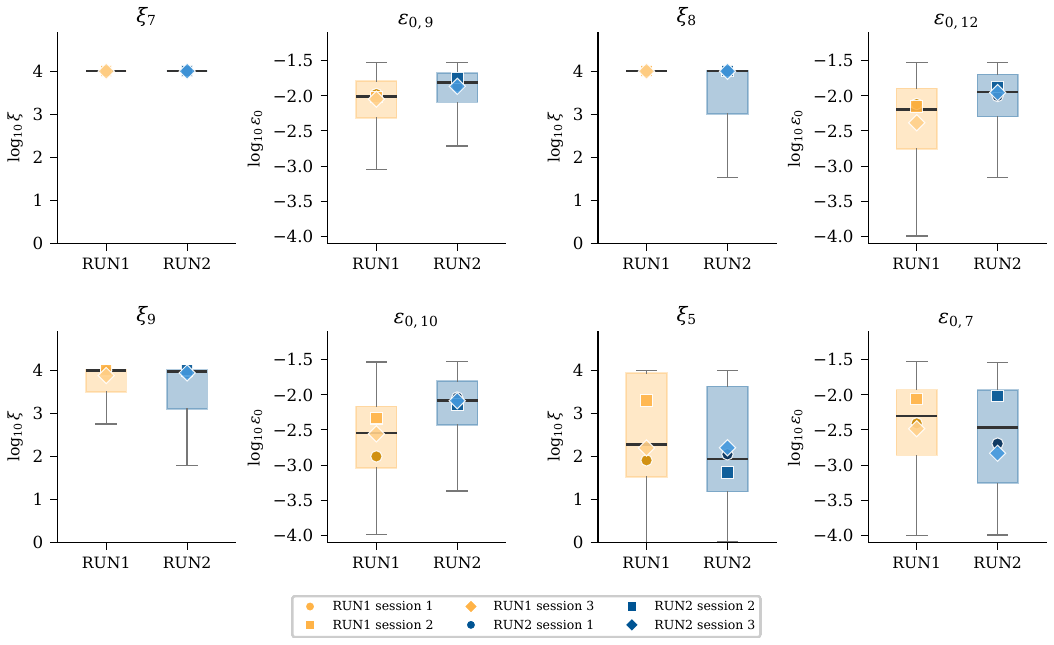}
\caption{}
\label{fig:top10pct_distributions}
\end{subfigure}

\vspace{0.6em}
\begin{subfigure}[b]{\linewidth}
\centering
\includegraphics[width=\linewidth]{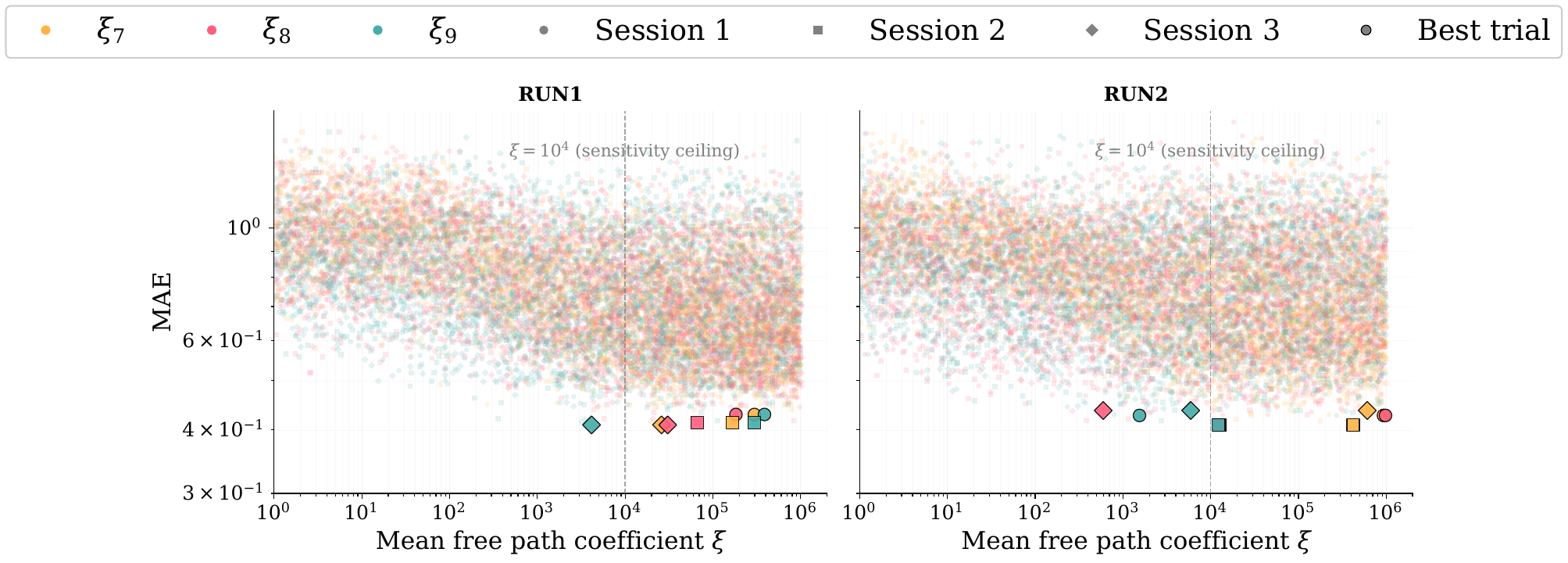}
\caption{}
\label{fig:xi_convergence}
\end{subfigure}
\caption{\rev{Parameter sensitivity across the six-session ensemble.
    (a) Distributions of the eight highest-importance parameters among the top-10\% trials} (by \revaug{objective-function} MAE) across six independent optimization sessions.
    Each panel shows a separate parameter; box-and-whisker plots span the interquartile range within a session and the horizontal line is the median.
    \rev{Individual session medians for RUN1 (orange) and RUN2 (blue) are shown as circles, squares, and diamonds for sessions~1, 2, and~3, respectively.}
    \revaug{Trials within a session are adaptively generated and are not statistically independent, so no significance test on the pooled trials is reported; the comparison between configurations rests on the three run-medians per configuration. The distributions describe the low-error region explored by the optimizer and are not parameter uncertainties.}
    \rev{(b) Scatter of all valid optimization trials in the ($\xi_j$, MAE) plane} for the three highest-importance mean free path parameters ($\xi_7$, $\xi_8$, $\xi_9$).
    Left subpanel: RUN1; right subpanel: RUN2.
    \rev{Different colours correspond to different $\xi$ parameters; different marker shapes (circles, squares, diamonds) distinguish the three independent optimization sessions.
    Large markers with black edges identify the best trial (minimum MAE) in each session.}
    The vertical dashed line at $\xi = 10^4$ marks the sensitivity ceiling above which the MAE is insensitive to the $\xi$ value.}
\label{fig:sensitivity}
\end{figure}

\paragraph{Parameter importance across sessions.}
\revaug{Figure~\ref{fig:importance_dotplot} shows a rank-correlation measure of parameter importance for all 19 free parameters across the six sessions. For each session and each parameter we evaluate the Spearman rank correlation between the parameter value and the objective function value over all valid trials of that session, and report its absolute value $|\rho_{\mathrm{Spearman}}|$. The mean free path coefficients are clamped at the sensitivity ceiling $\xi = 10^4$ (Fig.~\ref{fig:sensitivity}b) before the correlation is evaluated, so that the measure is not diluted by the region above the ceiling in which the objective does not respond to $\xi$ at all. The measure quantifies how strongly and how monotonically the objective responds to each parameter over the sampled region of parameter space; the 19 values do not sum to unity and are not fractions of the objective function variance. It cannot detect a non-monotonic dependence, a limitation we return to below.}
Individual session scores are shown as dots and the per-configuration arithmetic mean as a horizontal bar.
The ranking is strikingly consistent across the three independent runs within each configuration: the Spearman rank correlation between any two same-configuration sessions exceeds 0.44 in all pairwise combinations, and Kendall's concordance coefficient is $W = 0.716$ (RUN1) and $0.805$ (RUN2).

The dominant parameters differ systematically between the two shock configurations.
\revaug{In the RUN1 sessions the two highest-ranked parameters are the mean free path coefficients $\xi_7$ and $\xi_8$, with $|\rho_{\mathrm{Spearman}}| = 0.58$ and $0.47$ against 0.31 for the highest-ranked injection efficiency.}
\revaug{In the RUN2 sessions $\xi_7$ still ranks first (0.49), but it is followed immediately by the injection efficiencies $\epsilon_{0,9}$, $\epsilon_{0,10}$ and $\epsilon_{0,12}$ (0.44, 0.35 and 0.33), while $\xi_8$ falls to 0.23.}
The reversal in dominant parameter type is reproduced in all three independent runs of each configuration, confirming that it is a robust feature of the respective optimization landscapes and not a sampling artefact.
Only $\xi_7$ and $\epsilon_{0,9}$ appear consistently in the top-5 of both configurations.
These correspond to the field line that maintains the longest cumulative magnetic connection to the multi-spacecraft set during the event and to the longitudinal bin in which the shock--field-line intersection of those well-connected field lines repeatedly resides.
This pattern is not a property of the CME geometry per se but reflects the configuration of the observing spacecraft: the parameters most universally constrainable in a given event are those that affect the simulated fluxes at the largest number of spacecraft for the longest duration.
Different spacecraft configurations or events would therefore be expected to yield a different set of highest-importance parameters.

\paragraph{Top-10\% parameter distributions and inter-configuration differences.}
Figure~\ref{fig:top10pct_distributions} shows the distributions of the eight highest-importance parameters among the top-10\% trials (by MAE) across the six sessions.
Box-and-whisker plots summarize within-session distributions; individual run-median markers are overlaid to display intra-configuration scatter.

For $\xi_7$ and $\xi_8$, all three RUN1 run-medians converge to the upper region of the search space with coefficient of variation $\approx 0$\revaug{; the $\xi$ values entering panel (a) are clamped at the sensitivity ceiling $\xi = 10^4$, so a vanishing coefficient of variation here means that the run-medians all lie at or above the ceiling, not that they coincide at a particular finite value. It records a reproducible insensitivity of the fit to $\xi$ along these near-nose western field lines}.
For the injection efficiency parameters $\epsilon_{0,9}$, $\epsilon_{0,10}$, and $\epsilon_{0,12}$, the three RUN2 run-medians cluster within a narrow range (geometric standard deviation $< 1.4$) while the three RUN1 medians are consistently lower\revaug{, with no overlap between the RUN1 and RUN2 run-median ranges for any of the three parameters}.
\revaug{Trials within a session are generated adaptively by the TPE sampler, conditioned on the trials already evaluated, so they are neither mutually independent nor samples from a posterior distribution over the parameters, and pooling them into a two-sample test would amount to pseudo-replication. We therefore take the session, not the trial, as the unit of replication, attach no formal significance level to this difference, and describe it as a reproducible configuration-dependent shift: it has the same sign and a comparable magnitude in all three independent sessions of each configuration, and the run-median ranges of the two configurations do not overlap. For the same reason, the top-10\% trial distributions of Figure~\ref{fig:top10pct_distributions} describe the low-error region of parameter space explored by the optimizer and are not parameter uncertainties.}
Physically, the slower shock in RUN2 (lower $\mathit{MAGFLUX}$) requires the optimizer to increase the injection efficiencies near the CME nose to reproduce the observed particle fluxes; the mechanism is discussed in detail in the Sensitivity of Injection Efficiency to CME Magnetic Flux subsection.

\paragraph{Sensitivity ceiling of the mean free path coefficient.}
Figure~\ref{fig:xi_convergence} plots, for each of the three highest-importance $\xi$ parameters ($\xi_7$, $\xi_8$, $\xi_9$), every valid trial in the plane of ($\xi_j$, MAE) for RUN1 (left) and RUN2 (right).
The most prominent feature is a \emph{sensitivity ceiling} at $\xi \approx 10^4$: for $\xi_j > 10^4$, the MAE shows no systematic improvement with increasing $\xi$, and the scatter is uniformly distributed.
This plateau demonstrates that the simulated flux profiles become insensitive to the mean free path coefficient above $\sim$$10^4$\revaug{. We state this as an empirical property of the fit and do not infer from it that the transport has become scatter-free along the whole field-line path: the local mean free path in equation~(\ref{eq:mfp_rg}) scales as $1/|B(z)|$ and is therefore orders of magnitude shorter in the inner heliosphere than the value quoted at 1~au, so a single number quoted at 1~au does not characterize the whole trajectory}.
Consequently, the single best-fit values reported in Table~\ref{tab:best_params} for the western field lines (e.g., $\xi_7 = 3.0 \times 10^5$ for RUN1) should be interpreted with caution: physically, any $\xi_7 \gtrsim 10^4$ is equally consistent with the multi-spacecraft observations for that field line.
This caveat applies most strongly to $\xi_9$ and $\xi_{10}$, whose single-trial best-fit values lie far apart between RUN1 and RUN2 ($\sim 10^5$ vs.\ $\sim 10^2$). \revaug{The constraint on these two coefficients is weak over the whole search range: the objective function is flat with respect to $\xi_{10}$ throughout (Supplementary Fig.~S3), and once the ceiling clamp is applied the run-medians of both coefficients lie at or above the ceiling in both configurations. The apparent inter-configuration difference therefore reflects the weakness of the constraint rather than a physically distinct transport state.}

\subsection*{\revaug{Comparison with a Spatially Uniform Mean Free Path}} \label{subsec:collapse_test}

\revaug{The results presented so far show that the optimized $\xi_j$ vary strongly with longitude, but a spread in fitted values does not in itself establish that the spatial parameterization is required by the data. We therefore compared the 19-parameter model directly against a spatially uniform baseline on the objective function itself. Starting from the RUN1 best-fit solution, the seven free mean free path coefficients $\xi_4$--$\xi_{10}$ were replaced by a single global value $\xi_g$, the twelve injection efficiencies were held at their best-fit values, and the objective function was re-evaluated on a nine-point logarithmic grid spanning $\xi_g = 10^2$ to $10^6$, using identical evaluation windows, energy channels and mask. Two consistency checks on this re-evaluation are reported in the caption of Supplementary Table~S4.}

\revaug{The best global value, $\xi_g \approx 3.2\times10^{3}$, yields an MAE of \revsep{0.703}, against \revsep{0.458} for the 19-parameter solution. The degradation, \revsep{$\Delta\mathrm{MAE} = +0.245$}, is \revsep{more than twenty times both the session-to-session scatter of the optimization (standard deviation 0.011 across the six sessions) and the difference between the two MHD configurations (0.012)}.}

\revaug{The per-spacecraft breakdown identifies the mechanism (Supplementary Fig.~S2 and Supplementary Table~S4). As $\xi_g$ increases from $10^2$ to $10^6$, the MAE at PSP, the only spacecraft magnetically connected to the eastern field lines, rises monotonically from 0.50 to 1.06, whereas the MAE at SOHO falls monotonically from 1.15 to 0.40 and that at BepiColombo from 1.65 to 0.50. STEREO-A and Solar Orbiter are non-monotonic, with optima near $\xi_g \approx 10^{3}$ and $10^{3.5}$ respectively. The eastern and western observers therefore demand mean free paths that move in opposite directions, and no single value satisfies both: at the best compromise the MAE still reaches 0.92 at PSP, 0.91 at BepiColombo and 0.73 at STEREO-A, against 0.45, 0.51 and 0.40 for the 19-parameter solution.}

\revaug{Two caveats apply. First, because the injection efficiencies were held at values that had been optimized jointly with the spatially varying $\xi_j$, the quoted $\Delta\mathrm{MAE}$ is an upper bound on the degradation that a fully re-optimized global-$\xi$ model would suffer. Even with $\epsilon_{0,k}$ fixed, however, three of the five spacecraft come close to the 19-parameter solution somewhere on the grid: at $\xi_g = 10^6$ SOHO reaches 0.400 against 0.401, and BepiColombo 0.499 against 0.505, while Solar Orbiter comes within \revsep{0.04 of its reference value of 0.545}. The failure is confined to STEREO-A and PSP, whose residuals at the best compromise exceed their reference values by 0.33 and 0.47 and whose responses to $\xi_g$ move in opposite directions. Since $\epsilon_{0,k}$ rescales the injection amplitude in each longitude bin, whereas the conflict between the eastern and western observers appears in the shape of the time profiles, re-optimizing the injection efficiencies would not be expected to remove it. A countervailing bias acts in the opposite direction: because the evaluation mask depends on the simulated intensity, the poorer global-$\xi$ fits have more points removed from the sum ($N$ falls from 10\,480 for the 19-parameter solution to between 9\,538 and 10\,171 across the grid; Supplementary Table~S4), so their largest residuals are partly excluded and the quoted degradation is understated on that account. Second, this test addresses the spatial parameterization of $\xi_j$ only. The longitude dependence of $\epsilon_{0,k}$ is a diagnostic degree of freedom by construction (see the Free Parameters subsection of the Methods), so a global-$\epsilon_0$ baseline would measure a change in fit quality without testing the purpose for which that freedom was introduced. Taken together, the two caveats delimit what the comparison shows: within a fixed set of injection efficiencies, a spatially uniform mean free path cannot reproduce the five spacecraft simultaneously. Establishing predictive skill, as distinct from hindcast capability for this event, would require a reduced model in which the injection efficiencies are re-optimized alongside a single global coefficient, and a held-out validation such as leave-one-spacecraft-out; both are deferred to future work on a larger sample of events.}

\section*{Discussion} \label{sec:discussion}

\subsection*{Physical Interpretation of the Optimized Parameters} \label{subsec:phys_interp}

The two types of free parameters in our model (mean free path coefficients $\xi_j$ and injection efficiencies $\epsilon_{0,k}$) were designed to play fundamentally different roles in the optimization (see the Free Parameters subsection).
Here we discuss the physical implications of the optimized values and their sensitivity to the MHD background.

\paragraph{Mean free path coefficients $\xi_j$ as physical degrees of freedom.}
The optimized $\xi$ values in both MHD configurations exhibit a pronounced east--west contrast (Table~\ref{tab:best_params}): field lines on the eastern side ($j = 4$--6) settle at $\xi \sim 10^{2\text{--}3}$, suggestive of strong pitch-angle scattering, whereas the \rev{near-nose} western field lines ($j = 7, 8$) favour $\xi \sim 10^{5\text{--}6}$, \revaug{that is, values at or above the sensitivity ceiling, for which the fit determines only a lower bound on the mean free path}.
This spatial structure is physically plausible.
The CME shock in this event propagated roughly along the Sun--Earth line, so the eastern field lines are connected to the \revaug{quasi-parallel flank of the shock, where accelerated particles escape upstream along the magnetic field and amplify resonant Alfv\'{e}n waves that subsequently scatter them, whereas the western lines are connected to the nose or to the quasi-perpendicular flank, where upstream escape is inhibited, the self-generated wave level is correspondingly lower, and the scattering environment is weaker (see the Longitude Profile of the Mean Free Path and Shock Obliquity subsection)}.
\revaug{Whether a spatially varying mean free path is needed is examined directly by the comparison against a spatially uniform baseline reported in the Comparison with a Spatially Uniform Mean Free Path subsection: collapsing $\xi_4$--$\xi_{10}$ onto a single global value degrades the objective function by \revsep{0.245}, far beyond the session-to-session scatter of the optimization, because the eastern and western observers require mean free paths that differ in opposite directions.}
\revaug{The same east--west distinction is visible in the shape of the objective function itself, independently of any treatment of the optimization trials as samples. Evaluated over the uniformly sampled QMC stage, the median objective responds to $\xi_5$ and $\xi_6$ with a clear interior optimum, most sharply for $\xi_6$, whereas for the near-nose western coefficients $\xi_7$ and $\xi_8$ it decreases monotonically and then flattens above $\xi \approx 10^3$--$10^4$ (Supplementary Fig.~S3). Short mean free paths are therefore actively selected on the eastern side, while on the western side the objective simply ceases to discriminate between values.}

For the western field lines the comparison is less direct.
At the level of single best-fit values, RUN1 and RUN2 yield $\xi$ that differ by orders of magnitude (e.g., $\xi_{10}$: $4.4 \times 10^5$ in RUN1 vs.\ $5.5 \times 10^1$ in RUN2).
However, as established by the multi-run sensitivity ceiling at $\xi \approx 10^4$ (Fig.~\ref{fig:xi_convergence}), individual best-fit values for the western field lines, where multiple spacecraft are connected and pitch-angle scattering has limited effect on the time-integrated flux, are not strongly constrained above the ceiling.
The longitude-profile analysis with ceiling clamp (Fig.~\ref{fig:xi_longitude_profile}) confirms that, after accounting for this insensitivity, the run-median $\xi$ profiles for the western field lines in RUN1 and RUN2 are largely consistent within their three-run ranges.
The dominant MHD-dependent compensation in this event therefore manifests not in $\xi$ but in the injection efficiencies $\epsilon_{0,k}$, as discussed below.

\paragraph{Injection efficiency $\epsilon_{0,k}$ as a diagnostic degree of freedom.}
As argued in the Free Parameters subsection, the injection efficiency was introduced not as a genuinely physical free parameter but as a diagnostic probe of the MHD-derived shock conditions.
The optimized $\epsilon_{0,k}$ values indeed exhibit the behaviour expected from this design.
In both configurations, the values span nearly the full search range ($10^{-4}$--$3 \times 10^{-2}$) and show no smooth longitudinal trend that could be attributed to a simple physical gradient (Table~\ref{tab:best_params}).
Moreover, the spatial patterns differ substantially between RUN1 and RUN2: for instance, $\epsilon_{0,4}$ takes a value of $2.9 \times 10^{-2}$ in RUN1 but $1.2 \times 10^{-4}$ in RUN2, a difference exceeding two orders of magnitude at the same longitudinal bin.
Unlike the apparent $\xi$ differences discussed above, the $\epsilon_{0,k}$ values are not subject to a sensitivity ceiling, so these inter-configuration differences \revaug{reflect compensation for the configuration-dependent change in the shock compression ratio, velocity, and magnetic obliquity supplied by \texttt{SUSANOO}, together with any residual mismatch that $\epsilon_{0,k}$ absorbs}.
The corresponding inter-configuration comparison, based on the six-session ensemble, is presented in the Sensitivity of Injection Efficiency to CME Magnetic Flux subsection.

This degeneracy between $\epsilon_{0,k}$ and the MHD background is an inherent limitation of single-event optimization and motivates the joint MHD--SEP parameter estimation discussed in the Prospects for Operational Forecasting subsection.

The diagnostic utility of this approach is most apparent in the RUN2 results, where several bins yield $\epsilon_{0,k} > 10^{-2}$, values that exceed the range typically adopted in DSA models \cite{2000JGR...10525079Z, 2003JGRA..108.1082L, 2017JGRA..12210938H}.
Such high injection efficiencies are consistent with a residual mismatch at those longitudes between the shock conditions supplied by the MHD background and those required by the observations, which the optimizer can offset only by inflating $\epsilon_{0,k}$. Part of the mismatch may also reflect the injection prescription itself and the transport physics omitted by the assumed geometry, which are absorbed into the same parameters.
Identifying these longitudes provides targeted feedback for improving the MHD model or its coupling with the particle transport solver.

\revaug{One plausible major contributor is the uncertainty in the reconstructed compression ratio. As set out in the Model Coupling subsection, the shock-surface quantities delivered to the injection model at the inner boundary are not read directly from the MHD solution but reconstructed by extrapolating inward from $0.15$--$0.47$~au, and among the reconstructed quantities the compression ratio is both the least well determined, varying between about 2 and 5 across the fitting window, and the one to which the source term responds most steeply, both through the fifth-power dependence of equation~(\ref{eq:eff_model}) on $\beta - 1$ and, through $\beta$ itself, through the shape of the injected momentum spectrum in equation~(\ref{eq:source_term}). A compression ratio uncertain by that amount corresponds on its own to a factor of about 20, or 1.3 decades, in the injected particle number at fixed $\epsilon_{0,k}$, against the 2.5 decades spanned by the search range for $\epsilon_{0,k}$. A substantial part of the bin-to-bin scatter in the optimized injection efficiencies is therefore what the known uncertainty of a single reconstructed quantity would already produce. A scalar injection efficiency absorbs only the normalization; the accompanying change in the spectral index is not absorbed, so an error in the compression ratio propagates into the simulated spectral shape as well as into the overall intensity. This also indicates where an improvement would pay off most, namely in a determination of the compression ratio near the inner boundary that does not depend on an inward extrapolation.}

\paragraph{Parameter importance analysis.}
As introduced in the Multi-Run Reproducibility subsection of the Results, \revaug{the rank-correlation importance scores of Fig.~\ref{fig:importance_dotplot} quantify how strongly and how monotonically the objective function responds to each free parameter.}
Here we focus on the physical interpretation; the ensemble analysis across six independent sessions is presented in Fig.~\ref{fig:importance_dotplot}.

In the RUN1 sessions, the \revaug{highest-ranked} parameters are the mean free path coefficients $\xi_7$ and $\xi_8$.
These correspond to western field lines whose Parker-spiral footpoints map close to the spacecraft observing during the rising and peak phases of the event, making the simulated fluxes most sensitive to the scattering environment along these lines.
In the RUN2 sessions, by contrast, the dominant parameters are predominantly injection efficiencies ($\epsilon_{0,9}$, $\epsilon_{0,10}$, $\epsilon_{0,12}$), with $\xi_7$ and $\xi_8$ also present in the top positions across all three independent runs (see the Multi-Run Reproducibility subsection of the Results).

This reversal in dominant parameter type between the two configurations ($\xi$-dominated in RUN1 versus $\epsilon$-dominated in RUN2) illustrates an important caveat of \revaug{this importance analysis}.
The ranking is conditional on the MHD background and the specific distribution of trial samples; it reflects the sensitivity structure of the objective landscape explored, not an absolute physical hierarchy.
Accordingly, we interpret these results as informative guidance for understanding the optimization landscape, but caution that the numerical importance scores should not be over-interpreted and the ranking order may not be robust across different MHD realizations or sample sets.
\revaug{A second caveat is intrinsic to a rank correlation: it measures monotonic dependence only, and a parameter with a well-defined interior optimum receives a low score even though the objective responds strongly to it. Among the free parameters, $\xi_6$ (footpoint $-7.5^{\circ}$) is such a case. It has the lowest importance score of the 19, yet the median objective value evaluated over the uniformly sampled QMC stage varies by 0.19 across the search range and has a clear minimum near $\xi_6 \approx 10^2$, rising on both sides (Supplementary Fig.~S3). Its low rank-correlation score therefore reflects the shape of the dependence rather than the absence of one, and the eastern coefficients should not be read as uniformly unconstrained on the basis of Fig.~\ref{fig:importance_dotplot} alone.}

\subsection*{Longitude Profile of the Mean Free Path and Shock Obliquity} \label{subsec:xi_longitude}

The multi-run ensemble allows the longitude dependence of the transport mean free path to be characterized and compared across the two MHD configurations considered here.
Figure~\ref{fig:xi_longitude_profile} shows \revaug{the median of the top-10\% $\xi_j$ values within each session, clamped at $\xi = 10^4$ and then combined across the three independent sessions of each configuration as a geometric mean,} as a function of field-line index~$j$ (the sensitivity ceiling established in Fig.~\ref{fig:xi_convergence}); shaded bands indicate the three-run median range.

A pronounced east--west dichotomy is apparent and consistent across both RUN1 and RUN2 and all six independent sessions.
The eastern field lines ($j = 4$--$6$, $t = 0$ footpoint longitudes $-37.5^{\circ}$ to $-7.5^{\circ}$) consistently yield small median $\xi$ values ($\sim$$10^2$--$10^3$), indicating strong pitch-angle scattering.
The western field lines ($j = 7$--$9$, footpoints at $+7.5^{\circ}$ to $+37.5^{\circ}$) converge to or above the sensitivity ceiling ($\xi \gtrsim 10^4$), \revaug{where the fit no longer discriminates between values and constrains the mean free path only from below}.

This pattern is consistent with the relationship between the local shock obliquity angle~$\theta_{Bn}$ (the angle between the shock normal and the upstream magnetic field) and the interplanetary scattering environment.
In the Parker spiral geometry, the azimuthal magnetic field makes systematically different angles with the CME shock normal at different longitudes along the shock front.
Computed directly from the \texttt{SUSANOO} shock output at this event, $\theta_{Bn}$ ranges from $\approx 10^{\circ}$--$40^{\circ}$ (quasi-parallel) at the eastern flank to $\approx 50^{\circ}$--$63^{\circ}$ (quasi-perpendicular) at the western flank and near the CME nose, consistent with the general geometry established by observational and modelling studies \cite{2005ApJ...625..474T, 2025ApJ...995...77T}.

At a quasi-parallel shock, accelerated particles can stream freely upstream along the magnetic field lines, exciting resonant Alfv\'{e}n waves through a beam-plasma instability \cite{1983JGR....88.6109L}.
Self-consistent transport models show that these self-generated waves rapidly reduce the mean free path from its ambient value ($\sim$1~au) to values as small as $\sim$0.01~au along the entire field-line path between the shock and the observing spacecraft \cite{2003ApJ...591..461N}.
At a quasi-perpendicular shock, by contrast, particles cannot escape upstream efficiently; the foreshock wave generation is suppressed and the mean free path remains close to its ambient, large value \cite{2005ApJ...625..474T, 2006ApJ...646.1319T, 2013A&A...558A.110B}.
Forward-modelling studies with the iPATH code demonstrate that this mechanism produces a systematic longitude dependence of the diffusion coefficient, with the strongest scattering on the eastern (quasi-parallel) flank and the weakest scattering on the western (quasi-perpendicular) side \cite{2017JGRA..12210938H, 2022JGRA..12730343D, 2023JGRA..12831502D}.

We emphasize \revaug{five} important caveats.
First, $\xi_j$ in \texttt{SEPIF} is a time-independent, spatially uniform parameter for each field line, representing an event-average effective transport coefficient rather than the self-consistently evolving instantaneous mean free path.
Second, the single scalar $\xi_j$ absorbs contributions from all phases of the event (impulsive, plateau, and decay), which physically involve different turbulence levels; the optimizer necessarily compromises across phases.
Third, a degeneracy exists between $\xi_j$ and $\epsilon_{0,k}$ at a spacecraft connected to only one or two field lines, so we cannot fully exclude the possibility that part of the east--west contrast in $\xi$ reflects compensation for longitude-dependent MHD errors rather than a true transport difference.
Fourth, among the five spacecraft analysed only PSP is magnetically connected to the eastern field lines; consequently, $\xi_4$, $\xi_5$, and $\xi_6$ are constrained by far less observational information than their western counterparts and exhibit lower \revaug{rank-correlation} parameter importance (Fig.~\ref{fig:importance_dotplot})\revaug{, with the qualification noted above that this measure understates the constraint on $\xi_6$}.

\revaug{Fifth, the magnetic connectivity that underlies this longitudinal assignment is itself uncertain. The footpoint longitude of a Parker spiral reaching 1~au scales inversely with the assumed solar wind speed: for an observer at 1~au it is $61^{\circ}$ at $V_{sw} = 400$~km~s$^{-1}$ and $82^{\circ}$ at 300~km~s$^{-1}$. The background speeds observed at the four spacecraft shortly before shock arrival lie above 300~km~s$^{-1}$, so the latter is a worst case; even then the displacement of the footpoint ranges from $7^{\circ}$ at the heliocentric distance of BepiColombo to $21^{\circ}$ at that of SOHO, and the larger values exceed both our field-line spacing of $15^{\circ}$ and the $10^{\circ}$ width of the injection bins. A single uniform $V_{sw}$ can therefore assign a spacecraft to a neighbouring field line or bin, so that the absolute longitude attached to a given $\xi_j$ or $\epsilon_{0,k}$ is uncertain at that level. The measured speeds also change through the event, approaching the lower end of that range at the later times, so a single constant $V_{sw}$ cannot represent the connection equally well at all times; the field-line assignment of STEREO-A in particular could be affected. We do not, however, regard this as undermining the east--west contrast itself. At any given time the displacement acts in the same sense at every observer, so the ordering of the five spacecraft in connection longitude is preserved; and the quantity on which the contrast actually rests, namely the separation in footpoint longitude between PSP, the only observer connected to the eastern field lines, and the near-Earth spacecraft, changes only from $62^{\circ}$ to $67^{\circ}$ over the same range of assumed wind speed. A shift of that size cannot reverse a contrast of four to five orders of magnitude in $\xi$ across $90^{\circ}$ of longitude. What the uncertainty affects is the absolute longitude labelling of the profiles in Fig.~\ref{fig:longitude_profiles}, not the qualitative east--west structure. Defining the Parker connection from the \texttt{SUSANOO} solar wind speed at each observer would not straightforwardly reduce this uncertainty here. The modelled ambient wind carries its own error and can depart from the measured speed at the observer positions, so substituting it would replace a controlled assumption with an independent error source, in a study whose object is the transport calculation and the optimization procedure rather than the ambient wind model. For a forecasting application the situation is reversed, since in situ speeds are not available in advance; there the predicted \texttt{SUSANOO} wind would be the appropriate choice, at the cost of repeating the transport calculation for every trial.}
Taken together, these caveats place the east--west pattern at the level of a qualitative feature of this event. We characterize it as \emph{consistent with} the foreshock turbulence mechanism described above, without claiming to have established causality.

Finally, the outermost western field line $\xi_{10}$ ($j = 10$, footpoint $+52.5^{\circ}$) \revaug{shows the largest run-to-run scatter of any of the mean free path coefficients in RUN2 (geometric standard deviation 8.9), with appreciable scatter in RUN1 as well (1.8)}.
This field line is magnetically connected primarily to SOHO only during the early rising phase of the event, when fluxes are still small and the cumulative scattering effect is not yet apparent in the time--intensity profile.
Consequently the optimization cannot effectively distinguish different $\xi_{10}$ values, and this parameter remains essentially unconstrained by the available data. \revaug{Supplementary Fig.~S3 shows this directly: the objective function is flat with respect to $\xi_{10}$ across the whole search range, in contrast to the marked response it shows to $\xi_7$ and $\xi_8$.}

\subsection*{Sensitivity of Injection Efficiency to CME Magnetic Flux} \label{subsec:eps_magflux}

\begin{figure}[htbp]
\centering
\begin{subfigure}[b]{0.67\linewidth}
\includegraphics[width=\linewidth]{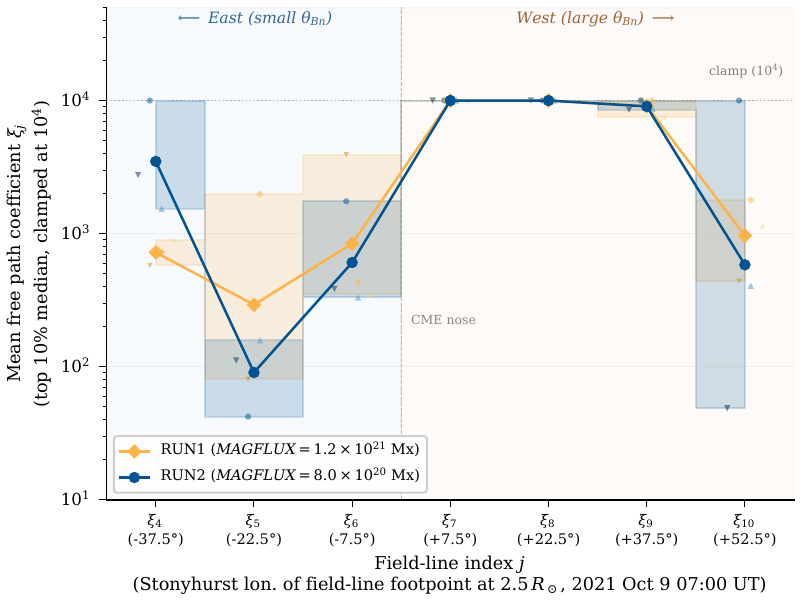}
\caption{}
\label{fig:xi_longitude_profile}
\end{subfigure}

\vspace{0.5em}
\begin{subfigure}[b]{0.67\linewidth}
\includegraphics[width=\linewidth]{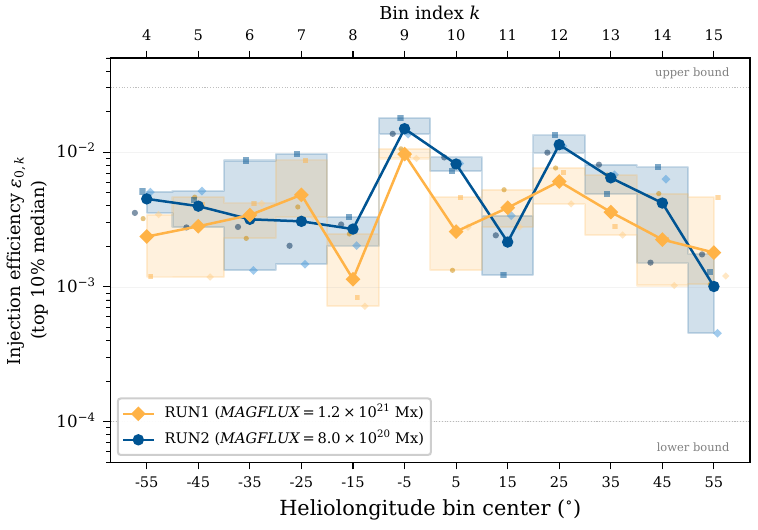}
\caption{}
\label{fig:epsilon_profile}
\end{subfigure}
\caption{\rev{Longitude profiles of the optimized transport and injection parameters.
    (a) Median mean free path coefficient $\xi_j$ as a function of field-line index~$j$} for RUN1 (orange, diamonds) and RUN2 (blue, circles), computed from the top-10\% of trials and clamped at $\xi = 10^4$.
    Shaded bands span the range of the three independent run-medians per configuration.
    The $x$-axis tick labels indicate the field-line index~$j$ and, below, the Stonyhurst footpoint longitude at $r = 2.5\,R_\odot$ at $t = 0$ (9 Oct 2021 07:00~UT).
    The horizontal dotted line marks the sensitivity ceiling $\xi = 10^4$ (see Fig.~\ref{fig:xi_convergence}).
    Blue (left) and orange (right) background shading indicate the quasi-parallel (eastern) and quasi-perpendicular (western) shock regions, respectively; the dashed vertical line marks the approximate CME nose.
    \rev{(b) Median injection efficiency $\epsilon_{0,k}$ as a function of longitude bin index~$k$} for RUN1 (orange, diamonds) and RUN2 (blue, circles).
    Medians are computed over the top-10\% of trials; shaded bands span the three-run median range per configuration.
    \rev{Bin centres in the simulation frame are indicated on the $x$-axis; the upper $x$-axis shows the corresponding bin index~$k$.
    The reversal at bin $k = 11$ ($+15^{\circ}$) is discussed in the text.}}
\label{fig:longitude_profiles}
\end{figure}

Because RUN1 and RUN2 differ only in the CME magnetic flux parameter ($\mathit{MAGFLUX}$; factor of 1.5), the six-run ensemble enables a direct, single-variable test of how the optimized injection efficiencies respond to a change in the MHD side of the model chain.
Figure~\ref{fig:epsilon_profile} shows \revaug{the per-session median of the top-10\% $\epsilon_{0,k}$ values, combined across the three sessions of each configuration as a geometric mean,} as a function of longitude bin index~$k$ for both configurations, with three-run range bands superimposed.

A clear and consistent pattern is apparent: in 8 out of 12 active longitude bins, RUN2 (lower $\mathit{MAGFLUX}$, slower shock) yields higher injection efficiencies than RUN1.
For three bins near the CME nose ($\epsilon_{0,9}$ at $\approx -5^{\circ}$, $\epsilon_{0,10}$ at $\approx +5^{\circ}$, and $\epsilon_{0,12}$ at $\approx +25^{\circ}$), the three-run median ranges do not overlap between RUN1 and RUN2\revaug{, so the difference is reproduced in every pairing of an independent RUN1 session with an independent RUN2 session}.
The geometric mean ratios (RUN2/RUN1) for these three bins are $1.54\times$, $3.17\times$, and $1.88\times$, respectively.

This systematic elevation of $\epsilon_{0,k}$ in RUN2 is a compensation effect.
\rev{The particle source term in the transport model contains two $\beta$-dependent factors: the effective injection efficiency $\epsilon \propto \epsilon_{0,k}\,[(\beta{-}1)/3]^{-5}$ (equation~(\ref{eq:eff_model}) in Methods), which controls the total number of injected particles, and the spectral shape $p^{-\beta}$, which distributes them in momentum.
The higher internal magnetic pressure of RUN1 resists CME deceleration and sustains a systematically higher radial shock propagation speed (approximately 6\% in $V_{\rm shk}$), from direct comparison of the MHD shock-parameter outputs.
The lower-$\mathit{MAGFLUX}$ RUN2 therefore has a smaller compression ratio $s$ and a correspondingly larger spectral index $\beta = 3s/(s{-}1)$.
The dominant consequence is that the steep fifth-power dependence $\epsilon \propto (\beta{-}1)^{-5}$ sharply reduces the total injected particle number at a given $\epsilon_{0,k}$; the accompanying steepening of the $p^{-\beta}$ spectrum further reduces the high-energy flux, but is a secondary effect.
The radial shock speed $V_{\rm shk}$, which enters the source term linearly, differs by only $\sim$6\% between the two configurations and likewise contributes at a subdominant level.
The optimizer therefore compensates primarily by raising $\epsilon_{0,k}$ in RUN2 by a factor of $\sim$1.5--3.2 across the three near-nose bins.}

One bin stands out as an anomaly: $\epsilon_{0,11}$ ($k = 11$, $\approx +15^{\circ}$) is consistently \emph{higher} in RUN1 than in RUN2 (geometric mean ratio $0.56\times$; RUN1 $>$ RUN2 in all three run pairs), opposite to the dominant trend.
This reversal suggests that the local shock--field-line intersection geometry at this longitude responds to the MAGFLUX change in a qualitatively different manner from the adjacent bins, possibly due to a non-monotonic dependence of the 3-D shock curvature on the CME magnetic configuration at that longitude.
Identifying such anomalous bins provides targeted guidance for refining the injection model or the MHD shock parameterization at specific heliolongitudes.

\subsection*{Comparison with Previous Multi-Spacecraft Studies} \label{subsec:palmerio}

The 9 October 2021 event was previously modelled by Palmerio et al.\cite{2024JSWSC..14....3P}, who coupled the WSA-Enlil MHD model with the SEPMOD test-particle code to reproduce SEP observations at the same five spacecraft.
Although both studies address the same event with physics-based modelling chains, the underlying philosophies differ in important respects, and a direct numerical comparison of output parameters is not straightforward.

Palmerio et al.\cite{2024JSWSC..14....3P} adopted a ``fixed source'' injection approach in which \revaug{the spatial form of the source is prescribed at the heliospheric inner boundary ($21.5\,R_\odot$) as a circular patch with an axisymmetric injection profile}, with the injection parameters scaled empirically from flare and CME observables.
Their focus was on demonstrating the feasibility of multi-point SEP reproduction using an empirically calibrated source term, without performing parameter optimization against the observed time profiles.
\revaug{In their framework, the injection amplitude and spectral shape are set from flare and CME observables through empirical scaling relations, and the interplanetary transport is scatter-free, so that no scattering parameter enters the model. The agreement with each spacecraft is quantified after the fact, including by dynamic time warping, rather than minimized as an objective function during the modelling.}

Our approach differs in two principal ways.
First, we replace the \revaug{prescribed, axisymmetric} fixed source with longitude-dependent injection efficiencies $\epsilon_{0,k}$ that are individually optimized for each heliolongitudinal bin, allowing the model to capture (and diagnose) the non-uniform particle injection imposed by the real shock geometry.
Second, we employ an automated black-box optimization that minimizes the multi-spacecraft MAE across a 19-dimensional parameter space, thereby determining the best-fit parameter values from the observations themselves.

These differences make the two studies complementary.
The fixed-source approach of Palmerio et al.\cite{2024JSWSC..14....3P} is computationally inexpensive and well suited for rapid event analysis, but it does not reveal spatial structure in the injection or constrain transport parameters against observations.
Our optimization-based approach provides this spatial diagnostic capability at greater computational cost.
In particular, the longitude-dependent $\epsilon_{0,k}$ patterns identified in the Best-Fit Parameter Values subsection highlight specific heliolongitudes at which the coupled MHD--injection--transport model shows residual mismatch with the observations, information that is not accessible from a fixed-source injection model with a prescribed spatial form.

As noted in the Methods (MHD model subsection), \revaug{Palmerio et al.\cite{2024JSWSC..14....3P} inserted the CME into WSA--Enlil as an unmagnetized hydrodynamic pulse}, whereas the present study uses a spheromak-type magnetic flux rope within \texttt{SUSANOO}, with the spheromak geometric parameters set to approximate first-order correspondence with the GCS reconstruction.
The choice of CME model affects the downstream shock parameters (compression ratio, magnetic obliquity, shock speed) and thus the SEP acceleration conditions. \revaug{The presence of an internal magnetic field bears directly on the present comparison: the two configurations examined here differ only in the CME magnetic flux, and the reproducible differences we find between them appear in the optimized injection efficiencies.}
A detailed comparison of how these different MHD treatments propagate into the optimized SEP parameters is beyond the scope of this paper, but the fact that both approaches achieve reasonable multi-spacecraft reproduction of this event lends confidence that the 9 October 2021 event is well suited as a benchmark for physics-based SEP modelling. 

\rev{More broadly, the picture that emerges here, namely that the multi-spacecraft fluxes reflect each observer's magnetic connection to different regions of an inhomogeneous CME-driven shock, has recently received independent support. Liu et al.\cite{2026arXiv260602445L} modelled the 24 February 2023 multi-spacecraft event by coupling a global MHD simulation (\texttt{AWSoM-R}) to a field-line-based SEP transport model and found that order-of-magnitude flux differences between nearby spacecraft arise from their connection to different parts of a distorted CME-driven shock, with the most efficiently accelerating high-compression region not necessarily coinciding with the nominally best-connected observer. This is the same physical effect that our longitude-dependent $\epsilon_{0,k}$ are designed to capture and diagnose.}

\subsection*{Temporal Evolution of the Mean Free Path} \label{subsec:tau}

One persistent challenge in our modelling is the accurate reproduction of the decay phase of the SEP event.
While the rising phase and peak intensities can be satisfactorily matched by optimizing the injection and transport parameters described in the Methods, the prolonged post-peak decay tends to fall off too rapidly under the time-independent mean free path formulation adopted in our primary analysis, leading to underestimation of the flux during this phase.

The \texttt{SEPIF} model supports an extended formulation in which the mean free path evolves with time via an $e$-folding timescale $\tau_j$ for each field line $j$:
\begin{equation} \label{eq:mfp_tau}
    \lambda_j(p, z, t) = \xi_j \left(\frac{p}{p_{\max}}\right) \left(\frac{B_{1\,\mathrm{au}}}{B(z)}\right) r_{g,\max}\; e^{-t/\tau_j},
\end{equation}
where a finite positive $\tau_j$ causes the mean free path to decrease with time, mimicking the growth of magnetic turbulence in the wake of the CME-driven shock, and the limit $\tau_j \to \infty$ recovers the time-independent equation~(\ref{eq:mfp}).
This parameterization is motivated by Minoshima et al.\cite{2026EP&S...78...64M}, who report that a temporally decreasing mean free path is required to reproduce the decay phase of certain SEP events.

We conducted supplementary experiments in which $\tau_j$ was treated as an additional free parameter alongside $\xi_j$.
These experiments confirmed that allowing finite $\tau_j$ values reduces the overall MAE compared to the time-independent baseline.
However, the improvement is spatially non-uniform: the fits for STEREO-A and SOHO, which sample particles over an extended post-peak period, are substantially improved, while the fits for spacecraft with different magnetic connectivity, notably Solar Orbiter, are degraded.
This suggests that applying a uniform temporal decay of the mean free path across all field lines is insufficient to simultaneously improve the reproduction at all observation points.
Moreover, even with optimized $\tau_j$ values, the prolonged decay tails observed by STEREO-A and SOHO at late times cannot be fully reproduced. These late-phase profiles are influenced by large-scale structures such as corotating interaction regions (CIRs)\rev{ \cite{2004SSRv..111..267R}} that are not captured by the present model; \revaug{for this event a stream interaction region preceded the CME-driven shock at the observers \cite{2022ApJ...934...55L, 2023ApJ...950..172W}.} Disentangling such event-specific effects from the intrinsic transport physics requires application to a broader sample of events.

A systematic decrease of the mean free path with time following a CME passage is not a universal feature of SEP transport.
The applicability of equation~(\ref{eq:mfp_tau}) depends on the specific magnetic and turbulence conditions prevailing along each field line, which vary considerably from event to event.
Our current implementation does not capture this spatial heterogeneity, and we therefore cannot assert that the exponential decay form is physically justified for all field lines in this particular event.
We regard the present result as a suggestion that a time-varying mean free path may play a role in the decay phase of this event, and defer a systematic investigation of the $\tau$ parameter to future work with a larger sample of SEP events.

\subsection*{Prospects for Operational Forecasting} \label{subsec:forecasting}

SEITEN was designed primarily for post-event analysis of a single SEP event, but its structure provides a foundation for several extensions towards operational forecasting.
We outline three directions that we consider most promising.

\paragraph{Multi-event application and prior distributions.}
The present study optimizes 19 parameters for a single event under two MHD backgrounds, yielding parameter estimates that are inherently event-specific.
To transition from event-specific fitting to a predictive capability, the same framework must be applied to a statistically meaningful sample of events.
By accumulating optimized parameter sets across diverse events (differing in CME speed, flare longitude, heliospheric conditions, and multi-spacecraft geometry), one can construct empirical prior distributions for the mean free path coefficients $\xi_j$.
Such priors would provide physically informed initial guesses for new events, reducing the number of trials needed to achieve an acceptable fit and enabling quasi-real-time parameter estimation.
Equally important, the scatter in the diagnostic injection efficiencies $\epsilon_{0,k}$ across multiple events can reveal systematic biases in the MHD model: if certain heliolongitudes consistently require anomalously high $\epsilon_{0,k}$ values, this provides targeted guidance for improving the shock parameterization in \texttt{SUSANOO} or refining the DSA injection model.
In a companion study, Y. Kato et al. (submitted) optimize the MHD input parameters of \texttt{SUSANOO} (such as the initial CME speed, magnetic flux, and spheromak width) by matching the simulated CME-driven shock arrival times and in situ magnetic field and plasma signatures against multi-spacecraft observations.
Combining these independently optimized MHD parameters with the SEP transport parameters constrained in the present work would constitute a step towards a fully joint MHD--SEP parameter estimation, which could further reduce the systematic uncertainties identified in this study.

\paragraph{Dimension reduction via parameter importance.}
\revaug{The importance analysis in the Physical Interpretation of the Optimized Parameters subsection shows that the objective function responds strongly to only a small subset of the parameters: two or three parameters reach $|\rho_{\mathrm{Spearman}}| \gtrsim 0.35$ in each configuration, while the remaining sixteen fall below 0.25 and eleven fall below 0.13.}
This concentration suggests, as a hypothesis to be tested rather than a result, that the effective dimensionality of the optimization problem is lower than the nominal 19. \revaug{The rank-correlation measure cannot establish this on its own: $\xi_6$ scores low on it and yet has the largest marginal swing of any coefficient (Supplementary Fig.~S3), so a low score does not license fixing a parameter. Testing the hypothesis would require reduced-parameter optimizations in which candidate parameters are held fixed and the loss of fit quality is measured, which we have not carried out.}
If it holds, fixing the less important parameters at robust default values, informed by the multi-event priors described above, would reduce the search space and the number of trials required for convergence, and would help to bring the optimization within the time constraints of operational forecasting.

\paragraph{Observational constraints and systematic uncertainties.}
A fundamental assumption of our framework is that observations from five spacecraft can be combined into a single scalar objective function with \revaug{near-uniform} weights (equation~(\ref{eq:mae_final})).
\revaug{BepiColombo is down-weighted by a factor of 0.7 to partially offset its much smaller number of contributing data points, but the objective remains close to a per-spacecraft average.}
This assumption merits scrutiny.
The five instruments differ in energy coverage, angular acceptance, calibration accuracy, and contamination characteristics.
In particular, BERM's location on the MPO radiator panel constrains its look direction towards the anti-sunward hemisphere during cruise, so the instrument samples only a restricted range of the pitch-angle distribution, predominantly particles propagating against or transverse to the interplanetary magnetic field direction \cite{Pinto2022}. The simulated fluxes at the BepiColombo location, which are pitch-angle integrated intensities, are therefore not strictly the same physical quantity as what BERM observes, and a systematic offset may arise during phases when the underlying particle distribution is strongly anisotropic (e.g., the rising phase). While we have not applied any correction for this look-direction effect in the present study, future work could incorporate BERM's field-of-view geometry into the model--data comparison.
Furthermore, as future multi-spacecraft missions (such as proposed Sun--Earth Lagrange L4 mission concepts\rev{, together with near-Earth and sub-L1 monitors,} equipped with in-situ energetic particle detectors \cite{2021SpWea..1902777P, 2024JASS...41....1L, 2024aero.conf..202P, 2024SSRv..220...73L}) expand the heliospheric observation network, the increased number of constraint points will enhance the spatial resolution of the optimization and reduce the parameter degeneracies identified in this work.

\section*{Conclusion}

\subsection*{Summary}

We have developed and demonstrated SEITEN, a framework for the automated optimization of free parameters in a physics-based SEP acceleration and transport model, applied to the 9 October 2021 multi-spacecraft event.
SEITEN couples the global MHD solar wind and CME propagation model \texttt{SUSANOO} with the particle acceleration and transport code \texttt{SEPIF}, which solves the focused transport equation with diffusive shock acceleration along multiple magnetic field lines.
A two-stage black-box optimization combining quasi-Monte Carlo (Sobol' sequence) sampling and the Tree-structured Parzen Estimator, implemented within the Optuna framework, is used to explore a 19-dimensional parameter space consisting of 7 mean free path coefficients ($\xi_j$) and 12 longitude-dependent injection efficiencies ($\epsilon_{0,k}$).
To our knowledge, this is the first study to apply systematic, automated parameter optimization to a coupled MHD--SEP transport model constrained simultaneously by five widely separated spacecraft. \revaug{The coupling acts through the shock source conditions supplied by \texttt{SUSANOO}; the interplanetary transport is solved on analytic Parker-spiral field lines with a uniform solar wind speed and without cross-field diffusion, so the optimized parameters reported below are effective parameters conditional on that transport geometry.}

The main findings of this study are summarized as follows:
\begin{enumerate}
    \item \textbf{Multi-point flux reproduction.}
    The optimized model reproduces the gross features of the proton differential intensity time profiles (onset, rise, and peak) at all five spacecraft (STEREO-A, SOHO, BepiColombo, PSP, and Solar Orbiter) across the energy range 1.5--60~MeV.
    The best-fit \revaug{objective-function} MAE in $\log_{10}$ space is \revsep{0.446} (RUN2) and \revsep{0.458} (RUN1), corresponding to a typical simulation-observation discrepancy of a factor of \revsep{$\sim$2.9} in differential intensity.
    Two independent MHD configurations yield closely similar overall fit quality (\revsep{$\Delta\mathrm{MAE} \approx 0.012$, comparable to the 0.011 scatter between sessions of the same configuration}). \revaug{This indicates a degeneracy between the MHD background and the SEP parameters, and an insensitivity of the objective function to that background, rather than robustness, and it motivates the joint MHD--SEP estimation discussed below.}

    \item \textbf{Spatial structure of transport parameters.}
    The optimized mean free path coefficients reveal a pronounced east--west contrast: field lines on the eastern side, to which only PSP is magnetically connected via the Parker spiral, settle at $\xi \sim 10^{2\text{--}3}$ (suggestive of strong scattering), while the western field lines connecting to the remaining four spacecraft favour much larger values. \revaug{The near-nose coefficients $\xi_7$ and $\xi_8$ lie at or above the sensitivity ceiling in both configurations, so the fit constrains only a lower bound there; $\xi_9$ and $\xi_{10}$ do so in RUN1 but are poorly determined in RUN2.}
    This pattern is qualitatively consistent with the obliquity-dependent turbulence picture in which the eastern (quasi-parallel) flank experiences stronger scattering than the western (quasi-perpendicular) flank.
    We caution, however, that the eastern $\xi$ values are constrained by far less observational information than their western counterparts, reflected in their lower \revaug{rank-correlation} importance scores, so the east--west pattern should be interpreted qualitatively rather than as a quantitatively constrained longitudinal profile.
    \revaug{Even so, a direct comparison against a spatially uniform baseline, in which $\xi_4$--$\xi_{10}$ are collapsed onto a single global value with the injection efficiencies held fixed, degrades the objective function from \revsep{0.458} to \revsep{0.703}, more than twenty times the session-to-session scatter of the optimization, because the eastern and western observers require mean free paths that differ in opposite directions. Within this comparison, in which the injection efficiencies were held fixed, no single global value reproduces the five spacecraft simultaneously, so the spatial parameterization of the mean free path is supported for this event, even though the individual western values are constrained only from below, with the exception of $\xi_9$ and $\xi_{10}$ in RUN2, which fall below the ceiling and are poorly determined. What the comparison establishes is hindcast capability for the 9 October 2021 event rather than predictive skill; a reduced model in which the injection efficiencies are re-optimized alongside a single global coefficient, together with held-out validation on a larger sample of events, is the natural next step.}

    \item \textbf{Injection efficiency as a diagnostic probe.}
    The longitude-dependent injection efficiencies $\epsilon_{0,k}$ exhibit large bin-to-bin scatter and differ substantially between the two MHD configurations, which is consistent with their intended role as diagnostic indicators of where the MHD-derived shock conditions are least accurate.
    Several bins in the RUN2 configuration yield $\epsilon_{0,k}$ values above the range typically adopted in DSA models ($> 10^{-2}$), pointing to specific heliolongitudes at which the shock conditions supplied by \texttt{SUSANOO} are least consistent with the observations.

    \item \textbf{Degeneracy with the MHD background.}
    Although the two MHD configurations produce nearly identical overall MAE, the individual best-fit parameters can differ substantially (e.g., $\epsilon_{0,4}$: $2.9 \times 10^{-2}$ in RUN1 vs.\ $1.2 \times 10^{-4}$ in RUN2, a difference of more than two orders of magnitude).
    Multi-run analysis shows that the injection efficiencies differ reproducibly between the two configurations at three near-nose longitude bins, with no overlap between the run-median ranges, while the western field lines are constrained too weakly for the corresponding differences in $\xi$ to carry physical meaning.
    These findings demonstrate that injection-side SEP parameters cannot be meaningfully interpreted without specifying the MHD context, and highlight the need for joint MHD--SEP parameter estimation.
    Moreover, because all runs in this study share the same ambient solar wind, with only the CME magnetic flux varied, the reported parameter spread represents a lower bound on the true systematic uncertainty; incorporating variations in the ambient solar wind model would likely increase it further.

    \item \textbf{Optimization efficiency.}
    The two-stage QMC+TPE strategy proves effective for this high-dimensional problem: the QMC stage provides broad coverage of the 19-dimensional space, and the subsequent TPE stage exploits the learned landscape to achieve further improvement.
    The best trials are found late in the optimization (trials \#1427 and \#1432 out of \revaug{1480}), and the top-5 trials cluster within a narrow MAE range, indicating convergence to a cluster of competitive low-error solutions.
\end{enumerate}

\subsection*{Outlook}

Several extensions of the present framework merit further investigation; here we highlight the most promising directions, with detailed physical motivation given in the corresponding Discussion subsections.

The most immediate priority is to apply the optimization to a statistically meaningful sample of SEP events spanning diverse CME speeds, flare longitudes, and heliospheric conditions.
Accumulated $\xi_j$ values would yield empirical prior distributions for new events, while systematic patterns in the diagnostic $\epsilon_{0,k}$ across events would expose persistent biases in the MHD model and target specific aspects of the shock parameterization in \texttt{SUSANOO} for refinement.

A closely related direction concerns the MHD input parameters themselves.
Y. Kato et al. (submitted) optimize the \texttt{SUSANOO} spheromak parameters by fitting the simulated CME-driven shock arrival times and in situ measurements at multiple spacecraft.
While that work does not treat the SEP transport parameters ($\xi_j$, $\epsilon_{0,k}$) optimized here, the two efforts are complementary, and the strong \revaug{SEP-parameter--MHD} degeneracy identified in this work motivates joint MHD--SEP estimation as an important next step.
Beyond the CME initial conditions, the ambient solar wind itself is a major source of uncertainty not explored in either study: variations in the input synoptic magnetogram or the coronal model can introduce CME-arrival uncertainties comparable to or exceeding those from the CME parameters \cite{2021SpWea..1902775R}, and ambient structures such as a high-speed stream adjacent to the CME-driven shock can substantially modify shock compression and obliquity, with direct consequences for the SEP intensity profiles \cite{2023ApJ...950..172W}.
\rev{The role of such large-scale structures of the background solar wind in shaping SEP propagation, including the effects of corotating interaction regions and perpendicular diffusion, has been the subject of growing simulation efforts \cite{2026ChA&A..50....1F}.}
Incorporating data-assimilated or ensemble-based ambient solar wind models into the optimization chain (alongside coronagraph-based GCS constraints \cite{Thernisien2011} that fix the CME geometry and reduce the MHD parameter dimensionality) is therefore essential for reducing the systematic uncertainties identified here.

On the model physics side, a particularly promising extension is, beyond the time-varying mean free path discussed in the Temporal Evolution of the Mean Free Path subsection, to model the enhanced scattering as a spatial effect tied to the approaching shock front (e.g., a mean free path that decreases towards the shock location), which would more naturally capture the turbulence amplification in the shock precursor region than the purely temporal exponential decay.

Finally, reducing the computational cost is essential for operational deployment.
\revaug{If the reduction in effective dimensionality suggested by the importance analysis is confirmed}, combining multi-event priors with surrogate-model-based optimization could substantially reduce the search space.
Together with the growing multi-spacecraft network (including proposed future missions to the Sun--Earth Lagrange L4 point\rev{, as well as near-Earth and sub-L1 monitors,} equipped with in-situ energetic particle instrumentation \cite{2021SpWea..1902777P, 2024JASS...41....1L, 2024aero.conf..202P, 2024SSRv..220...73L}), SEITEN has the potential to evolve into a quasi-real-time, physics-informed forecasting tool for the inner-heliospheric radiation environment.

A natural extension along this trajectory is sequential data-assimilation \cite{2026EP&S...78...64M}: refining the optimized parameters as new observations arrive during an ongoing event, so that the simulation transitions from a hindcast in early phases to a high-fidelity forecast for the remaining duration of the event.
This use case requires gradient information of the objective with respect to the free parameters, which the present black-box optimization obtains only through repeated forward simulations.
A natural complementary direction is therefore to make the transport solver itself differentiable, either by manual rewriting of \texttt{SEPIF} in a differentiable framework or by automatic-differentiation of a faithful emulator.
Ensemble Kalman or 4D-Var-style assimilation, gradient-based posterior sampling, and joint differentiable MHD--SEP optimization all become feasible once gradient access is available.
We regard this as a long-term direction and intend to explore it in dedicated follow-up work.

\section*{Methods} \label{sec:methods}

\subsection*{Models}

\subsubsection*{MHD model: \texttt{SUSANOO}} \label{subsec:susanoo}

To model the propagation of the solar wind and CME in the inner heliosphere, we used the \texttt{SUSANOO} model, a global three-dimensional MHD simulation of the heliosphere.
\rev{Because the governing MHD equations, boundary conditions, and numerical scheme of \texttt{SUSANOO} are described in full by Shiota et al.\cite{2014SpWea..12..187S} and Shiota \& Kataoka\cite{2016SpWea..14...56S}, we summarize here only the aspects relevant to the present study, whereas the \texttt{SEPIF} model, presented here for the first time, is described in detail in the following subsections.}
The model solves the solar wind plasma and magnetic field from the inner boundary at $25\,\slr{R}$ out to 2~au in three dimensions.
At the inner boundary, background solar wind plasma is injected and its propagation is calculated by solving the MHD equations.

The boundary conditions of the magnetic field are derived from coronal magnetic-field extrapolation based on the potential field source surface (PFSS) model using GONG magnetograms.
Using the PFSS results, the solar wind speed at the inner boundary is estimated with the empirical relation of Arge \& Pizzo\cite{2000JGR...10510465A}.
The plasma density and temperature are then derived from the solar wind speed using the empirical relations of Hayashi et al.\cite{2003JGRA..108.1102H}.
These empirical relations define the ambient (background) solar wind into which CMEs are launched. The same ambient solar wind is used for all ensemble runs in this study; only the CME spheromak parameters are varied across runs.
Recent studies have demonstrated that the accuracy of the ambient solar wind model can affect CME propagation as much as or more than the CME initial conditions themselves \cite{2021SpWea..1902775R, 2026arXiv260202300S}, and that ambient solar wind structures such as high-speed streams can substantially modify the CME-driven shock properties and the resulting SEP intensity profiles \cite{2023ApJ...950..172W}.
\rev{For the present event, Bamba et al.\cite{2026JGRA..13134744B} simulated the same 9 October 2021 CME with \texttt{SUSANOO} after superposing an idealized fast stream on the background wind, and found that the resulting interaction distorts the shock front and tilts the local shock normal away from the radial direction, most markedly near BepiColombo.}
Improving the fidelity of the ambient solar wind representation, \rev{including optimization of the background solar wind itself,} is an active area of research; in the present work, we rely on the established empirical approach described above and defer the exploration of data-assimilated or higher-fidelity solar wind models to future studies.

CMEs are introduced into the simulation domain as spheromak-type magnetic flux ropes. Spheromak-type CME injection in \texttt{SUSANOO} is described by a set of ten parameters, including, for example, the initial CME speed and the magnetic flux.
These spheromak parameters are determined with reference to flare observations and LASCO coronagraph observations, following the procedures described in Shiota \& Kataoka\cite{2016SpWea..14...56S} and Iwai et al.\cite{2021EP&S...73....9I}.

In the present study, rather than performing only a single simulation run based on one parameter set inferred from observations, we carried out a number of ensemble runs.
To select the most suitable run from among these ensemble members, we used not only in situ observations from multiple spacecraft but also IPS observations.

IPS is a radio-wave scattering phenomenon caused by density fluctuations in the solar wind.
From the three-dimensional plasma data produced by \texttt{SUSANOO}, synthetic IPS data can be generated by solving the scattering equation of radio waves along the line of sight to each radio source \cite{1971ApJ...168..543Y}.
We calculated synthetic IPS data for a number of ensemble members, compared them with the IPS observations at 327 MHz obtained by ISEE, Nagoya University, and selected the run that showed the best agreement with the observations based on Iwai et al.\cite{2019EP&S...71...39I}.
\rev{In this comparison, we assumed that the lines of sight showing enhanced IPS amplitude, expressed as an increased g-level, correspond to radio-source lines of sight intersecting the CME-driven shock and sheath region, and used the spatial and temporal distribution of these enhancements to evaluate the agreement between the synthetic and observed IPS signatures.}

The spheromak parameters used in this study are summarized in Table~\ref{tab:spheromak}.
\begin{table}[ht]
\centering
\caption{Spheromak input parameters for the two MHD configurations used in this study. All parameters except the magnetic flux (\textit{MAGFLUX}) are identical between RUN1 and RUN2. Parameter definitions follow Shiota \& Kataoka\cite{2016SpWea..14...56S}. \label{tab:spheromak}}
\begin{tabular}{lccl}
\hline
\textbf{Parameter} & \textbf{RUN1} & \textbf{RUN2} & \textbf{Description} \\
\hline
Launch time (UT) & \multicolumn{2}{c}{2021-10-09 06:19} & CME launch time \\
HCLONG ($^{\circ}$) & \multicolumn{2}{c}{5.0} & Stonyhurst heliographic longitude of CME centre \\
HCLATI ($^{\circ}$) & \multicolumn{2}{c}{9.0} & Stonyhurst heliographic latitude of CME centre \\
CMEVEL (km~s$^{-1}$) & \multicolumn{2}{c}{770} & Initial CME propagation speed \\
\textit{MAGFLUX} (Mx) & $1.2\times10^{21}$ & $8.0\times10^{20}$ & Total magnetic flux of the spheromak \\
RWIDTH & \multicolumn{2}{c}{3.0} & Aspect ratio of the spheromak cross-section \\
AWIDTH ($^{\circ}$) & \multicolumn{2}{c}{90.0} & Full angular width of the spheromak \\
TILTAN ($^{\circ}$) & \multicolumn{2}{c}{40.0} & Tilt angle of the internal field axis \\
CHIRAL & \multicolumn{2}{c}{$-1$} & Chirality ($+1$ = right-handed; $-1$ = left-handed) \\
DENSIT & \multicolumn{2}{c}{1.0} & Relative density of the CME plasma \\
\hline
\end{tabular}
\end{table}
For both MHD configurations (RUN1 and RUN2), all spheromak parameters except the magnetic flux were held fixed at the same values, guided by the graduated cylindrical shell (GCS) reconstruction of the same CME by Palmerio et al.\cite{2024JSWSC..14....3P}.
The initial CME speed was set to $v_{\mathrm{CME}} = 770$~km~s$^{-1}$, consistent with the GCS-derived speed.
The angular width was set to $w_A = 90^\circ$ (full angular width, following the definition in Shiota \& Kataoka\cite{2016SpWea..14...56S}), corresponding to a half-width of $45^\circ$, which lies within the GCS half-angular-width range of $28^\circ$--$49^\circ$ reported for this event.
A value near the upper end of this range was chosen to ensure robust initial magnetic connectivity between the CME-driven shock and the near-Earth spacecraft.
The tilt angle was set to $\theta_{\mathrm{tilt}} = 40^\circ$, matching the GCS value.
The aspect ratio was fixed at $R_W = 3$ and the chirality at $-1$ (left-handed).
The ejection direction was set to $(\theta, \phi) = (9^\circ, 5^\circ)$ in Stonyhurst heliographic coordinates, consistent with the GCS-derived propagation direction, placing the CME nose approximately $13^\circ$ west of the source active region (N18E08).
\revaug{Palmerio et al.\cite{2024JSWSC..14....3P} inserted the CME into WSA--Enlil as an unmagnetized hydrodynamic pulse, and their GCS reconstruction was performed independently of any flux-rope model, so the spheromak parameters listed above are not strictly equivalent to the GCS parameters.}
In particular, the spheromak angular width $w_A$ \rev{specifies the full angular width (full opening angle) of the initial flux-rope cross-section} on the inner boundary sphere (see Shiota \& Kataoka\cite{2016SpWea..14...56S}, Eq.~6), whereas the GCS half-angular width describes the lateral extent of the CME shell; similarly, the spheromak tilt angle $\theta_{\mathrm{tilt}}$ governs primarily the orientation of the internal magnetic field axis rather than the geometrical elongation of the envelope.
Despite these definitional differences, the adopted values provide a reasonable first-order correspondence with the observationally constrained GCS parameters.

Among the spheromak parameters, only the magnetic flux was varied across the ensemble runs.
From the ensemble, two configurations were selected for subsequent SEP optimization:
RUN1 (magnetic flux $= 1.2 \times 10^{21}$~Mx), which showed the best agreement with the IPS observations;
and RUN2 (magnetic flux $= 8.0 \times 10^{20}$~Mx), which best reproduced the CME-driven shock arrival time at Earth.
\rev{The simulated solar-wind plasma profiles for the two configurations are compared with in-situ observations at the five spacecraft in Fig.~\ref{fig:spacecraft_config}(b). }In terms of the CME time of arrival (ToA), RUN1 reproduces the arrival time at the location of BepiColombo reasonably well, whereas \rev{its shock arrives at Earth several hours earlier than observed (RUN1 was selected for its agreement with the IPS observations rather than the Earth arrival time, while RUN2 was selected to match the latter)}.
The possible cause and implications of this discrepancy are discussed in the Discussion.

From the MHD simulation results, the shock surface is extracted, and the shock parameters on that surface, including the angle between the shock normal and the magnetic field, are evaluated.
These parameters are then used as inputs to solve the particle equations described in the following sections.

\subsubsection*{SEP Acceleration and Transport Model: \texttt{SEPIF}} \label{subsec:sep_trans}
\rev{In \texttt{SEPIF} (the Solar Energetic Particle Injection and Focused transport model), }the transport of energetic particles from the shock front into interplanetary space is modelled by numerically solving the focused transport equation \cite{1995ApJ...442..861R,1998ApJ...509..415L,2026EP&S...78...64M},
\begin{eqnarray}
\pdif{f_j}{t} &=& -\pdif{}{z}\left[\mu v + \left\{ 1- \left(\frac{\mu v}{c}\right)^2 \right\}V_{sw}^c\right]f_j \nonumber \\
&& -\pdif{}{\mu} \left[-\frac{\left(1-\mu^2\right)v}{2} \pdif{\ln |B|}{z} \left\{1+\mu \left(\frac{V_{sw}^c}{v}-\frac{vV_{sw}^c}{c^2}\right)\right\} - \mu \left(1-\mu^2\right) \pdif{V_{sw}^c}{z}\right]f_j \nonumber \\
&& -\pdif{}{p}\left[p\left\{\frac{1-\mu^2}{2}V_{sw}^c \pdif{\ln |B|}{z} -\mu^2 \pdif{V_{sw}^c}{z} \right\}\right]f_j \nonumber \\
&& + \pdif{}{\mu} D_{\mu \mu} \pdif{}{\mu} \left(1-\frac{\mu v V_{sw}^c}{c^2}\right)f_j + Q_j, \label{FTE} 
\end{eqnarray}
where $f_j(z,\mu,p,t)$ denotes the particle distribution function; $z$ is the spatial coordinate along the mean magnetic field; $\mu$ is the cosine of the pitch angle; $v$ and $p$ are the proton velocity and momentum measured in the local solar wind frame; $c$ is the speed of light; $V_{sw}^c$ is the solar wind velocity in the frame co-rotating with the Sun; $|B|$ is the magnetic field strength; and $Q_j(z,\mu,p,t)$ represents the source term of energetic particles accelerated at the shock front.
The equation is solved independently along multiple magnetic field lines, which are labelled by the index $j$.
In equation \eqref{FTE}, the interplanetary magnetic field and the solar wind velocity are simplified by the steady-state Parker spiral field model,
\begin{eqnarray}
 B_r &=& B_0 \left(\frac{\slr{R}}{r}\right)^2, \label{eq:3} \\
 B_{\phi} &=& B_r \frac{\slr{\Omega}r}{V_{sw}} \left[\left(\frac{\slr{R}}{r}\right)^2-1\right], \\
 V_{sw}^c &=& V_{sw} \frac{|B|}{B_r}, \label{eq:5}
\end{eqnarray}
where $\slr{R}$ and $\slr{\Omega}$ are the solar radius and angular velocity, respectively.
We assume a uniform and constant solar wind speed of $V_{sw}=400 \; {\rm km \; s^{-1}}$\revaug{, chosen as a representative value close to the background solar wind speed actually observed at the four spacecraft shortly before the arrival of the CME-driven shock, which lies in the approximate range 300--400~km~s$^{-1}$ (Fig.~\ref{fig:spacecraft_config}(b)); it is not taken from the \texttt{SUSANOO} solution}, and the heliocentric distance $r$ is expressed as a function of $z$ using the relation $dr/dz = B_r/|B|$ with $z=0$ at $r=2.5 \slr{R}$.

The deviation from the motion along the mean field is modelled by the pitch-angle diffusion, while the diffusion across field lines is neglected.
We employ the pitch-angle diffusion coefficient $D_{\mu \mu}$ derived from a standard quasi-linear theory including an ad-hoc term to allow finite scattering through $\mu=0$,
\begin{eqnarray}
D_{\mu \mu} (\mu) &=& \frac{3 v}{2 \left(4-q\right) \left(2-q\right) \lambda_{j}} {\rm max}\left(|\mu|,\frac{\Delta \mu}{2}\right)^{q-1} \left(1-\mu^2\right), \label{eq:dmumu}\\
\lambda_{j} &=& \xi_j \frac{v}{\Omega_p}, \label{eq:mfp_def}
\end{eqnarray}
where $\Delta \mu$ is the grid spacing in $\mu$ space, $q=5/3$ corresponds to the spectral index of the power spectrum density of magnetic field fluctuations, and $\lambda_{j}$ denotes the mean free path parallel to the magnetic field.
The mean free path is assumed to be proportional to the local cyclotron radius, with the factor $\xi_j$ treated as a free parameter for each field line. \revaug{This linear proportionality to the gyroradius differs from the exponent that quasi-linear theory would imply for the same $q$; the choice is discussed in the Free Parameters subsection.}

We adopt an injection model to describe particle acceleration at the shock \cite{2017JGRA..12210938H}.
For each field line $j$ at each time $t$, let $\phi_j(t)$ denote the heliolongitude at which field line $j$ intersects the CME-driven shock surface, and let $k(j,t)$ denote the heliolongitudinal bin containing $\phi_j(t)$ (the bin specification is given later in this subsection).
The shock-derived quantities supplied by \texttt{SUSANOO} (the radial shock speed $V_{\mathrm{shk}}$, the upstream and downstream flow speeds in the shock rest frame $V_{\rm up}$ and $V_{\rm down}$, the upstream plasma density $n_{\mathrm{up}}$, and the angle between the shock normal and the magnetic field $\theta_B$) are functions of longitude on the shock surface and are sampled at $\phi_j(t)$ to evaluate the source term acting on field line $j$.
Quantities derived from these (such as the spectral index $\beta$, the injection momentum $p_{\rm inj}$, the maximum momentum $p_{\rm max}$, and the normalization constant $C$ defined below) inherit this longitude dependence.
The free injection efficiency varies stepwise across heliolongitudinal bins; the value used on field line $j$ is $\epsilon_{0,k(j,t)}$, abbreviated as $\epsilon_{0,k}$ when the dependence on $j$ and $t$ is clear from context.
The same convention applies to the shock mean free path coefficient $\xi^{\mathrm{shk}}_k$ introduced below.

When the shock front is located between $r_{i-1}$ and $r_i$, energetic particles are injected at $z_i$ on field line $j$ based on diffusive shock acceleration theory,
\begin{eqnarray}
Q_j(z_i,\mu,p,t) = C\, \epsilon\, n_{\mathrm{up}} \left|\frac{B_{1\,\mathrm{au}}}{B_i}\right| \mu^2 p^{-\beta} \frac{V_{\mathrm{shk}}}{r_i-r_{i-1}}, \;\;\; {\rm for \; } p \in [p_{\rm inj},p_{\rm max}], \mu \in [0,1], \label{eq:source_term}
\end{eqnarray}
where the index $j$ on $Q_j$ indicates that this is the source term acting on field line $j$, and all shock-derived quantities on the right-hand side are sampled at $\phi_j(t)$ as defined above.
The spectral index is given by $\beta = 3s/(s-1)$, where $s=V_{\rm up}/V_{\rm down}$ is the shock compression ratio.
The normalization constant is defined as $C=3(\beta-1)/[ p_{\rm inj}\{1-(p_{\rm max}/p_{\rm inj})^{1-\beta}\}]$.
The factor $1/|B_i|$ represents the cross-sectional area of a given flux tube to convert the volume density into the simulation unit of the line density.
The injection momentum $p_{\rm inj}$ is determined from the relation proposed by Li et al.\cite{2012AdSpR..49.1067L}\revaug{, for which a fuller derivation and a discussion of its range of validity are given by Ding et al.\ \cite{2023JGRA..12831502D}},
\begin{eqnarray}
 p_{\rm inj}=m V_{\rm up}\left[\left(\beta-1\right)+\tan\theta_B\right], \label{eq:p_inj}
\end{eqnarray}
where $m$ is the proton mass.

\revaug{Two features of equation~(\ref{eq:source_term}) require comment. First, the sign convention: $z$ increases outward from the Sun along each field line, so that $\mu > 0$ denotes motion away from the Sun and $\mu < 0$ motion towards it. The restriction $\mu \in [0,1]$ therefore means that particles are injected only into the outward-propagating hemisphere. This represents the population that escapes upstream from the shock and travels ahead of it along the field line towards the observer; particles injected with $\mu < 0$ would propagate back into the downstream region, which the present model does not follow, so no injection is applied there. In the implementation the restriction is enforced by a factor $[1+\mathrm{sgn}(\mu)]/2$ multiplying the angular term. Second, the angular dependence: the injected distribution is not isotropic over the outward hemisphere but is weighted as $\mu^{m_{\rm inj}}$ with $m_{\rm inj} = 2$, so that injection is maximal for field-aligned escape ($\mu = 1$) and vanishes smoothly at $\mu = 0$. It was chosen as the simplest increasing function of $\mu$ that vanishes smoothly at $\mu = 0$, reflecting the expectation that the escaping population is field-aligned. The angular term is normalized so that its integral over $\mu \in [0,1]$ is unity, that is, it is implemented as $(m_{\rm inj}+1)\mu^{m_{\rm inj}}$ with the constant absorbed into $C$; the total number of injected particles is therefore set by $\epsilon\, n_{\rm up} V_{\rm shk}$ and is independent of the assumed anisotropy, which redistributes those particles in pitch angle rather than changing their number. The mean pitch-angle cosine of the injected population is $(m_{\rm inj}+1)/(m_{\rm inj}+2) = 3/4$. Restricting the injection to $\mu > 0$ is likewise not expected to be a strong assumption: particles injected with $\mu < 0$ would in part be returned by magnetic mirroring in the converging field, so admitting them would not greatly change the population that ultimately propagates outward.}
The maximum momentum $p_{\rm max}$ is estimated from the model equation of diffusive shock acceleration \cite{1983RPPh...46..973D},
\begin{eqnarray}
 \dif{p_{\rm max}}{t} = \frac{V_{\rm up}^2}{\beta \kappa}p_{\rm max}, \;\;\;  p_{\rm max}(t=0)=p_{\rm inj}, \label{eq:pmax}
\end{eqnarray}
where the parallel spatial diffusion coefficient around the shock is given by
\begin{eqnarray}
 \kappa = \frac{\xi^{\mathrm{shk}}_k v^2 \cos^2\theta_B}{3 \Omega_p}, \label{eq:kappa}
\end{eqnarray}
and $\xi^{\mathrm{shk}}_k$ is a free parameter that controls the level of particle confinement at the shock.
\revaug{Equation~(\ref{eq:kappa}) retains only the parallel contribution to the diffusion coefficient along the shock normal. The full expression would be $\kappa = \kappa_\parallel \cos^2\theta_B + \kappa_\perp \sin^2\theta_B$, so that omitting $\kappa_\perp$ makes $\kappa$ vanish in the limit $\theta_B \to 90^{\circ}$ and underestimates it wherever the shock is quasi-perpendicular. In the present event this applies to the western flank and the region near the CME nose, where $\theta_B$ reaches $50^{\circ}$--$63^{\circ}$. Because $\kappa$ enters the model only through equation~(\ref{eq:pmax}), where $\mathrm{d}p_{\max}/\mathrm{d}t \propto 1/\kappa$, an underestimated $\kappa$ leads to an overestimated maximum momentum: at quasi-perpendicular geometry the model accelerates particles to high energies more readily than it should. The magnitude of the omission can be bounded. Adopting the strong-scattering, Bohm diffusion limit $\kappa_\perp/\kappa_\parallel = r_g/\lambda_\parallel$, and noting that $\lambda_\parallel = \xi^{\mathrm{shk}}_k r_g$ at the shock, the full coefficient becomes $\kappa = (v^2/3\Omega_p)\,(\xi^{\mathrm{shk}}_k \cos^2\theta_B + \sin^2\theta_B)$, which for the value $\xi^{\mathrm{shk}}_k = 1$ used throughout this study reduces to $\kappa = v^2/3\Omega_p$, independent of obliquity. Relative to equation~(\ref{eq:kappa}) this is larger by $1/\cos^2\theta_B$, that is by at most a factor of 4.9 at the largest obliquity encountered here. The consequence for the present results is correspondingly limited, because $\kappa \propto \xi^{\mathrm{shk}}$ and the preliminary sensitivity analysis described in the Free Parameters subsection found the simulated fluxes in the energy channels used here ($\lesssim 60$~MeV) to be insensitive to $\xi^{\mathrm{shk}}$ over orders of magnitude: a bounded factor of about five lies well inside the range over which the simulated profiles do not respond. The omission would nevertheless matter for a study extending to higher energies, where $p_{\max}$ becomes a binding constraint. We note it as a limitation of the acceleration description rather than of the transport description. We further note that the injection prescription of equation~(\ref{eq:p_inj}) is documented as applicable for shock-normal angles below about $85^{\circ}$, beyond which cross-field diffusion by field-line meandering governs the injection and the perpendicular diffusion coefficient must be retained explicitly \cite{2023JGRA..12831502D}. The largest obliquity encountered in the present event, $\theta_B \approx 63^{\circ}$, lies inside that range. The separate omission of cross-field diffusion in the interplanetary transport is discussed in the Model Coupling subsection and in the Discussion.}

In this approach, the number of accelerated particles depends on the injection efficiency $\epsilon$ appearing in $Q_j$, which is determined by the population of seed particles.
Since the knowledge of the seed particles remains limited, we adopt the model used in \cite{2017JGRA..12210938H}, which is derived under the assumption of a power-law distribution for seed population \cite{2012AdSpR..49.1067L},
\begin{eqnarray}
 \epsilon = \epsilon_{0,k} \min\left[1.0,\left(\frac{p_{\rm inj}}{p_{{\rm inj},\parallel}}\right)^{-5}\right], \label{eq:eff_model}
\end{eqnarray}
where $p_{{\rm inj},\parallel} = 3mV_{\rm up}$ is the injection momentum for strong parallel shocks $(\beta=4,\theta_B=0)$, $\epsilon_{0,k}$ is the reference injection efficiency in the heliolongitudinal bin $k=k(j,t)$ containing $\phi_j(t)$, and the right-hand side is evaluated on field line $j$ following the convention introduced above.

The phase space distribution is defined along 10 individual magnetic field lines co-rotating with the Sun, and it is discretized in $(z,\mu,p)$ space.
The computational domain spanning $z \in [0,2\;\mathrm{au}]$ $(r \in [0.0125\;\mathrm{au},1.5\;\mathrm{au}])$, $\mu \in [-1,1]$, and $p \in [0.015,0.46]$ $(E \in [0.1 {\rm MeV}, 100 {\rm MeV}])$ is sampled with 256, 32, and 16 grid points, respectively.
A logarithmically spaced grid is employed in momentum space.
The base longitude of each magnetic field line $j$ ($j = 1, \dots, 10$) at $r=2.5\slr{R}$ is given by $\phi_j = \phi_j(0)+\slr{\Omega} t$, and $\phi_j(0) \in [-82.5^{\circ},52.5^{\circ}]$.
Each field line is treated as representative of a $15^{\circ}$-wide longitudinal slab ($\pm 7.5^{\circ}$ around its central footpoint), implicitly assuming a uniform energetic-particle distribution within that slab.
Separately from the field line grid, the DSA injection parameters (the reference injection efficiency $\epsilon_{0,k}$ and the shock mean free path coefficient $\xi^{\mathrm{shk}}_k$) are specified on a heliolongitudinal grid of 18 bins ($k = 1, \dots, 18$) of width $\Delta\phi = 10^{\circ}$, spanning $\phi \in [-90^{\circ}, +90^{\circ}]$ in the simulation frame.
At each time step, the heliolongitude at which the MHD shock surface intersects a given field line $j$ determines the bin index $k$, and the corresponding values of $\epsilon_{0,k}$ and $\xi^{\mathrm{shk}}_k$ are used in the injection calculation for that field line.
For comparison with spacecraft observations, the omni-directional particle flux at an arbitrary radius and longitude is obtained by interpolating the simulated distribution function,
\begin{eqnarray}
F_j(z,p,t)= v \left|\frac{B}{B_{1\,\mathrm{au}}}\right| \int_{-1}^{1} f_j d\mu,
\end{eqnarray}
where the factor $v|B|$ is multiplied to convert the simulation unit into the observation unit of the particle flux.

\subsubsection*{Model Coupling}
\begin{figure}[htbp]
\centering
\includegraphics[width=0.95\linewidth]{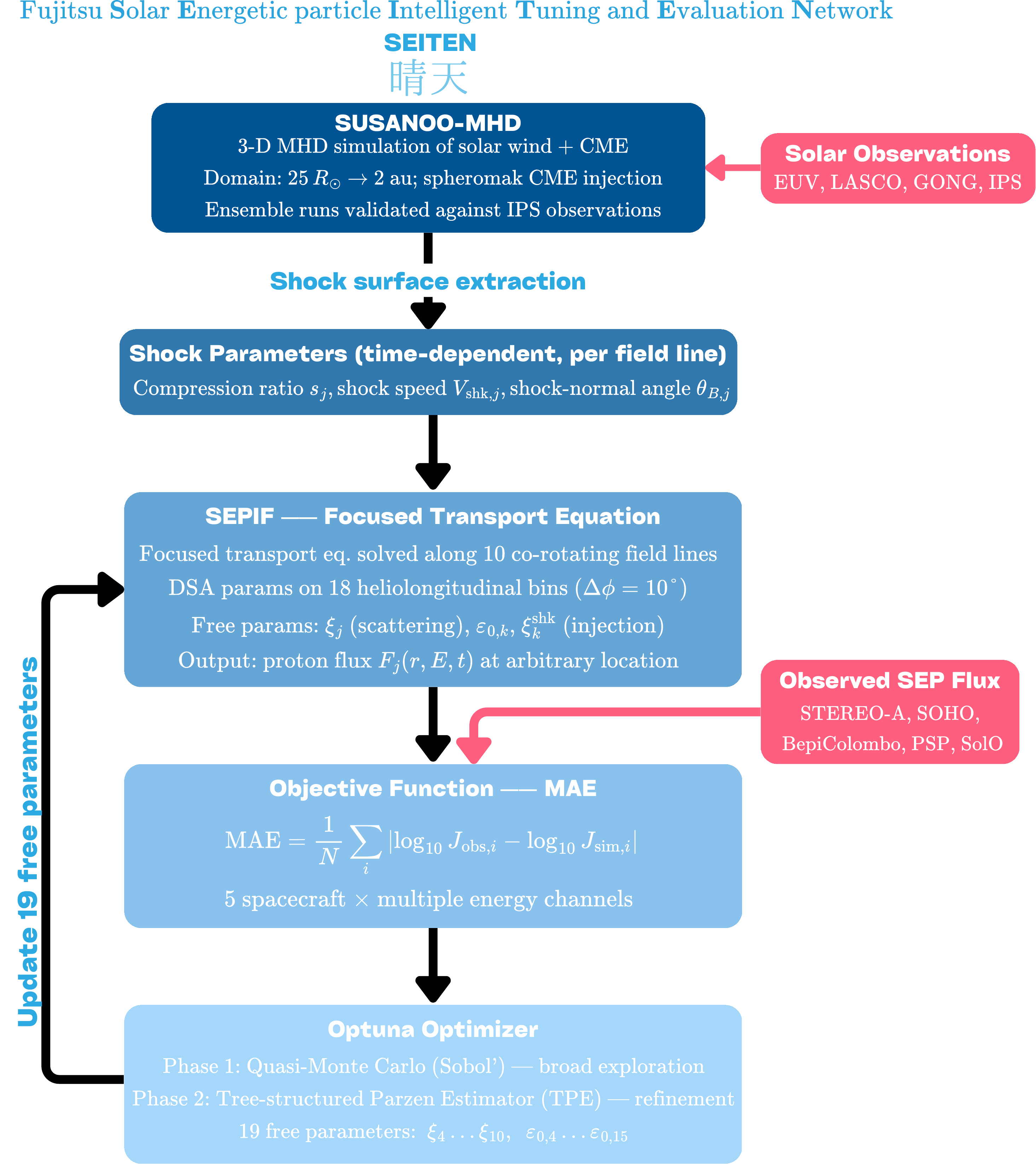}
\caption{%
Schematic overview of the model coupling and parameter optimization workflow.
The \texttt{SUSANOO} MHD simulation provides time-dependent shock parameters, which are passed to the \texttt{SEPIF} focused transport code.
\texttt{SEPIF} solves the particle transport along 10 co-rotating magnetic field lines, with DSA injection specified on 18 heliolongitudinal bins.
The simulated proton flux is compared with multi-spacecraft observations via the mean absolute error (MAE), and the 19 free parameters are iteratively updated by the Optuna optimizer using a two-phase strategy (Quasi-Monte Carlo followed by TPE).
}
\label{fig:model_coupling}
\end{figure}

We couple \texttt{SUSANOO} and \texttt{SEPIF} through a four-step pipeline (Fig.~\ref{fig:model_coupling}):
(1) extraction of shock-front quantities from the 3D MHD cube,
(2) shock detection on ecliptic slices,
(3) inward extrapolation from the \texttt{SUSANOO} inner boundary ($25\,\slr{R}$) to the \texttt{SEPIF} injection boundary ($2.5\,\slr{R}$),
and (4) transformation into the input variables required by the focused transport and DSA modules.

First, slice data are generated from the \texttt{SUSANOO} 3D outputs.
Shock positions are then detected at each time and longitude using the radial-velocity and density gradients.
For grid indices $(i_r, i_\phi, i_t)$ along the radial, longitudinal, and time axes, we evaluate
\begin{eqnarray}
\Delta v_r(i_r,i_\phi,i_t) &=& v_r(i_r+1,i_\phi,i_t)-v_r(i_r,i_\phi,i_t), \\
\Delta \rho_r(i_r,i_\phi,i_t) &=& \frac{\rho(i_r+1,i_\phi,i_t)-\rho(i_r,i_\phi,i_t)}{\rho(i_r+1,i_\phi,i_t)}.
\end{eqnarray}
A shock candidate is identified where both $\Delta v_r<-0.1$ and $\Delta \rho_r<-0.1$, and the most negative $\Delta v_r$ is selected as the shock location.
A local second-difference sign-change interpolation is then applied to obtain sub-grid shock radius $R_{\rm shk}$.
Upstream and downstream quantities are sampled around the detected front to construct time-dependent shock-surface parameters.

Because \texttt{SUSANOO} starts at $25\,\slr{R}$ while \texttt{SEPIF} starts at $2.5\,\slr{R}$, we perform the longitude-by-longitude inward extrapolation.
\revaug{All fits and averages of the MHD quantities that enter this reconstruction are taken over the heliocentric range $0.15$--$0.47$~au. The inner limit is set by the requirement that the shock be cleanly identifiable in the MHD output, which it is beyond about $0.15$~au; the outer limit of $0.47$~au ($100\,\slr{R}$) was chosen by inspection of the simulated profiles, as the range below which a linear or single power-law fit to the shock-surface quantities remains reasonable. The reconstruction therefore extrapolates the shock-surface quantities from this window inward to $2.5\,\slr{R} = 0.0125$~au, an order of magnitude below the inner edge of the window, and the shock is assumed not to spread in azimuth over that interval.}
The shock trajectory is fitted as
\begin{eqnarray}
R_{\rm shk}(t)=a_0 t^{a_1}+1\,\slr{R},
\end{eqnarray}
\rev{where the constant offset is fixed to $1\,\slr{R}$, assuming that the shock begins to propagate from the solar surface at the time of the flare onset. This constraint near the solar start condition stabilizes the inner-corona behaviour of the fit.}
\revaug{For the present event the fitted exponent lies in the range $a_1 \approx 0.8$--$0.85$ across longitudes, that is, a mildly decelerating shock.}
From this trajectory, the radial shock speed is evaluated and smoothed.
Additional quantities are reconstructed as follows:
\rev{for the radial flow components, the upstream flow is linearly regressed, while the downstream flow is determined from the velocity jump across the shock, $(V_{\rm shk}-V_{\rm up})/(V_{\rm shk}-V_{\rm down})$,} \revaug{whose median over the fitting window, measured in the shock rest frame, is applied unchanged at all smaller radii}\rev{; the azimuthal flow components are assumed uniform, as they are typically much slower than the radial components;}
$B_r$, $B_\phi$, and $n_{\mathrm{up}}$ are extrapolated inward using weighted fits (power-law behaviour in log space)\revaug{, the power-law form being motivated by the Parker solution assumed for the ambient field; for this event the fitted indices lie in the ranges $-3$ to $-1$ for $B_r$, $-2$ to $0$ for $B_\phi$, and $-2.5$ to $-1.5$ for $n_{\mathrm{up}}$, the first two showing considerably more scatter across longitudes than the third, as is to be expected where a spheromak flux rope has been injected and the field does not necessarily follow a Parker solution};
downstream density is \revaug{calculated from the extrapolated upstream density and the median of the compression ratio over the fitting window, rather than from a power-law extrapolation of $n_{\mathrm{down}}$ itself, which sometimes yields unphysically small or large compression ratios};
and shock-normal geometry is re-evaluated from neighbouring longitudes to maintain a smooth shock surface.
If the shock disappears at later times, $R_{\rm shk}$ is reset to the non-shock state and associated quantities are set to zero to avoid nonphysical tails.

\revaug{The uncertainties attached to this reconstruction are not uniform across the quantities involved, and we order them here by their expected effect on the SEP solution. The dominant one is the compression ratio. Its determination is the least well constrained step of the procedure, since it rests on a median taken over a window whose outer limit is itself set by inspection, and across the window it varies between roughly 2 and 5. It is also the quantity to which the source term responds most steeply: the compression ratio fixes the spectral index through $\beta = 3s/(s-1)$, which sets the ratio of high- to low-energy injected particles through equation~(\ref{eq:source_term}), and the injection efficiency of equation~(\ref{eq:eff_model}) carries a fifth-power dependence on $\beta - 1$. The size of the resulting effect on the injected particle number, and its bearing on the longitude dependence of $\epsilon_{0,k}$, are quantified in the Discussion. Next in importance is the shock-normal geometry, whose direction carries a comparable relative uncertainty but which enters the model in fewer places and therefore has a more limited effect. The shock position itself is the most robust of the reconstructed quantities: provided the assumption of no azimuthal distortion is accepted, a nearly uniform propagation reproduces it well. Finally, the extrapolation extends inward over an order of magnitude in radius, so its error must be expected to grow towards the inner boundary; we have not attempted to quantify that growth, and we note it as a limitation.}

After extrapolation, all variables are converted to the normalized units used in \texttt{SEPIF}
(length by au, time by au$/c$, velocity by $c$, magnetic field by G),
and transformed into the quantities directly used in equations~(\ref{FTE})--(\ref{eq:eff_model}) and the DSA source term.
The start time index for SEP calculation is chosen automatically as the first time when the median shock radius exceeds $2.5\,\slr{R}$, ensuring consistency with the transport-domain inner boundary.

The coupling module thus converts global MHD shock evolution from \texttt{SUSANOO} into time-dependent boundary and source parameters for \texttt{SEPIF}, bridging the radial gap between $25\,\slr{R}$ and $2.5\,\slr{R}$ with a controlled extrapolation procedure.

\revaug{The scope of the coupling is limited in one important respect. The information passed from \texttt{SUSANOO} to \texttt{SEPIF} consists solely of the shock-surface quantities listed in Supplementary Table~S1, and these enter the transport calculation only through the source term of equation~(\ref{eq:source_term}) and the quantities derived from it. The MHD solution is not used as the interplanetary transport background. \texttt{SEPIF} propagates particles along analytic Parker-spiral field lines defined by equations~(\ref{eq:3})--(\ref{eq:5}) with a uniform, constant solar wind speed of $V_{sw} = 400$~km~s$^{-1}$, independently of the solar wind speed, density and magnetic field that \texttt{SUSANOO} computes at the same positions, and diffusion across field lines is neglected. \revaug{Both treatments are in current use. In the iPATH model the shock is obtained from an MHD simulation while the transport of the escaped particles is solved on a Parker field with a constant solar wind speed \cite{2017JGRA..12210938H}, and in the iPATH--EUHFORIA coupling the shock parameters are taken from the MHD solution while the transport module continues to use the Parker field \cite{2022A&A...668A..71D}. Other frameworks instead propagate the particles in the MHD solution itself, among them SEPMOD as applied to the present event \cite{2024JSWSC..14....3P} and PARADISE driven by EUHFORIA \cite{2020PhDT.........8W, 2019A&A...624A..47W}. \texttt{SEITEN} belongs to the former class.} \revaug{For the latter we note that Minoshima et al.\ \cite{2026EP&S...78...64M} bounded the azimuthal transport of an SEP event observed by STEREO-A and near-Earth spacecraft and obtained $\lambda_\perp \lesssim 0.17$~au, which they took to justify a field-line-by-field-line one-dimensional treatment. That argument does not transfer directly to the present event, in which the five observers span \revproof{approximately $50^{\circ}$} in longitude, and the omission of cross-field diffusion is correspondingly less well justified here. Reinstating cross-field diffusion would not by itself resolve the ambiguity: its principal signature is a redistribution of particles in longitude, which in this event would be difficult to separate from the uncertainty in the longitudinal structure of the CME itself, and which our $10^{\circ}$ injection bins and $15^{\circ}$ field-line spacing are in any case too coarse to resolve.} The coupling is also one-way: the particle solution does not feed back on the MHD background. The optimized $\xi_j$ and $\epsilon_{0,k}$ should therefore be understood as \emph{effective} parameters, conditional on this assumed transport geometry, rather than as direct measurements of the underlying scattering and injection physics. Transport physics that the geometry omits, in particular perpendicular diffusion and any longitudinal structure of the solar wind speed, is necessarily absorbed into their fitted values.}
\rev{The full set of shock-surface quantities passed from the coupling module to \texttt{SEPIF} is listed in Supplementary Table~S1.}

\subsection*{Parameter Optimization Method} \label{sec:optimization}

\subsubsection*{Free Parameters} \label{subsec:free_parameters}
The SEP acceleration and transport model \texttt{SEPIF} described above contains a number of free parameters that control particle injection, acceleration, and scattering along each magnetic field line.
In the model, we parameterize the following three types of parameters as free variables across 10 heliospheric magnetic field lines and 18 heliolongitudinal bins: %

\begin{enumerate}
    \item \textbf{Mean free path coefficient} 
    $\xi_j$ ($j = 1, \dots, 10$): the ratio of the parallel scattering mean free path to the local cyclotron radius along each of the 10 magnetic field lines. 
    The instantaneous mean free path at position $z$ and momentum $p$ along field line $j$ is given by
    \begin{equation} \label{eq:mfp}
        \lambda_j(p, z) = \xi_j \left(\frac{p}{p_{\max}}\right) \left(\frac{B_{1\,\mathrm{au}}}{B(z)}\right) r_{g,\max},
    \end{equation}
    where $r_{g,\max}$ is the cyclotron radius at the maximum momentum $p_{\max}$, $B_{1\,\mathrm{au}}$ is the magnetic field strength at 1~au, and $B(z)$ is the local field strength.  
    A larger $\xi_j$ yields a longer mean free path (weaker scattering).
    \revaug{Equation~(\ref{eq:mfp}) and the definition $\lambda_j = \xi_j v / \Omega_p$ given in the \texttt{SEPIF} subsection are algebraically identical. Writing $r_{g,\max} = p_{\max} c /(e B_{1\,\mathrm{au}})$ and $r_g(p,z) = p c /(e B(z))$, the factors $p_{\max}$ and $B_{1\,\mathrm{au}}$ cancel and equation~(\ref{eq:mfp}) reduces to}
    \begin{equation} \label{eq:mfp_rg}
        \lambda_j(p,z) = \xi_j \, r_g(p,z),
    \end{equation}
    \revaug{so that $\xi_j$ is the parallel mean free path measured in units of the \emph{local} proton gyroradius, and $\lambda_j \propto p$. For a 10~MeV proton at 1~au with $B = 5$~nT one has $r_g = 6.1\times10^{-4}$~au, so that $\xi_j = 10^2$, $10^3$, $10^4$, $10^5$ and $10^6$ correspond to $\lambda_\parallel = 0.061$, $0.61$, $6.1$, $61$ and $610$~au respectively. The sensitivity ceiling at $\xi \approx 10^4$ reported in the Results therefore corresponds to $\lambda_\parallel \approx 6$~au \emph{evaluated at that energy and heliocentric distance}. Because $\lambda_j \propto 1/|B(z)|$, the local mean free path is orders of magnitude shorter in the inner heliosphere than this figure, so a value quoted at 1~au does not characterize the whole field-line path. We note for context that a previous application of the same $\lambda_\parallel = \xi v/\Omega_p$ parameterization to a different SEP event \cite{2026EP&S...78...64M} inferred $\lambda_\parallel \approx 0.5$--$1.0$~au with $\xi$ of order $10^3$, consistent with the conversion above.}

    \revaug{The exponent that relates $\lambda_j$ to the gyroradius is a setting of the transport solver rather than a quantity derived from the turbulence spectrum, and the value adopted throughout this study, $\lambda_j \propto r_g^{\,1}$, is not the one that quasi-linear theory would impose. Quasi-linear theory applied to a power spectrum of index $q$ gives $\lambda_\parallel \propto r_g^{\,2-q}$, that is $\lambda_\parallel \propto r_g^{1/3}$ for the Kolmogorov value $q = 5/3$ that enters the pitch-angle dependence of $D_{\mu\mu}$ in equation~(\ref{eq:dmumu}). The two prescriptions are not equivalent. The rigidity dependence of the parallel mean free path inferred from solar energetic particle observations is not in general the quasi-linear one, in either magnitude or rigidity dependence \cite{1982RvGSP..20..335P, 1994ApJ...420..294B, 2024ApJ...971..105L}, so imposing quasi-linear consistency is not by itself a compelling constraint on the formulation. With a unit exponent, moreover, $\xi_j$ is dimensionless and is directly the mean free path in units of the local gyroradius, which is convenient for an optimization in which all free parameters are sampled in a normalized logarithmic space.}

    \revaug{Two consequences should be stated. First, because $\xi_j$ is held fixed across energy in this study, the momentum dependence of $\lambda_\parallel$ is prescribed rather than measured, and the formulation does not coincide with the quasi-linear one. Second, the same structure offers a natural extension: determining $\xi_j$ separately in each energy channel would make the rigidity dependence of $\lambda_\parallel$ an output of the optimization rather than an assumption, and would allow multi-spacecraft observations to be used as an observational constraint on transport theory. We regard this as the more promising direction and defer it to future work.}
    \item \textbf{Injection efficiency} 
    $\epsilon_{0,k}$ ($k = 1, \dots, 18$): the reference injection efficiency at each of the 18 heliolongitudinal bins.
    \rev{Each $\epsilon_{0,k}$ is a time-independent constant attached to bin $k$; the apparent time dependence of the injection on a given field line $j$ arises solely from the changing field-line-shock intersection longitude $\phi_j(t)$, which selects different bins (and hence different $\epsilon_{0,k}$ values) at successive times via the assignment $k = k(j,t)$ defined in the \texttt{SEPIF} subsection.}
    The actual injection efficiency at the shock is modulated by the local shock geometry as described in equation~(\ref{eq:eff_model}).
    The injected particle flux is proportional to $\epsilon_{0,k}\, n_{\mathrm{up}}\, V_{\mathrm{shk}}$, where $n_{\mathrm{up}}$ is the upstream density and $V_{\mathrm{shk}}$ is the shock speed.
    \rev{For reference, a fixed injection efficiency of order $1\%$ ($\epsilon_0 \sim 10^{-2}$) of the upstream thermal population is a standard assumption adopted in previous DSA-based SEP models \cite{2000JGR...10525079Z, 2017JGRA..12210938H}; here we instead let $\epsilon_{0,k}$ vary around this canonical level to absorb the longitudinal non-uniformity of the injection (see below).}
    \item \textbf{Shock mean free path coefficient} 
    $\xi^{\mathrm{shk}}_k$ ($k = 1, \dots, 18$): the ratio of the mean free path to the cyclotron radius at the shock front for each heliolongitudinal bin.  
    This parameter controls the maximum momentum attainable by DSA: a smaller $\xi^{\mathrm{shk}}_k$ corresponds to stronger scattering at the shock and hence more efficient acceleration to higher energies.
\end{enumerate}

In our reduced-parameter configuration (hereafter the ``19-parameter model''), we restrict the search space based on physical considerations and preliminary grid-based sensitivity analyses.
Specifically, the shock mean free path coefficient is fixed at $\xi^{\mathrm{shk}}_k = 1.0$ for all 18 longitudinal bins.
This choice is motivated by the low sensitivity of the simulated fluxes to $\xi^{\mathrm{shk}}$ within the energy range of the present study ($\lesssim 60$~MeV), \rev{as confirmed by preliminary sensitivity analyses in which varying $\xi^{\mathrm{shk}}$ over orders of magnitude produced negligible changes in the simulated proton flux profiles for the energy channels used in the optimization.} %
The mean free path coefficients $\xi_1$--$\xi_3$ are fixed at $10^6$ (corresponding to very large mean free paths on far-eastern field lines whose footpoints lie far from any of the five spacecraft considered, so that their $\xi$ values have negligible influence on the simulated fluxes), while $\xi_4$--$\xi_{10}$ (7 parameters) are treated as free variables.
For the injection efficiency, $\epsilon_{0,4}$--$\epsilon_{0,15}$ (12 parameters) are optimized, while $\epsilon_{0,1}$--$\epsilon_{0,3}$ and $\epsilon_{0,16}$--$\epsilon_{0,18}$ are \revaug{held fixed at the arithmetic midpoint of the search bounds in physical-value space, $(10^{-4} + 3 \times 10^{-2})/2 = 1.505 \times 10^{-2}$}.
\revaug{These six outer bins lie outside the longitude range over which either MHD input has a valid shock ($s>1$; $-48.6^{\circ}$ to $+54.9^{\circ}$ across RUN1 and RUN2), so their fixed value does not enter the particle source term and does not affect the reported optimization results.}
This yields a total of 19 free parameters to be optimized simultaneously.

The two types of free parameters serve fundamentally different roles in the optimization.
The mean free path coefficients $\xi_j$ represent genuine physical degrees of freedom: the scattering environment is expected to differ among magnetic field lines owing to spatial variations in the turbulence level, and $\xi_j$ captures this physical inhomogeneity.
In contrast, the injection efficiencies $\epsilon_{0,k}$ are introduced as \emph{diagnostic} degrees of freedom.
If the shock parameters supplied by the MHD model (compression ratio, shock speed, shock-normal angle) were perfectly accurate at every heliolongitude, the injection model (equation~(\ref{eq:eff_model})) would already account for the longitudinal variation, and a single global $\epsilon_0$ would suffice.
By allowing $\epsilon_{0,k}$ to vary independently across longitude bins, the optimization can compensate for residual errors in the MHD-derived shock parameters; the resulting spatial pattern of $\epsilon_{0,k}$ therefore serves as a diagnostic indicator of where the MHD background or the DSA injection model is least accurate.

Supplementary Table~S2 summarizes the search bounds for each parameter type.
\revaug{All 19 free parameters} are sampled in logarithmic space, which is natural given that SEP fluxes span many orders of magnitude and the physical quantities themselves range over several orders of magnitude.
For the injection efficiency, previous studies of diffusive shock acceleration\cite{2000JGR...10525079Z, 2003JGRA..108.1082L, 2017JGRA..12210938H} generally find values $\epsilon_0 \lesssim 10^{-2}$; we intentionally extend the upper bound to $3 \times 10^{-2}$ so that if the optimizer drives $\epsilon_{0,k}$ beyond the physically plausible range, the result serves as an alert that the upstream MHD conditions or the injection model require revision at that longitude.

\subsubsection*{Optimization Algorithm} \label{subsec:algorithm}

To efficiently explore the 19-dimensional parameter space, we employ a multi-stage sampling strategy implemented within the Optuna hyperparameter optimization framework \cite{2019KDD_Akiba_Optuna}.
All parameters are internally mapped to a normalized space $[0, 1]^{19}$ via a logarithmic transformation:
\begin{equation} \label{eq:normalize}
    n = \frac{\log_{10}(x) - \log_{10}(x_{\min})}{\log_{10}(x_{\max}) - \log_{10}(x_{\min})},
\end{equation}
where $x$ is the physical parameter value and $x_{\min}$, $x_{\max}$ are the bounds given in Supplementary Table~S2.
The inverse mapping recovers the physical value:
\begin{equation} \label{eq:denormalize}
    x = 10^{\,\log_{10}(x_{\min})\,+\,n \cdot [\log_{10}(x_{\max}) - \log_{10}(x_{\min})]}.
\end{equation}
This normalization ensures that all samplers operate in a homogeneous unit hypercube, preventing parameters with vastly different scales from biasing the search.

Our optimization proceeds in two sequential stages, each employing a different sampling algorithm suited to its role:

\paragraph{Stage 1: Quasi-Monte Carlo (QMC) warmup.}
The first stage uses a Sobol' low-discrepancy sequence \cite{2001MaCom_Joe_Sobol} to uniformly fill the 19-dimensional normalized space.
Unlike pseudo-random sampling, the Sobol' sequence provides a more even coverage of the search space, which is particularly advantageous in high-dimensional settings where random sampling can leave large unexplored regions.
We employ the scrambled Sobol' variant to break correlations between successive samples while preserving the space-filling property.
In this stage, 1000 QMC trials are executed.

\paragraph{Stage 2: Tree-structured Parzen Estimator (TPE) exploration.}
The second stage applies the TPE algorithm \cite{2011NIPS_Bergstra_TPE}, a sequential model-based optimization method.
TPE models the conditional probability distributions of parameters given ``good'' (low objective) and ``bad'' (high objective) evaluations, and proposes new sample points that favour regions associated with good evaluations.
We configure TPE with multivariate modelling enabled (\texttt{multivariate=True}), which allows the estimator to capture inter-parameter dependencies rather than treating each dimension independently.
The \texttt{constant\_liar} strategy is activated to enable parallel trial execution: when multiple processes evaluate trials simultaneously, pending results are temporarily replaced by a constant value to avoid redundant exploration.
A total of 480 TPE trials are executed, building upon the knowledge gathered during the QMC warmup.
The TPE budget was informed by preliminary experiments with synthetic observations, in which the running best objective typically stabilized within $\sim$500 TPE evaluations even when the budget was extended to 1000.
\revaug{The behaviour on the real event is more varied across the six sessions. In one RUN2 session the overall best solution was already found during the QMC stage (trial 916), so no TPE-stage improvement is defined for it; of the remaining five sessions, four reached 80\% of their TPE-stage improvement within the first 104--191 TPE trials, while one required 432. No session found its overall best solution later than TPE trial 453. The session that required 432 trials indicates that the 480-trial budget is not generous, and we retain it for that reason.}
The total number of trials is therefore 1480.

Because each trial requires a full forward simulation of the SEP transport model, the computational cost per trial is substantial.
We parallelize the optimization by running multiple independent simulations concurrently.
The number of parallel processes $n_{\mathrm{jobs}}$ is determined automatically based on the available CPU resources as $n_{\mathrm{jobs}} = \max(1, \lfloor N_{\mathrm{CPU}} / 8 \rfloor)$, where $N_{\mathrm{CPU}}$ is the number of virtual CPUs accessible to the process.
Each process is allocated $\lfloor N_{\mathrm{CPU}} / n_{\mathrm{jobs}} \rfloor$ OpenMP threads for the transport solver.
The trial-level optimization state is coordinated through an SQLite database managed by Optuna, which ensures consistency even under concurrent access.

\subsubsection*{Objective Function} \label{subsec:objective}
The objective function to be minimized in the optimization process is a mean absolute error (MAE) computed in logarithmic space of the differential intensity across multiple spacecraft.
For each trial, the 19 normalized parameters are transformed to physical values via equation~(\ref{eq:denormalize}), combined with the fixed parameters to form a complete 46-parameter set ($\xi_{1\text{--}10}$, $\xi^{\mathrm{shk}}_{1\text{--}18}$, $\epsilon_{0,1\text{--}18}$), and passed to the \texttt{SEPIF} solver. %

After the simulation completes, the omni-directional proton flux $F_j$ output by \texttt{SEPIF} is converted to the differential intensity (directional average per unit solid angle) via
\begin{equation} \label{eq:diff_intensity}
    J^{\mathrm{sim}} = \frac{F_j}{4\pi},
\end{equation}
in units of particles~cm$^{-2}$~s$^{-1}$~sr$^{-1}$~(MeV/nuc)$^{-1}$, which matches the unit convention of the spacecraft particle instruments.
The resulting differential intensity time profiles are compared against observations from the five spacecraft.
For each spacecraft $s$ and each energy channel $c$, the MAE is computed in $\log_{10}$ space:
\begin{equation} \label{eq:mae_channel}
    \mathrm{MAE}_{s,c} = \frac{1}{N_{s,c}} \sum_{i=1}^{N_{s,c}} \left| \log_{10} J^{\mathrm{obs}}_{s,c}(t_i) - \log_{10} J^{\mathrm{sim}}_{s,c}(t_i) \right|,
\end{equation}
where $J^{\mathrm{obs}}$ and $J^{\mathrm{sim}}$ denote the observed and simulated differential intensities in units of particles~cm$^{-2}$~s$^{-1}$~sr$^{-1}$~(MeV/nuc)$^{-1}$, respectively, $t_i$ are the time stamps of the 5-minute resampled data, and $N_{s,c}$ is the number of valid data points satisfying $J^{\mathrm{obs}} \geq 10^{-4}$ and $J^{\mathrm{sim}} \geq 10^{-4}$.
The use of logarithmic space is natural because differential intensities span several orders of magnitude over the course of an event. %

\revaug{The observations are used as delivered by the instrument teams and are not background-subtracted. The only processing applied is the renaming of channel columns, conversion of the time stamps to UT, and resampling to a 5-minute cadence; no quiet-time or pre-event level is removed at any stage. Two consequences follow. First, strictly pre-eruption samples cannot enter the objective at all. The objective is evaluated on the time axis common to the observations and the simulation, and the simulated output begins only when the reconstructed shock reaches the inner boundary of the transport domain at $2.5\,\slr{R}$. For this event that occurs between 06:55 and 07:35~UT on 9 October depending on the spacecraft, which is after the eruption onset at 06:19~UT, so no pre-eruption sample is present on the evaluation axis, irrespective of the observed level and irrespective of the mask. Second, between the eruption and the arrival of particles at a given spacecraft the observed intensity can still be dominated by the quiet-time level, which does exceed $10^{-4}$ in some of the channels used. Those samples are excluded by the condition $J^{\mathrm{sim}} \geq 10^{-4}$ rather than by the choice of evaluation axis, and they therefore form part of the trial dependence described next. The absence of an explicit background component in the simulated intensity is a limitation of the present treatment. The nominal selection windows quoted in Supplementary Table~S3 begin earlier than this; they define the interval from which observations are drawn, not the support on which the objective is actually evaluated.}

\revaug{The additional condition $J^{\mathrm{sim}} \geq 10^{-4}$ does, however, make the set of evaluated points depend on the trial: a trial that underpredicts the intensity at a given time removes that point from the sum rather than being penalized for it. For the best-fit trials, 550 (RUN1) and 757 (RUN2) points out of 11\,030 are removed in this way, the earliest of them at 06:55~UT on 9 October, and the number of evaluated points varies between 9\,986 and 10\,480 (RUN1) and between 9\,956 and 10\,273 (RUN2) across trials. To assess the consequences, we re-scored the five highest-ranked trials of each configuration, for which the simulated time series had been retained, with the simulated intensity floored at $10^{-4}$ rather than masked, so that the number of evaluated points is fixed at 11\,030 for every trial and underprediction is penalized instead of removed. The MAE increases by \revsep{0.021--0.078} (RUN1) and \revsep{0.048--0.096} (RUN2). The same trial remains the best of the five in both configurations, although in RUN2 the ordering of the remaining four is not preserved, these five trials being separated by less than 0.02 in the original objective. We note that this check is restricted to trials that were selected as best under the original objective, and therefore cannot exclude the possibility that a trial ranked lower under that objective would be preferred under the floored formulation. An alternative formulation, in which the mask is defined as $J^{\mathrm{obs}} \geq 10^{-4}$ together with $t$ later than the eruption onset, gives results identical to the floored formulation, since the evaluation axis contains no pre-eruption bin under either definition.}

The per-spacecraft MAE is obtained by averaging over all energy channels:
\begin{equation} \label{eq:mae_sat}
    \mathrm{MAE}_s = \frac{1}{N_c^{(s)}} \sum_{c=1}^{N_c^{(s)}} \mathrm{MAE}_{s,c},
\end{equation}
where $N_c^{(s)}$ is the number of energy channels available for spacecraft $s$.
Supplementary Table~S3 lists the energy channels and MAE evaluation time windows used for each spacecraft.

For each spacecraft, the MAE is computed only within a restricted time window that ends before or near the CME-driven shock passage at the observation point \rev{(Supplementary Table~S3)}.
This restriction reflects a physical limitation of the current model: our treatment of the mean free path as time-independent (see the Free Parameters subsection) and the simplified post-shock downstream conditions do not capture the complex turbulence evolution and particle trapping that occur after the shock transit.
\revaug{In particular, at grid points located downstream of the shock front the mean free path of equation~(\ref{eq:mfp}) is overwritten at every time step by the Bohm limit, $\lambda = r_g$, so that the downstream transport is prescribed rather than governed by $\xi_j$. The post-shock intensities are consequently a poor basis for inferring the free parameters. The distribution function is continuous across the shock passage, so the intensity immediately after the passage still reflects the upstream transport history and retains some sensitivity to $\xi_j$; that sensitivity is, however, progressively overwritten by the prescribed downstream description as the decay phase proceeds, and the two contributions cannot be separated with the present treatment.}
Consequently, the simulated decay-phase fluxes are not expected to match observations in the post-shock regime, and including these intervals would bias the optimization towards unphysical parameter values.
\revaug{The decay-phase discrepancies discussed in the Multi-Point Flux Reproduction subsection and in the Temporal Evolution of the Mean Free Path subsection are therefore shown diagnostically and do not contribute to the optimization; the intervals that do contribute lie within the observation-selection windows shaded in Figure~\ref{fig:best_fit} and listed in Supplementary Table~S3; the samples that enter the objective are further restricted to the time axis common to the observations and the simulation and by the evaluation mask.}
By restricting the MAE evaluation to the pre-shock interval, where the focused transport equation and the DSA injection model are most applicable, we ensure that the optimized parameters reflect the physically meaningful acceleration and transport processes.

The final scalar objective value returned to the optimizer is a weighted average of the per-spacecraft MAEs:
\begin{equation} \label{eq:mae_final}
    \mathcal{L} = \frac{\sum_{s} w_s \, \mathrm{MAE}_s}{\sum_{s} w_s},
\end{equation}
\revaug{where the weights $w_s$ were set to unity for STEREO-A, SOHO, PSP and Solar Orbiter, and to 0.7 for BepiColombo. The same weights were used in all six optimization sessions and in both MHD configurations, and were fixed before the optimization was run.}

\revaug{Because the objective averages the mean absolute error over spacecraft rather than over individual data points, each spacecraft contributes to $\mathcal{L}$ in proportion to its weight and independently of how many channel-time points it supplies. BepiColombo/BERM provides two proton channels against three to six for the other spacecraft, has the shortest observation-selection window of the five (Supplementary Table~S3), and is subject to the field-of-view restriction described in the BepiColombo subsection. It accordingly contributes 250 channel-time points to the objective function, against 4\,165 for STEREO-A. Under strictly equal spacecraft weighting the smallest and most restricted of the five data sets would exert the same leverage on the fit as the largest; the factor 0.7 is a partial correction for that imbalance.}

\revaug{For reference, the strictly unweighted mean of the five per-spacecraft MAE values is \revsep{0.461} for the RUN1 best-fit trial and \revsep{0.460} for the RUN2 best-fit trial, against \revsep{0.458 and 0.446} for the weighted objective. The difference arises almost entirely from BepiColombo, whose per-spacecraft MAE is 0.505 in RUN1 and 0.679 in RUN2. We have also re-scored the five highest-ranked trials of each configuration under strictly equal weighting. \revsep{The ordering is unchanged in both configurations: trials \#1432 and \#1427 remain the best of their respective five under equal weighting, with unweighted means of 0.461 and 0.460. The five trials remain tightly clustered under either weighting (0.458--0.482 and 0.446--0.480 weighted, 0.461--0.502 and 0.460--0.480 unweighted).} The RUN2 column of Table~\ref{tab:best_params} should nevertheless be read together with the sensitivity ceiling and the multi-run statistics presented in the Results rather than as a uniquely determined parameter set. We have not re-scored every trial under equal weighting, so we do not claim that the ensemble-level results would be unchanged under a different weighting; we note only that they are consistent across all six sessions under the weighting actually used.}

\revsep{One accounting point should be recorded here. In the optimization runs, the Solar Orbiter EPT channel at 5.83--6.13~MeV did not contribute a residual to equation~(\ref{eq:mae_channel}): the observed series for that channel was not correctly loaded, which we identified in checks carried out after the runs. Since equation~(\ref{eq:mae_sat}) averages over that spacecraft's four channels, Solar Orbiter entered the numerator of equation~(\ref{eq:mae_final}) with an effective coefficient of $3/4$. The objective that was minimized is well defined and the parameters reported here are its optimum; the departure from the nominal weighting is of the same order as the factor 0.7 applied deliberately to BepiColombo, and the equal-weight re-scoring described above shows that the leading solutions are not sensitive to weighting at that level. The objective values quoted throughout this paper are computed with the EPT channel compared against the observation and therefore differ from those minimized during the runs. The optimized parameters, the comparison against the spatially uniform baseline, and the longitudinal structure of the mean free path are unchanged. Figures showing objective values for individual trials are left as computed during the runs; the correction is a near-uniform offset of about 0.02--0.03.}

Trials that fail due to simulation errors (e.g., missing output files, numerical divergence) are assigned \rev{large penalty values} far exceeding any physically meaningful MAE.
\rev{These specific numerical values are arbitrary placeholder values chosen only to lie far above the range of any successful trial (MAE of order unity); the distinct values used for different error categories carry no quantitative meaning and serve only to identify the category of failure afterwards.}
These penalty values ensure that the optimizer is steered away from pathological regions of parameter space without requiring explicit constraint handling.

\subsection*{\rev{Use of AI-assisted tools}}
\rev{During the preparation of this manuscript, the authors used GitHub Copilot (GitHub, Microsoft) and Claude (Claude Opus, Anthropic) to assist with writing and refactoring the data-analysis and optimization code, and Claude (Claude Opus, Anthropic) to improve the language and clarity of the manuscript. All AI-generated code was validated against the authors' own implementations and reference results, and all AI-assisted text was reviewed and edited by the authors, who take full responsibility for the content of this article.}

\section*{Acknowledgements}

We thank Yoshizumi Miyoshi and Hideyuki Hotta for insightful discussions that helped shape the interpretation presented in this work.
This study utilized the \texttt{SUSANOO} heliospheric simulation, which is operated by the Institute for Space-Earth Environmental Research (ISEE), Nagoya University (Shiota et al.\cite{2014SpWea..12..187S}, \url{https://cidas.isee.nagoya-u.ac.jp/susanoo/}).
This work was carried out by the joint research programme of ISEE, Nagoya University. %

We acknowledge the instrument teams and data archives whose observations made this study possible. We thank the STEREO IMPACT (LET, HET and MFE) and PLASTIC instrument teams. Courtesy of the SOHO/COSTEP and SOHO/LASCO consortia; SOHO is a project of international cooperation between ESA and NASA. We thank the BepiColombo MPO-MAG team. We acknowledge the NASA Parker Solar Probe Mission and the SWEAP team led by J. Kasper for use of data. Thanks to the Integrated Science Investigation of the Sun (IS$\odot$IS) Science Team (PI: D. J. McComas, Princeton University). The FIELDS experiment on the Parker Solar Probe spacecraft was designed and developed under NASA contract NNN06AA01C. Solar Orbiter is a mission of international cooperation between ESA and NASA, operated by ESA. This paper uses EPD data, generated and maintained by the EPD team; we also thank the Solar Orbiter MAG and SWA teams. We thank the WIND MFI and SWE teams. Some of the data used here were obtained from NASA's Space Physics Data Facility (SPDF) via CDAWeb.

\section*{\rev{Funding}}

\rev{K.I. was supported by JSPS KAKENHI Grant Number 25K24625. D.S. was supported by JSPS KAKENHI Grant Number 25K01052. M.P. was funded by the Portuguese Foundation for Science and Technology (FCT) through the research grant no.\ UI/BD/154742/2023, under the framework of the project `Advanced Methods for Solar Energetic Particle Events Characterization in the Inner Solar System', and by the European Space Agency (ESA) through contract no.\ 4000137865/22/ES/JD (Expert Support to BERM \& RADEM units).}

\section*{Author contributions statement}

Y.K. conceived and initiated the project, with K.K. providing scientific guidance from the outset. H.T.I. led the overall study and developed the SEITEN framework. D.S. established and maintained the \texttt{SUSANOO} MHD simulation environment, enabling its use for the simulations performed in this study. T.M. developed \texttt{SEPIF}, supported by the joint research programme of Tokai National Higher Education and Research System; N.F. implemented model upgrades. The optimization framework was developed iteratively: Y.K. introduced Optuna and wrote the initial code, M.K. iterated on and refined the implementation, and H.T.I. consolidated and systematized these efforts into the final framework. For observational data, A.P.G. performed reduction, calibration, and management of the BepiColombo/BERM data as part of the instrument team, G.M. provided access to the data, and M.P. advised on its scientific use; Y.K. and M.K. collected public data from STEREO, SOHO, WIND, PSP, and Solar Orbiter; K.I. provided IPS data from the Nagoya ISEE group. H.T.I. led the optimization analysis, post-optimization evaluation, and interpretation in close discussion with N.F. and Y.K., and N.F. additionally performed supporting analyses underpinning key arguments. H.T.I. wrote the manuscript, with Y.K., T.M., and K.I. contributing assigned sections. K.K. supervised the study. C.M. provided project administration at Fujitsu and secured funding for the work. All authors reviewed and approved the manuscript.

\section*{Additional information}

\textbf{Competing interests:} The authors declare no competing interests.

\section*{Data and code availability}

This study used data from the following spacecraft instruments: STEREO-A (IMPACT/MFE, IMPACT/LET, IMPACT/HET, PLASTIC), SOHO (LASCO, COSTEP), WIND (MFI, SWE), BepiColombo (MPO-MAG, BERM), PSP (FIELDS, SWEAP/SPC, IS$\odot$IS), Solar Orbiter (MAG, SWA/PAS, EPD).

BERM data are available from the Planetary Science Archive (PSA: \url{https://psa.esa.int/psa/#/pages/home}). %

The software packages used in this study include: Optuna \cite{2019KDD_Akiba_Optuna}, SunPy v6.0.3 \cite{sunpy_community2020}, Astropy v6.1.1 \cite{astropy:2022}, SEPpy v0.1.16 \cite{Palmroos2022} used for PSP/IS$\odot$IS Level~2 1~minute cadence data with seppy.loader.psp and Solar Orbiter/EPD Level~2 data of 5~minute cadence (retrieved from \url{https://espada.uah.es/epd/data/archive} with solo\_epd\_loader version 0.3.7 \cite{Gieseler_solo-epd-loader_2024}), SOHO/COSTEP Level~3 data of 5~minute cadence  (\url{http://ulysses.physik.uni-kiel.de/costep/level3/l3i/}), STEREO Level~1 Standard and H data (\url{https://izw1.caltech.edu/STEREO/DATA/Level1/Public/ahead/1Minute/}), NumPy \cite{2020Natur.585..357H}, Matplotlib \cite{2007CSE.....9...90H}, pandas \cite{2010mcKinney_pandas}, SciPy \cite{2020NatMe..17..261V}.
\rev{The \texttt{SEPIF} transport solver is parallelized with OpenMP for shared-memory multi-threading, and the optimization trial history and state are managed through an SQLite database via Optuna.}

The IPS datasets used in this article are available in the data repository of ISEE, Nagoya University \url{http://stsw1.isee.nagoya-u.ac.jp/ips_data-e.html}, BepiColombo/MPO-MAG is available in the \url{https://psa.esa.int/psa/#/pages/search}.
The magnetic field and plasma data for STEREO-A IMPACT/MFE and PLASTIC, WIND MFI and SWE, PSP/FIELDS and SWEAP/SPC, Solar Orbiter/MAG and SWA/PAS were retrieved using PySPEDAS v1.7.28 \cite{Grimes2022_pyspedas}.

\revaug{The \texttt{SEITEN} optimization framework includes software components developed within Fujitsu Limited and is not publicly released. The optimization procedure is specified in full in this article: the two-stage sampling strategy and its settings, the normalization of the parameters (equations~(\ref{eq:normalize}) and (\ref{eq:denormalize})), the objective function together with its mask and spacecraft weights (equations~(\ref{eq:mae_channel})--(\ref{eq:mae_final})), the search bounds (Supplementary Table~S2), and the energy channels and observation-selection windows used for each spacecraft (Supplementary Table~S3). The framework can therefore be reimplemented from the description given here. The source code and the derived data required to reproduce the analyses reported in this article may be made available to researchers for non-commercial academic use, on reasonable request to the corresponding author. Access is subject to approval by Fujitsu Limited and to a written agreement that specifies the intended use and excludes commercial exploitation and redistribution. Requests will be considered individually and are normally answered within eight weeks.}

\bibliography{01_bib_sep_optimize}

\end{document}


\begin{center}
{\Large\bfseries Supplementary Information}\\[1.2em]
{\large for}\\[0.6em]
{\large\bfseries \revaug{Automated Optimization of Solar Energetic Particle Acceleration and Transport Parameters Against Five-Spacecraft Observations of the 9 October 2021 Event}}\\[1.4em]
Hiroyuki Tako Ishikawa$^{1,*}$, Naho Fujita$^{1}$, Yuta Kato$^{1,2}$, Miki Kurihara$^{1}$, Takashi Minoshima$^{3}$, Daikou Shiota$^{2,4}$, Kazumasa Iwai$^{2}$, Marco Pinto$^{5}$, Ant\'{o}nio Pessanha Gomes$^{5,6}$, Go Murakami$^{7}$, Kanya Kusano$^{2}$, and Chihiro Mitsuda$^{1,2}$\\[1.0em]
{\small
$^{1}$Space Data Frontiers Research Center, Fujitsu Research, Fujitsu Limited, Kanagawa, Japan\\
$^{2}$Institute for Space--Earth Environmental Research, Nagoya University, Aichi, Japan\\
$^{3}$Center for Mathematical Science and Advanced Technology, Japan Agency for Marine-Earth Science and Technology, Kanagawa, Japan\\
$^{4}$Radio Research Institute, National Institute of Information and Communications Technology, Tokyo, Japan\\
$^{5}$Laboratory of Instrumentation and Experimental Particle Physics, Lisboa, Portugal\\
$^{6}$Instituto Superior T\'{e}cnico, Universidade de Lisboa, Lisboa, Portugal\\
$^{7}$Institute of Space and Astronautical Science, Japan Aerospace Exploration Agency, Kanagawa, Japan\\[0.5em]
$^{*}$ishikawa.h.tako@fujitsu.com
}
\end{center}

\vspace{1em}

\setcounter{figure}{0}
\renewcommand{\thefigure}{S\arabic{figure}}
\setcounter{table}{0}
\renewcommand{\thetable}{S\arabic{table}}

\begin{figure}[ht]
\centering
\includegraphics[width=0.48\linewidth]{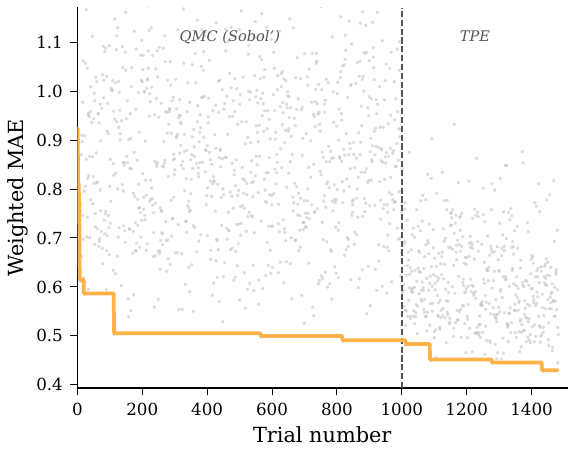}\hfill
\includegraphics[width=0.48\linewidth]{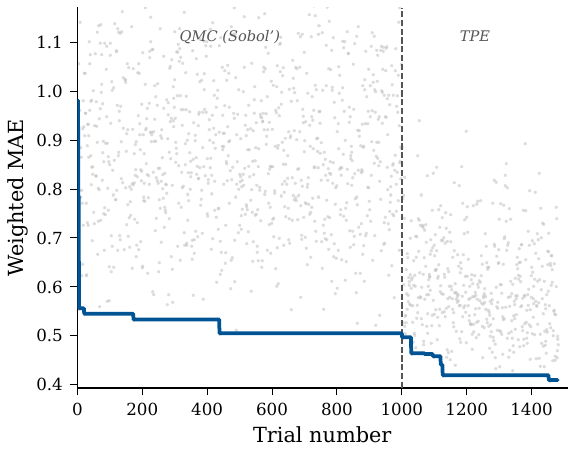}
\caption{Optimization history for the RUN1 (left) and RUN2 (right) MHD configurations. Each point represents the objective function value \revaug{(the weighted MAE of the main-text equation for $\mathcal{L}$, in $\log_{10}$ space)} of a single trial; the solid curve traces the running best. The vertical dashed line at trial 1000 marks the transition from QMC (Sobol') sampling to TPE.
\label{fig:opt_history}}
\end{figure}

\begin{figure}[ht]
\centering
\includegraphics[width=0.75\linewidth]{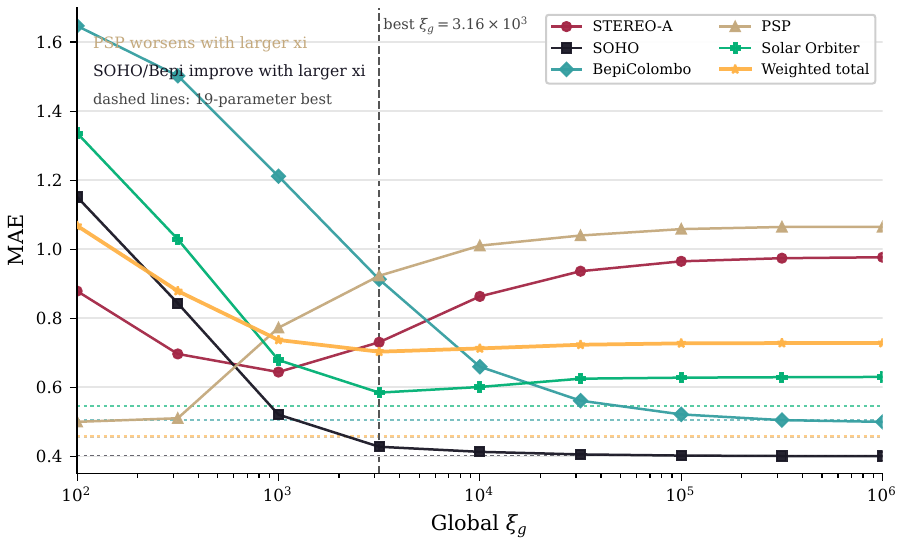}
\caption{\revaug{Response of the objective function to collapsing the seven free mean free path coefficients onto a single global value $\xi_g$, with the twelve injection efficiencies held at the RUN1 best-fit values. The horizontal axis is $\xi_g$ on a logarithmic scale; the vertical axis is the mean absolute error. Curves show the per-spacecraft MAE for the five observers and the weighted total. The horizontal dashed line marks the objective value of the 19-parameter best-fit solution and the vertical dashed line marks the best global value, $\xi_g = 3.2\times10^{3}$. PSP, the only spacecraft magnetically connected to the eastern field lines, degrades monotonically as $\xi_g$ increases, whereas SOHO and BepiColombo improve monotonically, so that no single value satisfies all five observers simultaneously.}
\label{fig:sweep_mae_vs_xi}}
\end{figure}

\begin{figure}[ht]
\centering
\includegraphics[width=\linewidth]{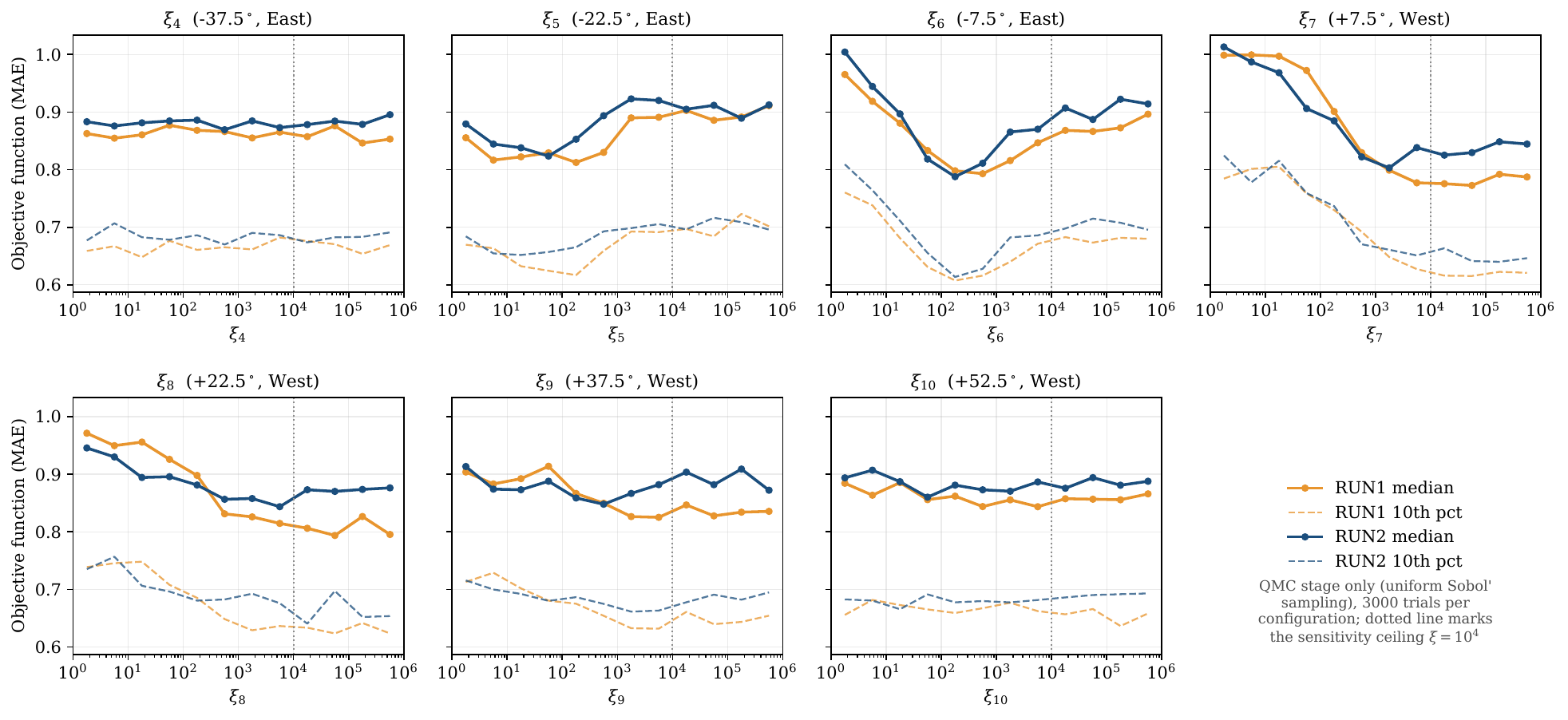}
\caption{\revaug{Marginal response of the objective function to each free mean free path coefficient. Each panel shows, for one coefficient $\xi_j$, the median objective value among all trials falling in a given logarithmic bin of $\xi_j$ (solid, circles) and the corresponding 10th percentile (dashed), for RUN1 (orange) and RUN2 (blue). Only the QMC stage is used, so that the trials are a uniform Sobol' sample of the 19-dimensional search space and the marginal is not biased by the adaptive sampling of the TPE stage; each configuration contributes 3000 trials. The dotted vertical line marks the sensitivity ceiling $\xi = 10^4$. The eastern coefficients $\xi_5$ and $\xi_6$ show interior optima, $\xi_7$ and $\xi_8$ decrease monotonically and then plateau above the ceiling, and $\xi_4$ and $\xi_{10}$ are flat across the entire search range. Values are marginalized over the remaining 18 parameters, so a flat curve indicates that the objective does not respond to that parameter on average, not that it is unconstrained under every configuration of the others.}
\label{fig:objective_vs_xi}}
\end{figure}

\begin{table}[htbp]
\centering
\caption{Shock parameters passed from the coupling module to \texttt{SEPIF}. All shock-surface quantities listed below are functions of heliolongitude $\phi$ on the shock front. When evaluating the source term $Q_j$ acting on field line $j$, these quantities are sampled at $\phi_j(t)$, the heliolongitude where field line $j$ intersects the shock at time $t$.}
\label{tab:model_coupling_params}
\begin{tabular}{llp{0.42\linewidth}}
\hline
\textbf{Quantity} & \textbf{Symbol used in this paper} & \textbf{Role in \texttt{SEPIF}} \\
\hline
Time coordinate & $t$ & Time axis for transport and injection \\
Heliolongitude on shock surface & $\phi$ & Field-line--shock intersection $\phi_j(t)$ and bin mapping $k(j,t)$ \\
Shock radius & $R_{\rm shk}$ & Shock position along each magnetic field line \\
Radial shock speed & $V_{\mathrm{shk}}$ & DSA source term in $Q_j$ \\
Upstream speed in shock frame & $V_{\rm up}$ & Injection momentum and spectral index \\
Downstream speed in shock frame & $V_{\rm down}$ & Compression and spectral index \\
Upstream density & $n_{\mathrm{up}}$ & Source normalization in $Q_j$ \\
Compression ratio & $s=V_{\rm up}/V_{\rm down}$ & $\beta=3s/(s-1)$ \\
Magnetic-field magnitude & $|B|$ & Transport coefficients and flux conversion \\
Shock obliquity angle & $\theta_B$ & Injection/acceleration efficiency terms \\
\hline
\end{tabular}
\end{table}

\begin{table}[ht]
\centering
\caption{Parameter bounds and scaling for the 19-parameter optimization. \label{tab:param_bounds}}
\begin{tabular}{lcccl}
\hline
\textbf{Parameter} & \textbf{$N_{\mathrm{free}}$} & \textbf{Min} & \textbf{Max} & \textbf{Scale} \\
\hline
$\xi$ (mean free path coefficient) & 7 & 1 & $10^6$ & log \\
$\xi^{\mathrm{shk}}$ (shock mean free path coefficient) & 0 (fixed) & \multicolumn{2}{c}{1.0} & --- \\
$\epsilon_0$ (injection efficiency) & 12 & $10^{-4}$ & $3 \times 10^{-2}$ & log \\
\hline
\end{tabular}
\end{table}

\begin{table}[ht]
\centering
\small
\caption{Energy channels and observation selection windows used for each spacecraft. \rev{The window end times were chosen to terminate before or near the local CME-driven shock passage at each observation point and therefore differ from the intervals adopted in other studies of this event (e.g., Palmerio et al.\ (2024, J.\ Space Weather Space Clim.\ 14, 3)). }\revaug{The windows listed here define the interval from which the observations are drawn. The objective function is evaluated on the time axis common to the observations and the simulation, and the simulated output begins only when the reconstructed shock reaches the inner boundary of the transport domain; the earliest sample that can therefore contribute lies between 06:55 and 07:35~UT on 9 October depending on the spacecraft, so no sample earlier than that enters the objective even though the windows nominally start at 00:00~UT.} \revsep{The last column gives the viewing direction of the data used: \emph{integrated} denotes a direction-integrated product, \emph{single} a single-aperture instrument, and the remaining entries the telescope selected.}\label{tab:channels_windows}}
\setlength{\tabcolsep}{4pt}
\begin{tabular}{llp{0.30\linewidth}lc}
\hline
\textbf{Spacecraft} & \textbf{Instrument(s)} & \textbf{Energy channels (MeV)} & \revaug{\textbf{Observation selection}} & \revsep{\textbf{Viewing}} \\
 & & & \revaug{\textbf{window (UTC, 2021)}} & \revsep{\textbf{direction}} \\
\hline
STEREO-A & LET, HET & 4.0--4.5, 4.5--5.0, 5.0--6.0, 14.9--17.1, 29.5--33.4, 40.0--60.0 & 9 Oct 00:00 -- 12 Oct 00:15 & \revsep{integrated} \\
SOHO & COSTEP/EPHIN & 4.3--7.8, 7.8--25, 25--40.9, 40.9--53.0 & 9 Oct 00:00 -- 12 Oct 00:52 & \revsep{single} \\
BepiColombo & BERM & 1.5--5.9, 5.9--9.1 & 9 Oct 00:00 -- 9 Oct 23:46 & \revsep{anti-sunward} \\
PSP & IS$\odot$IS/EPI-Hi & 13.5--16.0, 26.9--32, 45.3--53.8 & 9 Oct 00:00 -- 11 Oct 08:05 & \revsep{HET, A} \\
Solar Orbiter & EPD/EPT, HET & 5.83--6.13, 14.6--15.7, 29.4--32.0, 49.1--53.4 & 9 Oct 00:00 -- 11 Oct 01:52 & \revsep{sunward} \\
\hline
\end{tabular}
\end{table}
\begin{table}[ht]
\centering
\small
\caption{\revaug{Objective function of the spatially uniform baseline, in which the seven free mean free path coefficients $\xi_4$--$\xi_{10}$ are replaced by a single global value $\xi_g$ while the twelve injection efficiencies are held at the RUN1 best-fit values. Columns 2--6 give the per-spacecraft MAE, defined as the mean over that spacecraft's energy channels; column 7 gives the weighted mean over the five spacecraft using the weights adopted in the optimization (unity for STEREO-A, SOHO, PSP and Solar Orbiter, 0.7 for BepiColombo); column 8 gives the total number of channel-time points that pass the evaluation mask. The grid is logarithmic with a half-decade spacing. The row in bold is the best global value. The last row repeats the 19-parameter best-fit solution of the RUN1 representative session for reference; \revsep{a re-evaluation of that solution reproduced the objective value recorded by the optimizer, 0.4290, exactly; the value of 0.458 listed here is the same solution scored with the Solar Orbiter EPT channel corrected, as described in the Objective Function subsection of the Methods.} A confirmatory run in which the three fixed far-eastern coefficients $\xi_1$--$\xi_3$ were also set to $\xi_g$ at the best global value returned identical per-spacecraft and total values and an identical $N$, confirming that those field lines do not contribute to the objective function. Note that $N$ varies across the grid because the evaluation mask depends on the simulated intensity; the poorer fits have more points removed, so the degradation relative to the 19-parameter solution is if anything understated. \label{tab:global_xi_sweep}}}
\begin{tabular}{lcccccc r}
\hline
$\xi_g$ & STEREO-A & SOHO & BepiColombo & PSP & Solar Orbiter & Weighted total & $N$ \\
\hline
$10^{2.0}$ & \revsep{0.878} & \revsep{1.151} & \revsep{1.647} & \revsep{0.499} & \revsep{1.336} & \revsep{1.068} & \revsep{9538} \\
$10^{2.5}$ & \revsep{0.696} & \revsep{0.843} & \revsep{1.502} & \revsep{0.510} & \revsep{1.028} & \revsep{0.878} & \revsep{10129} \\
$10^{3.0}$ & \revsep{0.644} & \revsep{0.520} & \revsep{1.211} & \revsep{0.772} & \revsep{0.678} & \revsep{0.737} & \revsep{10171} \\
\revsep{\textbf{$10^{3.5}$}} & \revsep{\textbf{0.730}} & \revsep{\textbf{0.427}} & \revsep{\textbf{0.912}} & \revsep{\textbf{0.922}} & \revsep{\textbf{0.584}} & \revsep{\textbf{0.703}} & \revsep{9975} \\
$10^{4.0}$ & \revsep{0.863} & \revsep{0.412} & \revsep{0.659} & \revsep{1.010} & \revsep{0.600} & \revsep{0.712} & \revsep{9857} \\
$10^{4.5}$ & \revsep{0.936} & \revsep{0.405} & \revsep{0.561} & \revsep{1.039} & \revsep{0.624} & \revsep{0.723} & \revsep{9820} \\
$10^{5.0}$ & \revsep{0.965} & \revsep{0.402} & \revsep{0.521} & \revsep{1.058} & \revsep{0.627} & \revsep{0.727} & \revsep{9808} \\
$10^{5.5}$ & \revsep{0.974} & \revsep{0.400} & \revsep{0.504} & \revsep{1.064} & \revsep{0.629} & \revsep{0.728} & \revsep{9803} \\
$10^{6.0}$ & \revsep{0.976} & \revsep{0.400} & \revsep{0.499} & \revsep{1.064} & \revsep{0.629} & \revsep{0.728} & \revsep{9801} \\
\hline
19-parameter best & \revsep{0.399} & \revsep{0.401} & \revsep{0.505} & \revsep{0.454} & \revsep{0.545} & \revsep{0.458} & \revsep{10480} \\
\hline
\end{tabular}
\end{table}